\documentclass[twocolumn]{aastex7}

\usepackage{newtxtext,newtxmath}
\usepackage[T1]{fontenc}

\usepackage{graphicx}	
\usepackage{amsmath}	

\usepackage{graphicx}
\usepackage{amsmath}
\usepackage{subfigure}
\usepackage{multirow}
\usepackage{threeparttable}
\usepackage{array}
\usepackage{natbib}
\usepackage{xcolor}
\usepackage{hyperref}
\usepackage[version=4]{mhchem}
\usepackage{bm}
\usepackage{ulem}

\newcommand{\msun}{\mbox{${\rm M}_\odot$}}
\newcommand{\mstar}{\mbox{${M}_{\rm star}$}}
\newcommand{\cmjj}{\mbox{${\rm cm^{-2}}$}}
\newcommand{\cmjjj}{\mbox{${\rm cm^{-3}}$}}
\newcommand{\lya}{\mbox{${\rm Ly}\alpha$}}

\newcommand{\kms}{{\rm{km~ s}^{-1}}}

\newcommand{\dproj}{\mbox{$d_{\rm proj}$}}
\newcommand{\rvir}{\mbox{$r_{\rm vir}$}}

\newcommand{\nh}{\mbox{${n}_{\rm H}$}}
\newcommand{\Nh}{\mbox{${N}_{\rm H}$}}

\newcommand{\Nhi}{\mbox{${N}_{\rm HI}$}}

\newcommand{\bnt}{\mbox{${b}_{\rm NT}$}}
\newcommand{\Tpie}{\mbox{${T}_{\rm PIE}$}}

\defcitealias{Werk2013}{Werk13}
\defcitealias{Chen2023}{Chen23}

\begin{document}

\title{Kinematics of \ion{H}{1} and \ion{O}{6} Absorbers: Insights into the Turbulence Driver of the Multiphase Circumgalactic Medium}

\correspondingauthor{Zhijie Qu}
\email{quzhijie@tsinghua.edu.cn}

\author[0000-0002-2941-646X]{Zhijie Qu}
\affiliation{Department of Astronomy, Tsinghua University, Beĳing 100084, People’s Republic of China}
\affiliation{Department of Astronomy \& Astrophysics, The University of Chicago, 5640 S. Ellis Ave., Chicago, IL 60637, USA}
\email{quzhijie@tsinghua.edu.cn}

\author[0000-0001-8813-4182]{Hsiao-Wen Chen}
\affiliation{Department of Astronomy \& Astrophysics, The University of Chicago, 5640 S. Ellis Ave., Chicago, IL 60637, USA}
\email{hwchen@uchicago.edu}

\author[0009-0004-5840-1816]{Eliana Schiller}
\affiliation{Department of Astronomy \& Astrophysics, The University of Chicago, 5640 S. Ellis Ave., Chicago, IL 60637, USA}
\email{}

\author[0000-0002-6593-8820]{Jing Wang}
\affiliation{Kavli Institute for Astronomy and Astrophysics, Peking University, Beijing 100871, China}
\email{}

\author[0000-0003-2491-060X]{Max Gronke}
\affiliation{Zentrum für Astronomie der Universität Heidelberg, Astronomisches Rechen-Institut, Mönchhofstr. 12-14, 69120 Heidelberg}
\email{}



\begin{abstract}
We investigate large-scale gas kinematics in the multiphase circumgalactic medium (CGM) using the observed correlation between line width (Doppler $b$ parameter) and column density ($N$) for \ion{H}{1} and \ion{O}{6} absorbers.  Leveraging extensive public galaxy survey data at $z\lesssim0.1$, we construct a new galaxy sample based on the availability of background Quasi-Stellar Objects (QSOs) with far-ultraviolet spectra from the Far Ultraviolet Spectroscopic Explorer (FUSE). 
By combining this FUSE–galaxy sample with literature collections, we find that \ion{H}{1} absorbers exhibit a clear inverse correlation between Doppler width and column density over nearly five orders of magnitude in $N_{\rm HI}$,  from $N_{\rm HI} \approx 10^{13}\,\mathrm{cm^{-2}}$ to $N_{\rm HI} \approx 10^{18}\,\mathrm{cm^{-2}}$, while \ion{O}{6} absorption follows a positive correlation across $N_{\rm OVI}\approx 3\times10^{13}$–$10^{15}\,\mathrm{cm^{-2}}$.  We develop a model framework to interpret these contrasting trends and show that \ion{H}{1} absorbers are best described as systems of approximately constant total column density ($N_{\rm H}$), whereas \ion{O}{6} traces regions of roughly constant spatial density ($n_{\rm H}$ and $n_{\rm OVI}$).  Under the latter scenario, the observed $b_{\rm OVI}$–$N_{\rm OVI}$ relation maps directly to a velocity–size relation consistent with a Kolmogorov-like turbulent spectrum.  Together, these findings reveal a coherent physical picture in which \ion{H}{1} and \ion{O}{6} trace a continuous turbulent cascade spanning more than five orders of magnitude in spatial scale—from cool, photoionized clumps to warm, highly ionized halo gas---with accretion in the halo outskirts likely driving the turbulent energy injection that sustains the multiphase CGM.
\end{abstract}


\keywords{surveys -- galaxies: halos -- intergalactic medium -- quasars: absorption lines}



\section{Introduction}
The circumgalactic medium (CGM) is a dynamic gaseous reservoir that hosts gas inflows and outflows, regulating galaxy growth.
The CGM connects the interstellar medium (ISM) in galaxies with the intergalactic medium (IGM) on cosmological scales, preserving a record of accreted material and metal-enriched outflows \cite[e.g.,][]{Putman2012, Tumlinson2017}. 
In particular, thermodynamics plays a central role in shaping the physical conditions of the CGM, and determining the fate of the gas \citep[e.g.,][]{CAFG2023}.
Key parameters such as density, temperature, pressure, and non-thermal processes determine the ionization state of the gas and influence its interaction with galactic and intergalactic environments \citep[e.g.,][]{Chen2025}.

Turbulence and large-scale kinematics (e.g., bulk velocity field) play a pivotal role in redistributing energy, modifying thermodynamic states, and mixing multiphase gas in the circumgalactic space \citep[e.g.,][]{Ji2019, Tan2021, Gronke2022, Lv2024}.
These processes collectively shape the global properties of the CGM, including the spatial distribution of the multiphase gas and associated kinematics \cite[e.g.,][]{Voit2018, Buie2020,  Schmidt2021, Koplitz2023}.

Wide-field Integral Field Spectrographs (IFSs) provide an essential observational window into the cool ionized halo gas. These data enable the construction of velocity structure functions directly from observed velocity fields, offering a robust characterization of turbulence \citep[e.g.,][]{Li2020}. Yet, two barriers remain. First, limited surface-brightness sensitivity biases kinematic measurements toward the densest, highest-emissivity clumps \citep[e.g.,][]{Chen2019, Li2020, Li2023} or regions boosted by QSO fluorescence \citep[e.g.,][]{ChenC2023, Johnson2024}. Second, atmospheric seeing of $1"$ limits the effective resolution to $\gtrsim 5$ kpc at $z > 0.4$ \citep[see discussion in][]{ChenC2023}, obscuring the critical small-scale kinematics ($\lesssim 10$ kpc) driving the turbulent cascade.

High-spectral-resolution absorption spectroscopy offers a powerful means of resolving non-thermal motions in the absorbing gas \citep[e.g.,][]{Rauch1996, Rudie2019, Qu2022}. 
Species with different atomic masses exhibit different dependencies on thermal and non-thermal motions, making it possible to decompose their respective contributions to the observed linewidths.
Specifically, combining linewidth measurements of low-ionization species and \ion{H}{1} yields constraints for the non-thermal motions within cool clouds at small scales between $\sim 1$ pc and $\lesssim 1$ kpc, recovering a $1/3$ power-law slope between the observed turbulent velocity and clump size, which is consistent with a subsonic Kolmogorov turbulence \citep[][hereafter \citetalias{Chen2023}]{Chen2023}.
This approach has significantly advanced the understanding of small-scale turbulence, but constraints on large-scale turbulence and kinematics ($\sim 10$--100 kpc) in the CGM around normal galaxies are still lacking.

The multiphase CGM offers a unique opportunity to probe gas kinematics across a wide range of physical scales using absorption spectroscopy. In particular, the warm CGM at $\log T/{\rm K}\!\approx\!5$--5.5 plays a central role in regulating gas exchange---both accretion and feedback---between galaxies and the IGM. This phase is especially important because its radiative cooling timescale is short ($\sim\!100$ Myr for gas with density $\nh\sim10^{-4}~\cmjjj$ and metallicity $Z\sim0.3\,Z_\odot$; see e.g., \citealt{Sutherland1993}), substantially shorter than the characteristic dynamical time of the CGM ($\sim\!1$ Gyr; \citealt{White1978, Stern2020}).
In addition, species with high ionization potentials (e.g., \ion{O}{6}) probe large-scale kinematics in the CGM due to their sensitivity to relatively low densities \citep[e.g.,][]{spitzer1956, Sutherland1993} and their large physical sizes compared to low-ionization species \citep[e.g.,][]{savage2002}.
In this work, we explore the possibility of using the observed kinematics of single species, such as \ion{H}{1} and \ion{O}{6}, as tracers of the velocity field on large scales ($\gtrsim 1$ kpc) in the CGM of normal galaxies.
Throughout this study, we adopt a flat $\Lambda$ cosmology with $\Omega_{\rm M} = 0.3$, $\Omega_\Lambda = 0.7$, and $H_0 = 70 ~\kms~{\rm Mpc^{-1}}$.

\section{Data Reduction and Analysis}
\label{sec:data}

In this work, we leverage extensive low-redshift galaxy survey data from public archives at $z \lesssim 0.1$ to construct a sample of galaxies with suitable background QSOs. 
The selection is based on the availability of far-ultraviolet (FUV) absorption spectra of the QSOs obtained with the Far Ultraviolet Spectroscopic Explorer \citep[FUSE;][]{FUSE}, which provides wavelength coverage from 905 to 1187 Å with a full-width-at-half-maximum velocity resolution of ${\rm FWHM}\approx 15\,\kms$.
The focus on available FUSE spectra is driven by the need for observing the \ion{O}{6}\,$\lambda\lambda\,1031, 1037$ doublet at these low redshifts.
To improve the significance of analyses, we also include available literature galaxy-QSO samples.
The adopted data, reduction methods, and analysis are summarized in this section.

\subsection{The galaxy sample}
\label{sec:fusegal}
To construct the FUSE-galaxy sample for CGM studies, we first collect FUSE sightlines in four archival categories, including QSO, Seyfert, BL Lacertae, and emission galaxies.
This approach leads to a total of 275 FUSE sightlines.
Because most of these background sources are active galactic nuclei (AGN), we use AGN to refer to all of them.
Next, we exclude sightlines with low signal-to-noise (S/N $<2$) continuum or with significant geocoronal contaminating lines, reducing the sample to 204 FUSE archival sightlines.

We adopt a two-step approach to establish the FUSE AGN-galaxy sample by cross-matching the FUSE archival sightlines with galaxies in the SIMBAD archive \citep[][before January 2023]{Wenger2000}, 
First, we select the FUSE sightlines with known $z<0.13$ galaxies at projected distances less than 100 kpc, which leads to a sample of 31 FUSE archival sightlines summarized in Table \ref{tab:agn}.
In this table, we also present the cross-match result with the {\it Hubble Space Telescope} ({\it HST}) archive. Of the 31 FUSE AGN sightlines, 26 have associated FUV spectroscopic observations from {\it HST} providing additional spectral coverage for the associated Ly$\alpha$ transition.
In total, 41 galaxies at $z<0.13$ are found to occur within 100 kpc from the background AGN.

\begin{table*}[]
    \centering
    \caption{Summary of the FUSE and {\it HST} AGN sightlines}
    \label{tab:agn}
    \begin{tabular}{lcccrcr}
\hline
\hline
QSO            & RA           & DEC          & $z_{\rm QSO}$ & $\rm S/N_{FUSE}$ $^a$ & {\it HST}  & $\rm S/N_{HST}$ $^a$ \\
\hline
MRK 335         & $00:06:19.53 $ & $+20:12:10.3 $ & $0.025$ & $ 25.6$ & COS  & $ 32.6$ \\
PG 0026+129     & $00:29:13.81 $ & $+13:16:04.5 $ & $0.142$ & $  5.1$ & COS  & $ 17.1$ \\
PKS 0405-123    & $04:07:48.43 $ & $-12:11:36.7 $ & $0.573$ & $ 16.3$ & COS  & $ 44.3$ \\
PG 0832+251     & $08:35:36.00 $ & $+24:59:43.8 $ & $0.330$ & $  3.1$ & COS  & $ 12.0$ \\
PG 0838+770     & $08:44:45.26 $ & $+76:53:10.0 $ & $0.131$ & $  8.7$ & COS  & $ 21.9$ \\
3C 232          & $09:58:20.95 $ & $+32:24:02.4 $ & $0.530$ & $  1.8$ & COS  & $  8.9$ \\
PG 1004+130     & $10:07:26.10 $ & $+12:48:55.9 $ & $0.241$ & $  7.8$ & COS  & $  9.7$ \\
MRK 141         & $10:19:12.59 $ & $+63:58:02.7 $ & $0.042$ & $  2.7$ & $...$ & $...$ \\
PG 1048+342     & $10:51:43.86 $ & $+33:59:26.6 $ & $0.167$ & $  4.0$ & COS  & $ 19.3$ \\
HE 1115-1735    & $11:18:10.70 $ & $-17:52:00.0 $ & $0.216$ & $  5.9$ & $...$ & $...$ \\
MRK 734         & $11:21:47.11 $ & $+11:44:18.5 $ & $0.050$ & $  5.5$ & $...$ & $...$ \\
3C 263          & $11:39:57.04 $ & $+65:47:49.4 $ & $0.652$ & $ 12.7$ & COS  & $ 30.5$ \\
PG 1211+143     & $12:14:17.61 $ & $+14:03:12.7 $ & $0.081$ & $ 20.4$ & COS  & $ 20.1$ \\
TON 1480        & $12:15:09.21 $ & $+33:09:55.2 $ & $0.616$ & $  4.6$ & $...$ & $...$ \\
PG 1216+069     & $12:19:21.06 $ & $+06:38:38.6 $ & $0.331$ & $  5.2$ & COS  & $ 21.0$ \\
MRK 205         & $12:21:44.04 $ & $+75:18:38.3 $ & $0.071$ & $ 14.5$ & STIS & $ 14.7$ \\
3C 273          & $12:29:06.71 $ & $+02:03:08.9 $ & $0.158$ & $ 40.5$ & COS  & $ 72.1$ \\
QSO J1230+0115  & $12:30:50.00 $ & $+01:15:22.7 $ & $0.117$ & $  6.1$ & COS  & $ 47.7$ \\
PG 1229+204     & $12:32:03.62 $ & $+20:09:29.4 $ & $0.064$ & $  5.6$ & COS  & $ 16.8$ \\
PG 1259+593     & $13:01:12.92 $ & $+59:02:06.6 $ & $0.478$ & $ 32.9$ & COS  & $ 31.1$ \\
PKS 1302-102    & $13:05:33.02 $ & $-10:33:19.3 $ & $0.278$ & $ 16.6$ & COS  & $ 26.1$ \\
PG 1307+085     & $13:09:47.04 $ & $+08:19:49.5 $ & $0.154$ & $  9.2$ & COS  & $ 20.4$ \\
PG 1309+355     & $13:12:17.74 $ & $+35:15:20.6 $ & $0.183$ & $  3.0$ & COS  & $ 14.3$ \\
PG 1352+183     & $13:54:35.66 $ & $+18:05:17.5 $ & $0.151$ & $  3.0$ & COS  & $ 23.4$ \\
PG 1543+489     & $15:45:30.24 $ & $+48:46:08.9 $ & $0.400$ & $  3.5$ & STIS & $ 27.0$ \\
PG 1626+555     & $16:27:56.09 $ & $+55:22:32.0 $ & $0.133$ & $ 13.8$ & COS  & $ 24.6$ \\
3C 351          & $17:04:41.37 $ & $+60:44:30.5 $ & $0.372$ & $  4.9$ & STIS & $ 18.0$ \\
LEDA 63618      & $19:45:00.53 $ & $-54:15:03.0 $ & $0.019$ & $  4.6$ & $...$ & $...$ \\
PKS 2135-14     & $21:37:45.20 $ & $-14:32:55.0 $ & $0.200$ & $  2.9$ & COS  & $ 19.8$ \\
PHL 1811        & $21:55:01.50 $ & $-09:22:25.0 $ & $0.194$ & $ 10.6$ & COS  & $ 33.7$ \\
HE 2336-5540    & $23:39:13.27 $ & $-55:23:50.4 $ & $1.355$ & $  2.7$ & COS  & $ 12.8$ \\
\hline
\multicolumn{7}{l}{$^a$ The median signal-to-noise (SN) per resolution element.} \\
    \end{tabular}

\end{table*}

We further identify galaxy groups based on their close associations both in redshift with a corresponding velocity difference $\Delta\varv \lesssim 300~\kms$) and in projected distances with $\dproj \lesssim 200$ kpc.  This leads to 35 unique galaxies or galaxy groups with a FUSE sightline occurring at $< 100$ kpc from at least one of the galaxies.
While the galaxy environment may affect the observed CGM properties \citep[e.g.,][]{Johnson2015, Burchett:2016}, recent studies have also found that the observed CGM properties correlate most strongly with the galaxies found closest in projected distance to the absorber \citep[e.g.,][]{Qu2023}.  In the subsequent analysis, we associate the detected absorption features with the galaxies at the smallest $\dproj/\rvir$ for galaxy groups, where $\rvir$ is the virial radius of the host dark matter halo. 
This approach is guided by the findings for \ion{O}{6} absorbers by \citet{CUBSVII}.

To verify that the galaxies in our sample are indeed those at the smallest $\dproj/\rvir$ in each unique system, we search the public archives for luminous (and likely massive) galaxies projected out to 300 kpc in the selected 31 FUSE AGN fields.  These additional galaxies may potentially be the host galaxies that dominate the dark matter potential of their lower-mass neighbors at smaller \dproj. 
This effort has uncovered only one such case, where a low-mass dwarf ($\log \mstar/\msun \approx 7$) and a sub-$L^*$ galaxy ($\log \mstar/\msun \approx 10$) are found at $\dproj=80$ and 170 kpc, respectively.  We attribute the absorption features to the sub-$L^*$ galaxy rather than the dwarf, based on the smallest $\dproj/\rvir$ criterion.
In addition, we also find 56 isolated galaxies or galaxy groups without any members projected within 100 kpc.  These galaxies help expand the absorption probe to the outskirts of the CGM.
In total, our galaxy sample comprises 91 isolated galaxies or galaxy groups at $z<0.13$ and $\dproj \lesssim 300$ kpc from a FUSE AGN sightline.  

A summary of the galaxy and AGN pairs in our sample is presented in Table \ref{tab:sample}, which lists in columns (1) through (7) the galaxy ID, right ascension and declination coordinates, redshift, luminosity distance, the associated AGN sightline, and projected distance.  
Note that for the galaxies at $z<0.03$ (corresponding to a maximum distance of $\sim 100$ Mpc), we estimate the distance using the cosmic-flow3 model\footnote{https://edd.ifa.hawaii.edu/CF3calculator/} \citep{Tully2016, CF3_edd}, where $H_0$ is scaled to $70~\kms~\rm Mpc^{-1}$.
For $z > 0.03$ galaxies, we calculate the luminosity distance from redshift assuming the Hubble flow model.
To maintain a focus on the kinematics of the absorption features, we only report the distance information for the sample galaxies here.
Additional galaxy properties, including optical colors, $\mstar$, and star formation rate, have also been determined using available archival imaging data, including the Legacy Survey, Pan-STARRS, and Sloan Digital Sky Survey \citep[SDSS;][]{Legacy_survey, PS1, SDSS-IV}.  Stellar masses of these galaxies range between $\log \mstar/\msun \approx 8$ and 11, which will be presented in a separate paper (Schiller et al.\ in prep.).

\subsection{Absorption Spectroscopy}
\label{sec:abs}

Our AGN sample contains 31 sightlines, all of which are selected to have useful FUSE spectra and 26 of which have additional {\it HST} FUV spectra obtained using the Space Telescope Imaging Spectrograph \citep[STIS;][]{STIS} or the Cosmic Origins Spectrograph \citep[COS;][]{COS} to extend the wavelength coverage beyond $1187$ \AA.
These FUV spectra are processed and co-added using custom software to optimize the signal-to-noise ratio (S/N) and the wavelength calibration (see Appendix).
The median S/N of individual coadded FUSE, {\it HST}/COS, and STIS spectra is summarized in Table \ref{tab:agn}.
The final combined FUSE spectra exhibit a median S/N of $\approx 6$ per resolution element, while the median S/N of all {\it HST} spectra is $\approx 20$.

For each galaxy-QSO pair, absorption features are searched within a velocity window of $\pm 600~ \kms$ around the systemic redshift of the target galaxies.
This velocity threshold is about twice the virial velocity of the target galaxies, ensuring that no absorption features will be missed.  This is also supported by previous studies, which show that the bulk velocity of the absorbing gas exhibits a standard deviation of $\lesssim 150~ \kms$ \citep[e.g.,][]{Chen2010, Tumlinson2013, Huang2021, CUBSVII}.

To characterize the observed absorption features, we perform a multi-component decomposition using Voigt profiles to derive the column density ($N$), Doppler parameter ($b$), and velocity centroid ($\varv_{\rm c}$) of each kinematic component.
The number of fitted \ion{H}{1} components relies on a global inspection of all available low-ionization transitions, mainly the Lyman series, together with the metal lines of \ion{C}{2}-{\small III} and \ion{Si}{2}-{\small III}.
The decomposition of \ion{O}{6} is considered together with \ion{C}{4} and \ion{Si}{4}, which are in the hotter phase than \ion{H}{1} \citep[e.g.][]{Werk2016, Zahedy2019}.
Here, we adopt a minimum number of absorption components required to explain the structure in the line profiles beyond what a simple Gaussian can describe.
For components detected at a significance level of $>3 \sigma$, we report the median values and their associated $1\sigma$ uncertainties derived using a Bayesian framework described in \citet[][see also \citealt{Zahedy2021}]{Qu2022}. 
Meanwhile, saturated features result in 2-$\sigma$ lower limits for the column densities. 
In these cases, an upper bound of $\log N/\cmjj \approx 18-18.5$ is also inferred based on the absence of damping wings in most cases.

For non-detected transitions, we determine $2\sigma$ upper limits for the column density, assuming a fiducial Doppler parameter of $b = 30~\kms$, consistent with the typical broadening measured for \lya\ and \ion{O}{6} \citep[e.g.,][]{Danforth2016}.
For galaxies with neither \ion{H}{1} nor \ion{O}{6} detected, the line centroid is fixed to the galaxy systemic redshift.  For galaxies with detected \ion{H}{1} but no \ion{O}{6}, the line centroid is aligned with the velocity of the \ion{H}{1} line(s).
All measurements of individual components and non-detections are presented in Table \ref{tab:idv}.
In this study, we only adopted measurements with constrained Doppler $b$ parameters in the following kinematic analysis.

In addition to the properties of individual components, we also compute the line-of-sight (LOS) integrated quantities, including total column density (\(N_{\mathrm{tot}}\)) and velocity dispersion (\(\sigma_\varv\)). 
These values are derived using the formalism described in \citet{CUBSVII}, which accounts for the contributions of all detected components and their uncertainties:
\begin{eqnarray}
\sigma_{\varv}^2 = \frac{\sum N(\varv)(\varv-\varv_c)^2}{N_{\rm tot}}, 
\label{eq:sigmav}
\end{eqnarray}
where $N(\varv)$ is the column density of the modeled Voigt profile in each velocity bin, $\varv_{\rm c} = \frac{\sum N(\varv)\varv}{N}$ is the line centroid.
The LOS properties are summarized in columns (8) through (11) of Table \ref{tab:sample}.

\subsection{Literature samples}
\label{sec:literature}

In addition to our galaxy sample, we incorporate literature samples with absorption measurements of \ion{H}{1} and \ion{O}{6} to enhance the statistical significance of the following analysis.
These literature samples are selected to have component-resolved Voigt profile fits for \ion{H}{1} or \ion{O}{6}, ensuring compatibility with our kinematic study.
To maintain consistency in decomposition methodology, we visually inspect the profile fitting in these studies and evaluate their alignment with our multi-component Voigt profile analysis.
While most sightlines exhibit reasonable agreement with our decomposition routine, we re-measure a subset of sightlines, where discrepancies in component identification or parameterization could introduce systematic biases.
Examples of these cases include asymmetric absorption features or different noise characteristics, which necessitate revised component structures compared to original literature reports.

We first consider three samples with resolved \ion{H}{1} component measurements, which are from the COS-Halos survey \citep{Tumlinson2013}, the COS-LRG survey \citep{Chen2018, Zahedy2019}, and the CUBS program \citep{Chen2020}.
For the COS-Halos sample, we re-measure the column density $N$ and Doppler $b$ parameter for individual Voigt components to fully utilize all available data, including new spectra covering $1100-1150$ \AA~ obtained after \citet{Tumlinson2013}.
For the COS-LRG and CUBS programs, we adopt the reported column densities and Doppler parameters of individual components from \citet[][]{Qu2022}, which employ the same data reduction and analysis methodology as this work.

For \ion{O}{6} absorption, we also incorporate data from the COS-Halos survey \citep{Werk2013}, the COS-LRG survey \citep{Zahedy2019}, and the CUBS program \citep[][CUBS VII]{CUBSVII}.
For the CUBS survey, we adopt the \ion{O}{6} component measurements, as the analysis methodology is identical to ours.
For the COS-LRG galaxies, we visually inspect the Voigt-profile fits and confirm that the component-decomposition criteria are consistent with our approach.
For the COS-Halos sample, we re-analyze the \ion{O}{6} profiles, adopting the minimum number of components required in the decomposition to capture structure in the absorption profile. 
In most cases, our measurements agree with \citet{Werk2013}, but discrepancies occur in the number of identified components for certain sightlines.
For example, in the system at $z = 0.2270$ toward LBQS\,1340$-$0038, \citet{Werk2013} identified a single broad \ion{O}{6} component, whereas our decomposition reveals three narrower components to account for asymmetric features in the \ion{O}{6} doublet.
Therefore, our analysis yields narrower, kinematically distinct components for a few sightlines compared to the broader components reported in \citet{Werk2013}.

\begin{figure*}
    \centering
    \includegraphics[width=0.98\textwidth]{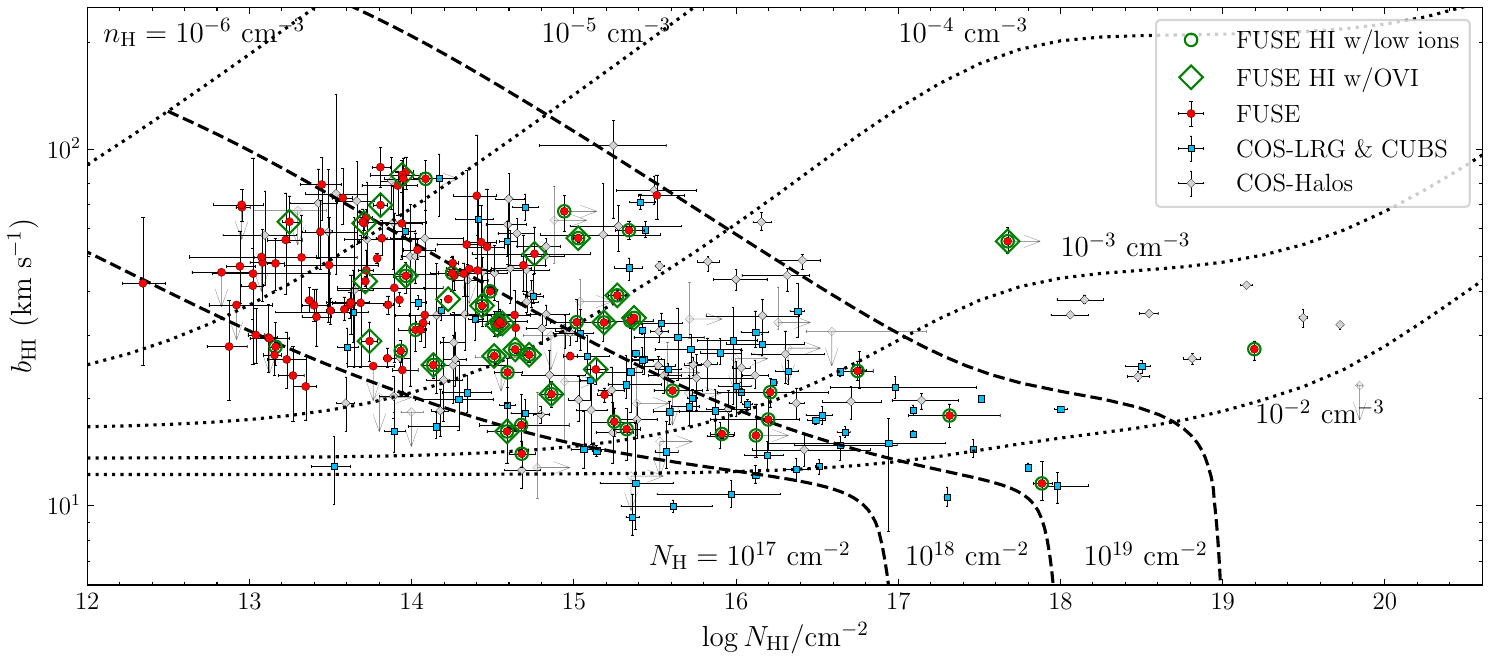}
    \caption{The $N$-$b$ relation for individual \ion{H}{1} components, showing a 8.0 $\sigma$ negative correlation.
    The adopted samples include all \ion{H}{1} components in the FUSE sample (this work, circles), COS-Halos \citep[][diamonds]{Werk2013}, COS-LRG \citep[][squares]{Chen2018, Zahedy2019}, and CUBS compiled in \citet{Qu2022}, which are summarized in Section \ref{sec:data}.
    For the FUSE sample, we also mark the \ion{H}{1} absorption components with corresponding low-ionization state metals (e.g., \ion{C}{2}-{\footnotesize III} and \ion{Si}{2}) using large circles and \ion{O}{6} using large diamonds.} 
    A kinematical model is presented as dashed (constant-$N_{\rm H}$) and dotted lines (constant-$n_{\rm H}$; see details in Section \ref{sec:h1}).
    Except for those systems with $\Nhi\gtrsim 10^{18}~\cmjj$, the majority of the CGM absorbers can be explained by our kinematic model in the range of $\log \nh/\cmjjj \approx -2$ to $-5$ and $\log \Nh/\cmjj \approx 17$-19.
    \label{fig:h1_Nb}
\end{figure*}

\section{Large-scale gas kinematics in \ion{H}{1} and \ion{O}{6}}

Combining our galaxy sample and the literature samples, here we examine the large-scale CGM kinematics traced by the \ion{H}{1} and \ion{O}{6} absorption transitions through their observed $N$-$b$ relation.
We first investigate the line broadening of individual \ion{H}{1} components under the assumption of photoionization equilibrium (PIE).
For \ion{O}{6}, we analyze both the line widths of individual components and the velocity dispersions among components residing in the same halos along each sightline. 

\subsection{Kinematics traced by individual \ion{H}{1} components}
\label{sec:h1}
\ion{H}{1} is the most sensitive tracer for cool gas, because of its abundance and large cross section (e.g., \citealt{Chen2001, Rudie2012, Tumlinson2013, Borthakur2015, Borthakur2024}).
Here, we investigate the kinematics in the CGM by exploring the $N_{\rm HI}-b_{\rm HI}$ relation.
This relation has been adopted to characterize the physical condition in the intergalactic medium (IGM), which is mainly probed by low-column-density absorbers with $\log N_{\rm HI}/\cmjj \lesssim 13.5$, whereas higher-column-density absorbers are primarily associated with collapsed halos \citep[e.g.,][]{Hui1997, Dave:1999, Schaye1999, Kim2002}.
The low-column IGM absorbers mainly show a positive correlation between column densities and Doppler $b$ parameters \citep[e.g.,][]{Kirkman1997, Dave2001}.
However, the \ion{H}{1} sample compiled in this study is dominated by high-column components as shown in Figure \ref{fig:h1_Nb}, which is expected because these absorbers are found in targeted searches in the vicinities of known galaxies.

Overall, a 8.0 $\sigma$ inverse-correlation is observed between $\log N_{\rm HI}$ and $b_{\rm HI}$ using a generalized Kendall $\tau$ test ($\tau=-0.307$ and $p=1.4\times 10^{-15}$).
This trend is driven by the components at $\log N_{\rm HI}/\cmjj \approx 13$--17.
There are also eleven components with $\log N_{\rm HI}/\cmjj \gtrsim 18$ that exhibit relatively broad width $b_{\rm HI} \gtrsim 20~ \kms$ and lie above the mean inverse correlation. 
These components are examined individually, and most of these strong absorption systems at $z\lesssim 0.2$ either do not exhibit sufficiently high-order Lyman series coverage to decompose the major components (see examples of decomposed $z\approx 0.5$ Lyman limit systems in \citealt{Zahedy2019}) or the available FUSE spectra do not offer sufficient S/N to resolve high-order Lyman series lines.

Here, we develop a kinematic model to reproduce the observed relation between $N_{\rm HI}$ and $b_{\rm HI}$ established from observations at $\log N_{\rm HI}/\cmjj \lesssim 17$.
The baseline assumption is such that the low column density gas is photoionized by the ultraviolet background (UVB), and that the observed line widths are driven by a combination of thermal and non-thermal broadening.

To model the thermal contribution, we compute PIE solutions for a range of gas density ($\nh$) and total hydrogen column ($\Nh$) using \texttt{Cloudy} \citep[version C23;][]{Chatzikos2023cloudy}.
The adopted UVB follows the \citet{FG20} model at $z = 0.1$, representative of the median redshift of the observational sample.
The systematic uncertainties between different UVB prescriptions are approximately 0.2 dex (\citealt{HM12, KS19}), comparable to the UVB redshift evolution of $0.2-0.3$ dex from $z=0.4$ to $0$ \citep{FG20}.
Therefore, we neglect redshift-dependent variations for simplicity.
Uncertainties in the UVB models are expected to result in an offset in the inferred lognH by up to 0.5 dex for a fixed $\Nh$ (see tests in \citealt{Zahedy2019}), which is likely contributing to the large scatter in the observed 
$\Nhi$-$b$ relation, but is insufficient to significantly alter the trend.

For the non-thermal contribution, we adopt the empirical turbulent velocity and size relation for cool CGM clumps from \citetalias{Chen2023}, which is Kolmogorov-like:
\begin{equation}
    \bnt = (18~\kms) (l/{\rm 1~kpc})^{1/3},
    \label{eqn:nt_motion}
\end{equation}
where the size $l$ is inferred from the \texttt{Cloudy} for a given pair of $N_{\rm HI}$ and $n_{\rm H}$.
Combining the inferred \Tpie\ and \bnt, we work out the predicted total line width following
\begin{equation}
    b_{\rm obs}^2 = \bnt ^2 + \frac{2k_B \Tpie}{m_i},
    \label{eq:1}
\end{equation}
where $m_i$ is the mass for a given element.

In Figure \ref{fig:h1_Nb}, we compare the expectations from our kinematic model with the observed $N_{\rm HI}$-$b_{\rm HI}$ relation.
The lines of constant-density models exhibit four major parts, driven by different mechanisms.
First, the equilibrium temperature $\Tpie$ sets the minimum $b_{\rm HI}$, varying from $\approx\!11~\kms$ to $23~\kms$ in the density range of $\log\,\nh/\cmjjj\!\approx\!-2$ to $-5$.
Then, $b_{\rm HI}$ increases as \bnt\ starts to dominate in larger clumps with higher $N_{\rm HI}$ in the optically thin regime, following a power-law slope of $\approx 1/3$. 
In the optically thick regime with $\log\,N_{\rm HI}/\cmjj\!\approx\!17$--19, self-shielding causes a rapid increase in $f_{\rm HI}$ with increasing $N_{\rm HI}$. This results in only a mild increase in $\Nh$ (and consequently in the size, $l$) for a fixed $n_{\rm H}$, producing a flattening of the constant-density curves.
At $\log N_{\rm HI}/\cmjj\!\gtrsim\!20$, the gas becomes predominantly neutral.  Further increases in $N_{\rm HI}$ correspond to larger physical path lengths ($l$), and $b_{\rm HI}$ once again increases with $\log\,N_{\rm HI}$ following a power-law trend similar to that seen in the low-$N_{\rm HI}$ regime.

In addition, we also show the constant-\Nh\ model curves in Figure \ref{fig:h1_Nb}, where $b_{\rm HI}$ decreases with increasing $N_{\rm HI}$.
This is understood from the expectation that a higher \nh\ is needed to produce an increased \Nhi\ while \Nh\ is held fixed, resulting in a reduced path length and a reduced \bnt. When $\Nhi$ approaches the total $\Nh$ in a predominantly neutral gas, the required $\nh$ can be high ($\log\,\nh/\cmjjj\!>\!0$), leading to a lower $\Tpie$ and consequently a sharp decline in the thermal line width and in the expected $b_{\rm HI}$ at $\Nhi\!=\!\Nh$.

Comparisons between observations and expectations from the kinematic model indicate that the observed $\Nhi$-$b_{\rm HI}$ relation is best reproduced by the constant-\Nh\ models.
In particular, a lower bound of $\log\,\Nh/\cmjj\!\approx\!17$ is consistently found across all samples \citepalias[see e.g.,][for discussions]{Chen2023}, which may reflect the characteristic cooling scale of multiphase gas \citep[see e.g.,][]{McCourt2018, Gronke2020, Liang:2020}.
We also identify a tentative upper bound at $\log \Nh/\cmjj \approx 18.5$--$19.0$, which may be explained by dynamical processes that suppress the survival of large coherent structures in the CGM environment (see e.g., \citealt{Keres2009}).
However, the physical explanations remain highly uncertain and require further investigation in future studies.
At the same time, some absorbers exhibit high column densities and broad line widths, $\log \Nhi/\cmjj \gtrsim 19$ and $b>20~ \kms$ in the inner halo.
These components may be unresolved multiple components under FUSE or {\it HST}/COS resolution and/or possibly trace the interface between the interstellar medium and CGM \citep[e.g.,][]{Wang2025}.

\begin{figure*}
    \centering
    \includegraphics[width=0.98\textwidth]{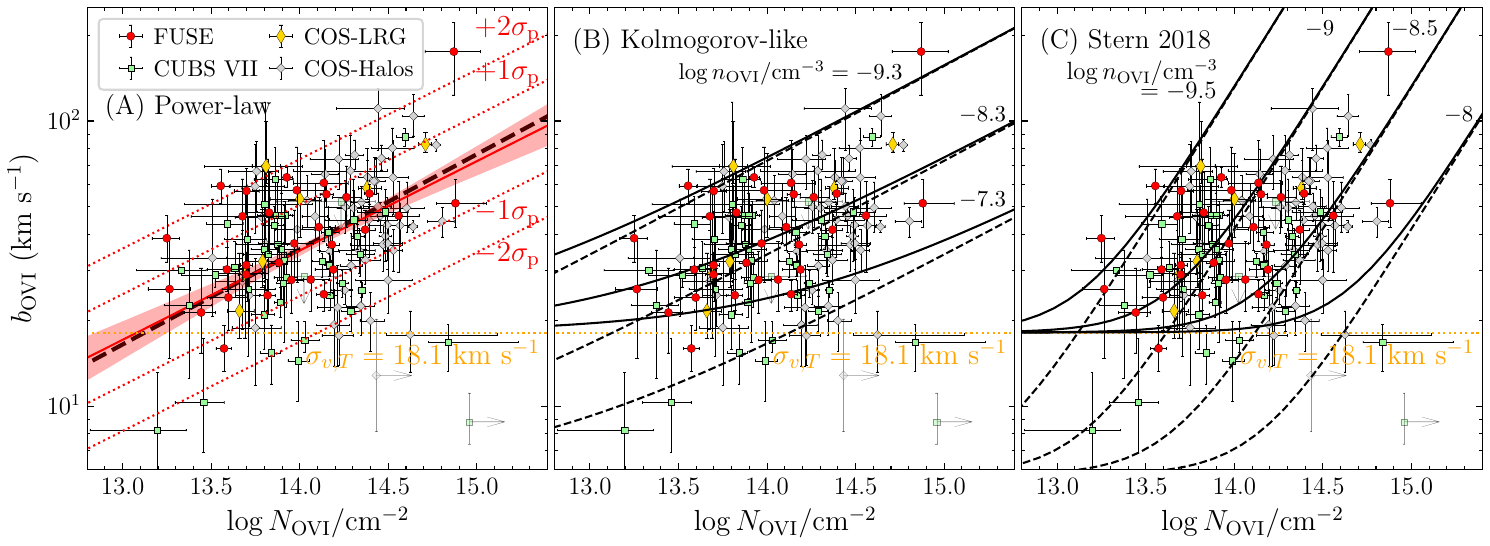}
    \caption{The $N$-$b$ relation for individual \ion{O}{6} components, showing a 5.1 $\sigma$ positive correlation.
    The adopted samples are from the FUSE sample (this work), COS-Halos \citep{Werk2013}, COS-LRG \citep{Zahedy2019}, and CUBS \citep{CUBSVII}.
    Panel (A): A power law fitting suggests a slope of $0.32\pm0.06$ and an intrinsic scatter of $0.16\pm 0.01$ dex.
    The dotted red lines show the 1\,$\sigma_{\rm p}$ and 2\,$\sigma_{\rm p}$ scatters associated with the best-fit model as the solid red line, while the dashed black line represents a power law model with a slope of $1/3$.
    Panel (B): The model-predicted $N$-$b$ relation assuming different $n_{\rm OVI}$ and two fiducial temperatures, $\log T/{\rm K} = 4.5$ under the assumption of PIE (dashed curves) and $\log T/{\rm K} = 5.5$ under the assumption of \ion{O}{6} originating in collisionally-ionized gas (solid curves). In addition, a Kolmogorov-like size-linewidth relation is adopted for modeling the non-thermal contribution to the observed $b_{\rm OVI}$ following a power-law slope of $1/3$ from \citetalias{Chen2023}. 
    Panel (C): Expectations from the \citet{Stern2018} model for \ion{O}{6}, illustrating the correlation between $N$ and $\bnt$. As in Panel (B), the solid curves represent collisionally ionized \ion{O}{6} with $\log T/{\rm K} = 5.5$, while the dashed curves correspond to photoionized gas with $\log T/{\rm K} = 4.5$. 
    In both panels (B) and (C), the flattening of the model curves toward low $b$ values reflects the diminishing contribution from non-thermal line width and the increasing dominance of thermal line widths determined by the gas temperature.
    The thermal contribution at $\log T/{\rm K}=5.5$ is shown as the horizontal dashed line.  For cooler temperatures, the expected thermal contribution to the \ion{O}{6} line width is lower.  We therefore interpret the observed $b_{\rm OVI}$ as driven primarily by non-thermal motions.}
    \label{fig:o6_Nb}
\end{figure*}

\subsection{Kinematics traced by individual \ion{O}{6} components}
\label{sec:o6}
In contrast to \ion{H}{1}, the $N_{\rm OVI}$-$b_{\rm OVI}$ relation exhibits a different trend.
Figure \ref{fig:o6_Nb} presents the measured $\log N_{\rm OVI}$ versus $b_{\rm OVI}$ for individual \ion{O}{6} components from the combined FUSE, CUBS \citep{CUBSVII}, COS-LRG \citep{Zahedy2019}, and COS-Halos samples \citep{Werk2013}.
The detected \ion{O}{6} components exhibit column densities varying in $\log N_{\rm OVI}\approx 13.5-15$, showing a $5.1 ~\sigma$ positive correlation with a generalized Kendall's rank coefficient of $\tau = 0.22$ and a random distribution probability of $p=4.0\times 10^{-7}$.

Here, we apply a linear fit to the observed $\log N$-$\log b$ relation for \ion{O}{6} under a Bayesian framework.
To account for measurement uncertainties in both $N$ and $b$, we adopt the likelihood formalism presented in \citet[][i.e., Equation 27 in Section 2.4]{Sharma2017}. 
This approach includes an intrinsic scatter ($\sigma_{\rm p}$) to capture physical variations in the model. It is determined simultaneously with the best-fit linear parameters: 
\begin{equation}
    \log(b_{\rm OVI}/\kms) = \alpha [\log(N_{\rm OVI}/\cmjj)-14] + C.
    \label{eqn:linear}
\end{equation}
We find that the data are best described by $\alpha = 0.32\pm0.06$, $C = 1.55\pm0.02$, and $\sigma_{\rm p} = 0.16\pm0.01$. 
The resulting relation is shown as the solid red line in Figure \ref{fig:o6_Nb}A, with the associated 1\,$\sigma$ uncertainties indicated by the red-shaded region and the 1\,$\sigma_{\rm p}$--2\,$\sigma_{\rm p}$ envelopes marked by the dotted lines.  For comparison, the expected thermal contribution at $\log T/{\rm K}=5.5$ is shown as the horizontal dashed line.  For cooler temperatures, the expected thermal contribution to the \ion{O}{6} line width is still lower.  We therefore interpret the observed $b_{\rm OVI}$ as driven primarily by non-thermal motions.

While the best-fit model reproduces the majority of the data points, a few narrow, strong \ion{O}{6} components appear as outliers in the lower-right portion of the diagram.
These strong features may arise in gaseous environments influenced by local ionizing radiation fields or non-equilibrium processes, being photoionized or exhibiting significant non-equilibrium ionization to produce strong \ion{O}{6} at a low temperature $T\lesssim 10^5$ K \citep[e.g.,][]{Gnat2017, Kumar2025}.

The observed relation between $N_{\rm OVI}$ and $b_{\rm OVI}$ can be interpreted in the context of two distinct physical scenarios: a Kolmogorov-like turbulent velocity field model and a radiative cooling flow model.
First, motivated by the observed positive $N$–$b$ correlation, we first model the \ion{O}{6} absorbers under a constant-density assumption with a Kolmogorov-like turbulent velocity field.  For optically-thin transitions, such a turbulent constant-density gas naturally produces a positive $N$–$\bnt$ correlation with a slope $\approx 1/3$, because the column density $N$ scales linearly with the path length $l$.  This differs from the constant-$N_{\rm H}$ assumption adopted to explain the inverse $N_{\rm HI}$-$b_{\rm HI}$ correlation shown in Figure \ref{fig:h1_Nb}.

Following the framework laid out for interpreting the observed \ion{H}{1} absorption properties, we employ Equation \ref{eqn:nt_motion} to characterize the non-thermal broadening.
In particular, the absorber size is calculated as $l = N_{\rm OVI}/n_{\rm OVI}\equiv N_{\rm OVI}/(n_{\rm H}\,Z_{\rm O}\,f_{\rm O^{5+}})$, where $Z_{\rm O}$ and $f_{\rm O^{5+}}$ are the oxygen abundance and ionization fraction of $\rm O^{5+}$, respectively.
For the thermal contribution, we consider two representative temperatures of $\log T/{\rm K} \approx 4.5$ and 5.5 corresponding to conditions expected for photo-ionized and collisionally-ionized gas, respectively. 
The photoionization and collisional ionization equilibrium conditions determine the expected ionization state (i.e., $n_{\rm H}$ and $T$), which in turn constrains \ion{O}{6} number densities, $n_{\rm OVI}$, for a given gas metallicity, $Z$, and $f_{\rm O^{5+}}$.  We find $\log n_{\rm H}/\cmjjj \approx -4.5$ to $-2.5$ and $\log n_{\rm OVI}/\cmjjj \approx -9.3$ to $-7.3$ (as shown in Figure \ref{fig:o6_Nb}B) for $Z\approx 0.3~Z_\odot$ and an ionization fraction of $f_{\rm OVI}\approx0.2$. The inferred $n_{\rm OVI}$ would be lower for gas of lower metallicity.

To reproduce the observed $\log N_{\rm OVI}$, however, the allowed range of $n_{\rm OVI}$ remains degenerate with the path length $l$, which is set by the turbulent velocity field.  Consequently, the observed \ion{O}{6} absorption properties alone do not uniquely constrain variations in the intrinsic turbulent velocity-size relation.  We will return to the implications of this limitation in Section \ref{sec:dis} below.
Nevertheless, this exercise shows that adopting a representative $\log n_{\rm OVI}/\cmjjj \approx -8.3$, individual \ion{O}{6} components with $\log N_{\rm OVI}/{\cmjj} \approx 14$ probe the kinematics on scales of $\sim 10$ kpc and longer path lengths for lower density gas or larger $N_{\rm OVI}$.

Next, we compare the observed $N$-$b$ relation with the \ion{O}{6} gaseous halo model of \citet{Stern2018}, which attributes the observed \ion{O}{6} absorption to either a high-pressure, collisionally ionized warm-hot phase ($\log T/{\rm K}\approx 5.5$) or a photoionized phase located beyond the accretion shock ($\log T/{\rm K}\approx 4.5$).
In these scenarios, the non-thermal broadening ($b_{\rm NT}$) arises from velocity shear within the absorbers driven by gravitational motions.
As shown in Figure \ref{fig:o6_Nb}C, such bulk flow models naturally produce a linear correlation between the column density and the velocity width $b_{\rm NT}$ for high column density components, while the thermal broadening dominates low column density components (\citealt{Stern2018}).

To assess the two scenarios, we analyze the residuals from comparing the observed $N_{\rm OVI}$ vs.\ $b_{\rm OVI}$ correlation with model expectations.
The residual is defined as $\Delta_\perp/\sigma_{\perp}$, where $\Delta_\perp$ represents the perpendicular distance of measurements from the median model line, and $\sigma_{\perp}$ is calculated as $\sigma_{\perp}^2 = (\alpha^2\sigma_{\log N}^2  + \sigma_{\log b}^2)/ (1+\alpha^2)$.
For both scenarios, the thermal broadening at $\log T/{\rm K} =4.5$ ($b=5.7~\kms$) is assumed to reproduce narrow line widths for weak components.
The resulting standard deviations of $\Delta_\perp/\sigma_{\perp}$ are 3.1, 2.5, and 3.4 for slopes of 0, 1/3, and 1, respectively.
By examining all potential slopes, the smallest scatter ($\Delta_\perp/\sigma_{\perp}$) occurs for a slope of $\alpha=1/3$, which is consistent with the best-fit power law model, favoring a Kolmogorov-like scenario.
This supports the assumption that \ion{O}{6} absorbers trace turbulent gas of constant density on halo scales.

\subsection{Inter-cloud kinematics traced by resolved \ion{O}{6} components}
\label{sec:o6s}

\begin{figure*}
    \centering
    \includegraphics[width=0.98\textwidth]{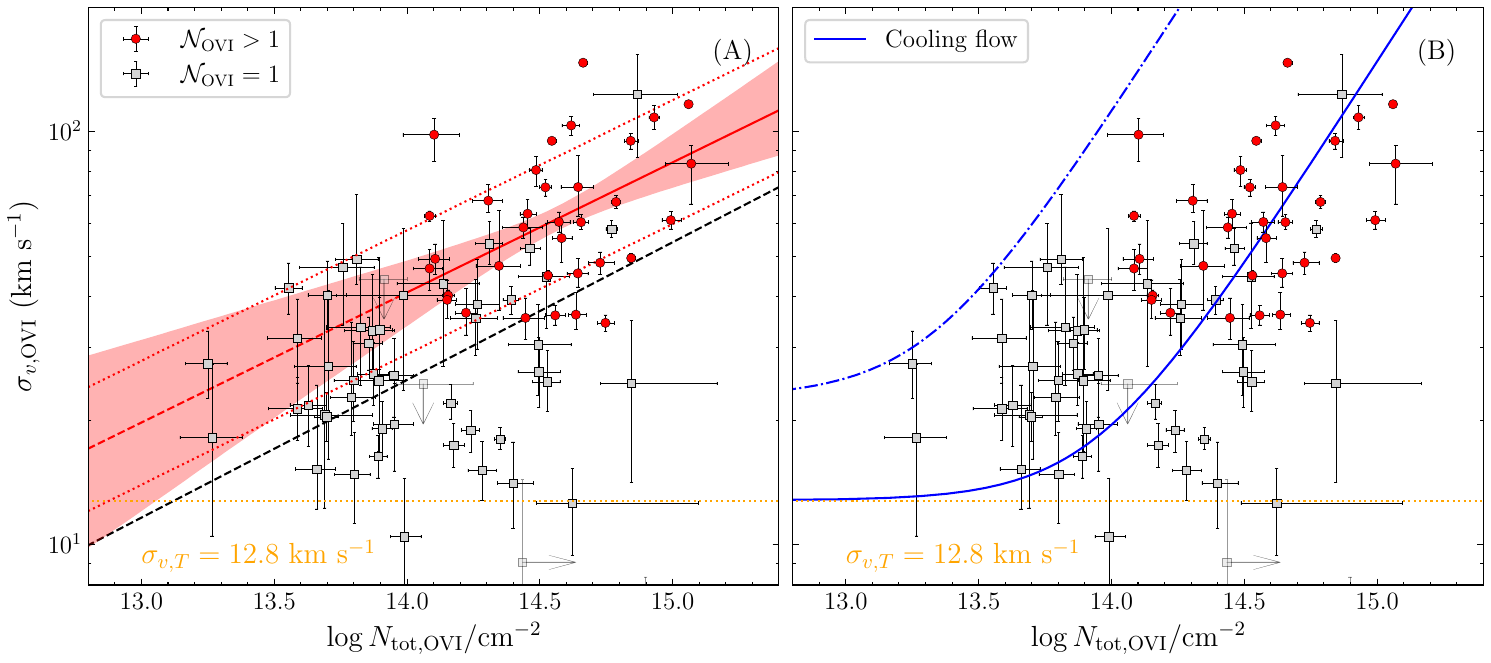}
    \caption{The observed line-of-sight \ion{O}{6} $N_{\rm tot}$-$\sigma_\varv$ relation integrated across all components.
    Circles represent the results from absorbers showing multiple \ion{O}{6} components, 
    while squares represent systems with a single detected component.
    Panel (A): A power-law fit to the multi-component sightlines (red circles) yields a best-fit slope of $\alpha=0.31_{-0.12}^{+0.13}$ (see Equation \ref{eqn:linear} for the definition of $\alpha$).
    The solid line and shaded region represent the best-fit model and its associated 1\,$\sigma$ uncertainties, while the dotted lines indicate the 1\,$\sigma_{\rm p}$ intrinsic scatter. 
    The best-fit model is shown as a red dashed line for $N_{\rm tot, OVI}<10^{14}\,\cmjj$ to indicate an extrapolation below the column density range covered by empirical observations for the multi-component systems. 
    The horizontal dotted line marks the expected thermal broadening at $\log T/{\rm K} = 5.5$.
    The best-fit model for single \ion{O}{6} components from Figure \ref{fig:o6_Nb}(A) is duplicated here as the black dashed line, 0.15 dex lower to account for the difference between $b_{\rm OVI}$ and $\sigma_{\varv, \rm OVI}$}. 
    At a given $N_{\rm tot}$, the observed $\sigma_\varv$ values are $\approx 0.2$ dex higher in multi-component absorbers than in single-component systems, implying a filling factor of 25$\%$ along individual sightlines (see details in Section \ref{sec:o6s}).
    For comparison, we also include the expectations from the cooling-flow models in Panel (B) with the solid and dashed-dotted lines showing model expectations for gas of $\log T/{\rm K} = 5.5$ and 6, respectively \citep[e.g.,][]{Heckman2002, Bordoloi2017}.
    \label{fig:N_sigma}
\end{figure*}

In the combined FUSE and literature sample, there are 34 sightlines with multiple \ion{O}{6} components, including 22 two-component, seven three-component, and five four-component sightlines.
The decomposition of components relies on the quality of the spectrum (i.e., continuum signal-to-noise ratio and spectral resolution) and the absorption strength.
Therefore, the observed single-component systems may also contain multiple components.
Here, we focus on these observed multi-component sightlines, which provide an opportunity to probe the kinematics at larger scales than individual components.
While individual \ion{O}{6} components show a power-law relation between $\log N$ and $\log b$, it remains unclear whether such a power-law relation holds for multiple components in single halos.

We investigate inter-cloud kinematics using the sightlines exhibiting multi-component structures.
For the 34 sightlines that exhibit multiple \ion{O}{6} components, the velocity dispersion ($\sigma_{\varv, {\rm OVI}}$) is computed along each sightline, weighted by the \ion{O}{6} column density profile (Equation \ref{eq:sigmav}). 
Figure \ref{fig:N_sigma} shows the multi-component systems in red with ${\cal N}_{\rm OVI}$ indicating the number of components, while single-component systems are shown in gray for comparison.
The total column densities $N_{\rm tot,OVI}$ of multi-component systems are $\log N_{\rm tot,OVI}/\cmjj \approx 14$--15, while the corresponding velocity dispersions range from $\approx 30$ to $100~\kms$.
The observed line-of-sight velocity dispersions are clearly larger than the intra-cloud line widths of individual components at a fixed column density.
The velocity dispersions of multi-component sightlines are dominated by the inter-cloud difference between absorbers, while the single-component sightlines only exhibit the line width of individual absorbers.
Therefore, comparison between single- and multi-component sightlines constrains the different velocity-dispersion-size relations, i.e., the intra-cloud and the inter-cloud kinematics.

To assess the significance of the observed $N_{\rm tot}$-$\sigma_\varv$ relation in the multi-component systems, we perform a generalized Kendall's $\tau$ test and obtain $\tau = 0.23$ with a probability of random correlation of $p = 1.2 \times 10^{-2}$. 
Fitting a power-law model of the form defined in Equation \ref{eqn:linear}, we find $\alpha_{\rm los} =0.31_{-0.12}^{+0.13}$ and $C_{\rm LOS} =1.62\pm0.08$ for the integrated line-of-sight velocity dispersion, $\sigma_{\varv, {\rm OVI}}$ (solid red line in Figure \ref{fig:N_sigma}A), and an intrinsic scatter of $0.16\pm0.02$ dex (dotted lines in Figure \ref{fig:N_sigma}A).
The best-fit power-law slope for multi-component absorbers is consistent with that of single-component systems, but single-component systems exhibit a lower amplitude by $\approx 0.2$ dex (black, dashed line in Figure \ref{fig:N_sigma}A; noting the 0.15 dex difference between the Doppler $b$ parameter and velocity dispersion $\sigma_{\varv, {\rm OVI}}$).

The offset in $C_{\rm LOS}$ between single- and multi-component systems suggests that the velocity differences among multiple \ion{O}{6} components within a halo trace the kinematics on larger scales than the absorbing path length inferred directly from $N_{\rm OVI}$.
The filling factor of \ion{O}{6} absorbing gas along each sightline is defined as the absorbing path length over the available path length in a halo and can, therefore, be calculated from this offset.
Applying the measured power-law slopes for both single- and multi-component sightlines (i.e., 1/3), a 0.2-dex difference in $\log\,\sigma_\varv$ corresponds to a scale difference of $0.2\times3 =0.6$ dex.
This implies a characteristic line-of-sight filling factor of $f_{\rm LOS} \approx 25\% $.
Because this value reflects only the fraction of space traced by \ion{O}{6} absorption, it should be regarded as an upper limit for the entire halo.
Such a path-length filling factor can constrain the volume filling factor together with the covering fraction, assuming that the path-length filling factor is constant in \ion{O}{6}-covering regions.
Adopting the typical \ion{O}{6} covering fraction of $\approx 60\%$ within galaxy halos \citep[e.g.,][]{Tchernyshyov2022, CUBSVII}, we infer an overall volume filling factor of \ion{O}{6}-bearing gas of roughly 10\text{–}20\% in the CGM.

The consistent power law slope of the $N$-$b$ (or $N$–$\sigma_\varv$) correlation---whether measured for individual components or for the total integrated LOS---supports a scenario in which the observed velocity width is driven by a turbulent cascade.
However, we also consider an alternative explanation based on the widely discussed radiative cooling flow model for the observed \ion{O}{6} properties \citep[e.g.,][]{Heckman2002, Bordoloi2017}. 
In a cooling flow framework, \ion{O}{6} absorption arises in radiatively cooling gas as the gas moves through a hot medium (either infalling through a hot halo or outflowing in a hot wind), and the characteristic size is expected to be $l\approx \tau_c\,\sigma_\varv$, where $\tau_c$ is the cooling time scale.  
This naturally leads to a linear relation between $N_{\rm OVI}$ and $\sigma_{\varv,\rm OVI}$ at high $N_{\rm OVI}$.
Figure \ref{fig:N_sigma}B shows the expectations from cooling flow models for gas with metallicity of 0.3 $Z_\odot$ and temperatures of $\log T/{\rm K} = 5.5$ and $6$.  These models also reproduce the observed $N_{\rm OVI}$-$\sigma_{\varv, {\rm OVI}}$ relation, while the scatter remains substantial, similar to the turbulence model, and more data is needed to eventually distinguish between models.

\section{Summary and Discussion}
\label{sec:dis}

\begin{figure}
    \centering
    \includegraphics[width=0.46\textwidth]{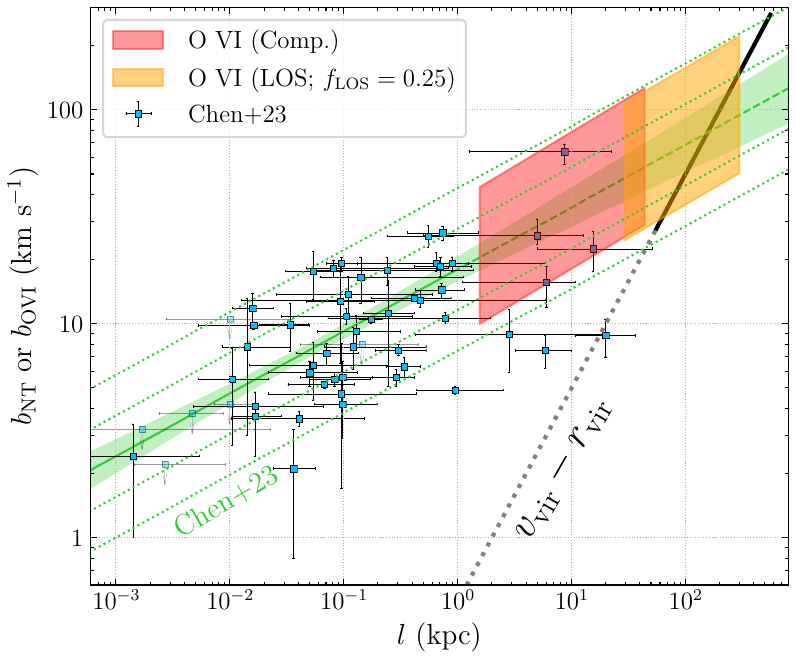}
    \caption{The observed relation between line width and spatial scale in the CGM, combining \ion{O}{6} measurements from this study (red and orange shaded bands) with previous results based on individual absorbing components from \citetalias{Chen2023} (cyan squares). The width and height of the red and orange bands represent the ranges in spatial scale and intrinsic scatter ({$\pm\,2\,\sigma_{\rm p}$}) for individual and integrated \ion{O}{6} absorbers, respectively (see discussion in \S\S\,\ref{sec:o6}–\ref{sec:o6s} and Figures \ref{fig:o6_Nb} and \ref{fig:N_sigma}}). The Kolmogorov-like turbulent model derived in \citetalias{Chen2023} is shown as a thin solid line for $l\!<\!1$ kpc and as a dashed line beyond this range, with the associated $1\,\sigma$ scatter indicated by dotted lines. For comparison, the thick solid line shows the expected correlation between projected virial velocity ($\varv_{\rm vir}$) and halo radius ($\rvir$) for gravitationally bound dark-matter halos with $\log M_{\rm halo}/\msun\!=\!10$--13. The relation is dotted below this mass range, where galaxies become too faint to be systematically detected in current surveys \citep[e.g.,][]{Chen2020}. The observed line widths on scales of $\sim\!100$ kpc are consistent with gravitational motions within halos, whereas on smaller scales the non-thermal motions follow a Kolmogorov-like turbulent cascade.
    \label{fig:joint}
\end{figure}

By combining the newly constructed FUSE-galaxy sample (\S\,\ref{sec:data}) with literature measurements (\S\,\ref{sec:literature}), we have analyzed CGM kinematics traced by \ion{H}{1} and \ion{O}{6} absorption lines through their observed relationships between line width and column density.
For both transitions, most absorbers exhibit line widths broader than expected from purely thermal motions and are therefore attributed to non-thermal processes.
While \ion{H}{1} exhibits a clear inverse correlation between Doppler width, $b_{\rm HI}$, and $N_{\rm HI}$ across nearly five orders of magnitude, from $N_{\rm HI} < 10^{13}\,\mathrm{cm^{-2}}$ to $N_{\rm HI} \approx 10^{18}\,\mathrm{cm^{-2}}$ (Figure \ref{fig:h1_Nb}), the \ion{O}{6} absorption is best described by a positive correlation between $b_{\rm OVI}$ and $N_{\rm OVI}$, spanning $N_{\rm OVI} \lesssim 3\times10^{13}$ to $\approx 10^{15}\,\mathrm{cm^{-2}}$ (Figures \ref{fig:o6_Nb} and \ref{fig:N_sigma}). In both species, substantial scatter is present about the mean trends.

To interpret these contrasting behaviors, we developed a model framework that provides a physical basis for the $N$--$b$ relations under two limiting cases: constant total column density and constant spatial density. We show that \ion{H}{1} absorbers are best characterized as structures selected by roughly constant $N_{\rm H}$, whereas \ion{O}{6} absorbers preferentially trace a constant $n_{\rm OVI}$. 
When a Kolmogorov-like turbulent velocity spectrum (see Equation \ref{eqn:nt_motion}) is incorporated, the observed scatter in the $N$--$b$ relation naturally arises from underlying small-scale fluctuations in gas density, metallicity, and turbulent velocity fields. 

In this picture, the differing $N$–$b$ relations of \ion{H}{1} and \ion{O}{6} arise naturally from their sensitivity to distinct physical parameters: while \ion{O}{6} preferentially traces a limited range of thermodynamic conditions in diffuse, warm gas, \ion{H}{1} traces the broader underlying gas distribution across multiple phases. The observed coexistence of these absorbers therefore reflects the multiphase structure of halo gas rather than a one-to-one correspondence between individual absorbing components. A more quantitative test of this framework, incorporating joint constraints from additional ionic species and expanded samples with improved spectral coverage, will be presented in future work.

Nevertheless, this framework enables an objective determination of the best-fit slope $\alpha$ for the observed \ $b_{\rm OVI}$ vs.\ $N_{\rm OVI}$ relation (Equation \ref{eqn:linear}), which maps directly to the slope of the underlying velocity-size relation for the \ion{O}{6}-bearing components. 
Indeed, a likelihood analysis that accounts for measurement uncertainties in both axes and includes an intrinsic scatter term yields a best-fit slope consistent with the Kolmogorov expectation (see Figure \ref{fig:o6_Nb}A). 
The intrinsic scatter likely reflects variations in ionization state and metallicity, corresponding to a range of O$^{5+}$ densities from $n_{\rm OVI}\approx 5\times 10^{-10}\,\cmjjj$ to $n_{\rm OVI}\approx 5\times 10^{-8}\,\cmjjj$ (see Figure \ref{fig:o6_Nb}B).

We summarize these results in Figure \ref{fig:joint}, which combines \ion{O}{6} measurements tracing the warm-hot gas from this study with previous results based on individual cool-gas absorbers from \citet[][cyan squares]{Chen2023}.
The red band highlights the range of inferred path length and non-thermal velocity width ($\pm 2\,\sigma_p$) for individual \ion{O}{6} components with a mean density of $n_{\rm OVI}\!\approx\!5\!\times\!10^{-9}\,\cmjjj$, which corresponds to a $\log\,\nh/\cmjjj
\!\approx\!-4.0$ for an ionization fraction of $f_{\rm O^{5+}}\!=\!0.2$ and metallicity of $Z_{\rm O}\!=\!0.3\,Z_{\rm O,\odot}$ \citep[typical of what is found for \ion{O}{6} absorbing gas; see e.g.,][]{Savage2014, Zahedy2019}.  
At this fixed density, the observed $N_{\rm OVI}$ values require path lengths of $l\approx 2 - 50$ kpc, and the intrinsic scatter $\sigma_p$ reflects variations in the turbulent energy among different components---where positive (negative) fluctuations correspond to more turbulent (quiescent) systems.  
Alternatively, assuming the same turbulent energy power (i.e., Equation \ref{eqn:nt_motion}), the broader (narrower) line widths could also arise from longer (shorter) path length, reproducing the same $N_{\rm OVI}$ with intrinsically lower (higher) $n_{\rm OVI}$ (see Figure \ref{fig:o6_Nb}). 
To illustrate this degeneracy between the scenarios, we present the \ion{O}{6} sample as shaded regions rather than individual points in Figure \ref{fig:joint}.

By incorporating individual \ion{O}{6} absorbing components, we extend the previously established velocity width-size relation to larger spatial scales. 
Across five orders of magnitude, from $\sim 1$ pc to $\sim 100$ kpc, the relation traced by different tracers, low- and intermediate-ionization species from \citetalias{Chen2023} together with highly-ionized \ion{O}{6}, is well described by a single power law consistent with a Kolmogorov-like turbulent cascade.  
Because different spatial scales are probed by gas of different densities, this continuity implies a strong dynamical coupling among the various CGM phases.  Indeed, some studies found that entrained cool gas and the surrounding hot medium exhibit similar velocity structure functions, differing by only $\approx 0.1$--0.2 dex in amplitude \citep[e.g.,][]{Gronke2022}, reinforcing the scenario that turbulent entrainment governs the kinematics of the multiphase CGM.

In addition to the non-thermal linewidths of individual \ion{O}{6} components, the LOS velocity dispersion (scaled up by $\sqrt{2}$) for sight lines with multiple \ion{O}{6} components is also considered in Figure \ref{fig:joint} (the orange band) for a line-of-sight filling factor of $f_{\rm LOS} = 0.25$ (see \S\,\ref{sec:o6s}). Assuming the same characteristic density of $n_{\rm OVI}\!\approx\!5\!\times\!10^{-9}\,\cmjjj$, the observed $N_{\rm tot, OVI}$ values require path lengths ranging from $l\approx 30$ kpc to $l\approx 300$ kpc. 
For comparison, we also include the expected correlation between virial velocity ($\varv_{\rm vir}$) and halo radius ($\rvir$) for gravitationally bound dark-matter halos for the range of halo masses $\log M_{\rm halo}/\msun\!=\!10$--13 probed by the galaxy samples considered in this study (thick solid line in Figure \ref{fig:joint}).  At spatial scales approaching the halo virial radius, the observed \ion{O}{6} velocity widths are comparable to the expected circular velocities, proportional to the velocity dispersion in the virialized halo. 

Taken together, our analysis supports a coherent picture in which the observed kinematic properties of \ion{H}{1} and \ion{O}{6} absorption trace a continuous turbulent cascade that spans more than five orders of magnitude in spatial scale—from cool, photoionized clumps to the warm, highly ionized halo gas.  
At large radii, the comparable amplitudes of \ion{O}{6} velocity dispersion and halo circular velocities suggest that gravity-powered processes in the halo outskirts could be a principal driver of turbulent energy injection in the CGM \citep[also see][]{Goldner2025}.
In this framework, more massive halos are expected to sustain higher turbulence amplitudes, although current observational evidence for such a dependence remains tentative.  The elevated turbulent energy observed in the CGM of quiescent galaxies \citep{Qu2022} may reflect this trend, but present models and samples remain limited in scope.  Expanding the absorber–galaxy sample across a wider range of environments and halo masses will be essential for testing whether turbulence acts as a universal regulator linking gas accretion, feedback, and the multiphase structure of the CGM.

~\\

This work is dedicated to the memory of Dr.\ Fakhri Zahedy, in recognition of his lasting contributions to the study of the circumgalactic medium and his enduring impact on our community.
We thank the anonymous referee for a careful review and valuable suggestions that significantly improved the presentation of this paper.
The authors thank Sean Johnson, Gwen Rudie, Jonathon Stern, and Fakhri Zahedy for their constructive suggestions on this work and Jerry Kriss for discussion on FUSE spectrum reduction.
ZQ and HWC acknowledge the support from NASA ADAP 80NSSC23K0479.
MG thanks the European Union for support through ERC-2024-STG 101165038 (ReMMU).

\appendix

\section{FUV Spectrum Reduction}
We retrieved all pipeline-reduced one-dimensional (1D) spectra from individual exposures from the Mikulski Archive for Space Telescopes (MAST; \dataset[doi: 10.17909/n7ff-8a08]{\doi{10.17909/n7ff-8a08}}). These 1D spectra are reduced using the latest version of the FUSE calibration pipeline, {\tt CalFUSE} 3.2 \citep{CalFUSE} and co-added following the procedures described here.  
The FUSE detector has eight segments (i.e., LiF1A/1B/2A/2B, SiC1A/1B/2A/2B), covering a wavelength range from 900 to 1180 \AA.
We coadd all useful exposures for individual segments first, where the potential wavelength offsets between different exposures are $\lesssim 5~\kms$ (see the corrections for individual exposures in \citealt{Wakker2006} using  {\tt CalFUSE} 2.1.6 and 2.4).
We noted that in most cases, segments SIC1A and SIC2B do not contribute signals to the coadded spectrum because of their low sensitivities, so these two segments are ignored in this study.
In some extreme cases, only segments LIF2A and LIF2B are useful.

Between segments, there are notable wavelength calibration discrepancies, which can be as large as $20-30~\kms$ in some cases.  Therefore, additional alignment between segments is necessary before combining different segments.
Following previous studies (e.g., \citealt{Wakker2006}), all useful segments are aligned between each other using low ionization state transitions, including \ion{N}{1}, \ion{O}{1}, \ion{Ar}{1}, \ion{Si}{2}, and \ion{Fe}{2} from the Milky Way (MW).
The absolute wavelength is calibrated by matching low-ionization absorption species with MW \ion{H}{1} 21 cm emission line obtained from the Leiden/Argentine/Bonn (LAB) Survey of Galactic \ion{H}{1} \citep{LAB_HI}.  This exercise improves the accuracy of wavelength calibration in the coadded spectrum to within $\pm 5~ \kms$.

For the 26 sightlines with {\it HST} FUV spectra, a similar set of procedures is performed to obtain a final combined spectrum per sightline.  COS spectra share a wavelength calibration issue similar to that of FUSE \citep[e.g.,][]{Wakker2015}. Following the steps described in \citet{Chen2018}, individual exposures are corrected for wavelength offsets before being combined for optimal S/N.
In contrast, STIS spectra do not show any detectable wavelength offset between spectral orders, and individual orders from individual exposures are co-added to form a final combined spectrum per sightline. 

\bibliographystyle{aasjournalv7}
\bibliography{ms} 

@ARTICLE{Liang:2020,
       author = {{Liang}, Cameron J. and {Remming}, Ian},
        title = "{On the model of the circumgalactic mist: the implications of cloud sizes in galactic winds and haloes}",
      journal = {\mnras},
         year = 2020,
        month = feb,
       volume = {491},
       number = {4},
        pages = {5056-5072},
          doi = {10.1093/mnras/stz3403},
archivePrefix = {arXiv},
       eprint = {1806.10688},
 primaryClass = {astro-ph.GA},
       adsurl = {https://ui.adsabs.harvard.edu/abs/2020MNRAS.491.5056L}
}

@ARTICLE{Dave:1999,
       author = {{Dav{\'e}}, Romeel and {Hernquist}, Lars and {Katz}, Neal and {Weinberg}, David H.},
        title = "{The Low-Redshift Ly{\ensuremath{\alpha}} Forest in Cold Dark Matter Cosmologies}",
      journal = {\apj},
         year = 1999,
        month = feb,
       volume = {511},
       number = {2},
        pages = {521-545},
          doi = {10.1086/306722},
archivePrefix = {arXiv},
       eprint = {astro-ph/9807177},
 primaryClass = {astro-ph},
       adsurl = {https://ui.adsabs.harvard.edu/abs/1999ApJ...511..521D}
}

@ARTICLE{Chen2010,
       author = {{Chen}, Hsiao-Wen and {Helsby}, Jennifer E. and {Gauthier}, Jean-Ren{\'e} and {Shectman}, Stephen A. and {Thompson}, Ian B. and {Tinker}, Jeremy L.},
        title = "{An Empirical Characterization of Extended Cool Gas Around Galaxies Using Mg II Absorption Features}",
      journal = {\apj},
         year = 2010,
        month = may,
       volume = {714},
       number = {2},
        pages = {1521-1541},
          doi = {10.1088/0004-637X/714/2/1521},
archivePrefix = {arXiv},
       eprint = {1004.0705},
 primaryClass = {astro-ph.CO},
       adsurl = {https://ui.adsabs.harvard.edu/abs/2010ApJ...714.1521C}
}

@ARTICLE{COS,
       author = {{Green}, James C. and {Froning}, Cynthia S. and {Osterman}, Steve and {Ebbets}, Dennis and {Heap}, Sara H. and {Leitherer}, Claus and {Linsky}, Jeffrey L. and {Savage}, Blair D. and {Sembach}, Kenneth and {Shull}, J. Michael and {Siegmund}, Oswald H.~W. and {Snow}, Theodore P. and {Spencer}, John and {Stern}, S. Alan and {Stocke}, John and {Welsh}, Barry and {B{\'e}land}, St{\'e}phane and {Burgh}, Eric B. and {Danforth}, Charles and {France}, Kevin and {Keeney}, Brian and {McPhate}, Jason and {Penton}, Steven V. and {Andrews}, John and {Brownsberger}, Kenneth and {Morse}, Jon and {Wilkinson}, Erik},
        title = "{The Cosmic Origins Spectrograph}",
      journal = {\apj},
         year = 2012,
        month = jan,
       volume = {744},
       number = {1},
          eid = {60},
        pages = {60},
          doi = {10.1088/0004-637X/744/1/6010.1086/141956},
archivePrefix = {arXiv},
       eprint = {1110.0462},
 primaryClass = {astro-ph.IM},
       adsurl = {https://ui.adsabs.harvard.edu/abs/2012ApJ...744...60G}
}

@ARTICLE{STIS,
       author = {{Woodgate}, B.~E. and {Kimble}, R.~A. and {Bowers}, C.~W. and {Kraemer}, S. and {Kaiser}, M.~E. and {Danks}, A.~C. and {Grady}, J.~F. and {Loiacono}, J.~J. and {Brumfield}, M. and {Feinberg}, L. and {Gull}, T.~R. and {Heap}, S.~R. and {Maran}, S.~P. and {Lindler}, D. and {Hood}, D. and {Meyer}, W. and {Vanhouten}, C. and {Argabright}, V. and {Franka}, S. and {Bybee}, R. and {Dorn}, D. and {Bottema}, M. and {Woodruff}, R. and {Michika}, D. and {Sullivan}, J. and {Hetlinger}, J. and {Ludtke}, C. and {Stocker}, R. and {Delamere}, A. and {Rose}, D. and {Becker}, I. and {Garner}, H. and {Timothy}, J.~G. and {Blouke}, M. and {Joseph}, C.~L. and {Hartig}, G. and {Green}, R.~F. and {Jenkins}, E.~B. and {Linsky}, J.~L. and {Hutchings}, J.~B. and {Moos}, H.~W. and {Boggess}, A. and {Roesler}, F. and {Weistrop}, D.},
        title = "{The Space Telescope Imaging Spectrograph Design}",
      journal = {\pasp},
         year = 1998,
        month = oct,
       volume = {110},
       number = {752},
        pages = {1183-1204},
          doi = {10.1086/316243},
       adsurl = {https://ui.adsabs.harvard.edu/abs/1998PASP..110.1183W}
}

@ARTICLE{Qu2023,
       author = {{Qu}, Zhijie and {Chen}, Hsiao-Wen and {Rudie}, Gwen C. and {Johnson}, Sean D. and {Zahedy}, Fakhri S. and {DePalma}, David and {Boettcher}, Erin and {Cantalupo}, Sebastiano and {Chen}, Mandy C. and {Cooksey}, Kathy L. and {Faucher-Gigu{\`e}re}, Claude-Andr{\'e} and {Li}, Jennifer I. -Hsiu and {Lopez}, Sebastian and {Schaye}, Joop and {Simcoe}, Robert A.},
        title = "{The Cosmic Ultraviolet Baryon Survey (CUBS) - VI. Connecting physical properties of the cool circumgalactic medium to galaxies at z {\ensuremath{\approx}} 1}",
      journal = {\mnras},
         year = 2023,
        month = sep,
       volume = {524},
       number = {1},
        pages = {512-528},
          doi = {10.1093/mnras/stad1886},
archivePrefix = {arXiv},
       eprint = {2306.11274},
 primaryClass = {astro-ph.GA},
       adsurl = {https://ui.adsabs.harvard.edu/abs/2023MNRAS.524..512Q}
}

@ARTICLE{Burchett:2016,
       author = {{Burchett}, Joseph N. and {Tripp}, Todd M. and {Bordoloi}, Rongmon and {Werk}, Jessica K. and {Prochaska}, J. Xavier and {Tumlinson}, Jason and {Willmer}, C.~N.~A. and {O'Meara}, John and {Katz}, Neal},
        title = "{A Deep Search for Faint Galaxies Associated with Very Low Redshift C IV Absorbers. III. The Mass- and Environment-dependent Circumgalactic Medium}",
      journal = {\apj},
         year = 2016,
        month = dec,
       volume = {832},
       number = {2},
          eid = {124},
        pages = {124},
          doi = {10.3847/0004-637X/832/2/124},
archivePrefix = {arXiv},
       eprint = {1512.00853},
 primaryClass = {astro-ph.GA},
       adsurl = {https://ui.adsabs.harvard.edu/abs/2016ApJ...832..124B}
}

@ARTICLE{Johnson2015,
       author = {{Johnson}, Sean D. and {Chen}, Hsiao-Wen and {Mulchaey}, John S.},
        title = "{On the possible environmental effect in distributing heavy elements beyond individual gaseous haloes}",
      journal = {\mnras},
         year = 2015,
        month = may,
       volume = {449},
       number = {3},
        pages = {3263-3273},
          doi = {10.1093/mnras/stv553},
archivePrefix = {arXiv},
       eprint = {1503.04199},
 primaryClass = {astro-ph.GA},
       adsurl = {https://ui.adsabs.harvard.edu/abs/2015MNRAS.449.3263J}
}

@ARTICLE{Wenger2000,
       author = {{Wenger}, M. and {Ochsenbein}, F. and {Egret}, D. and {Dubois}, P. and {Bonnarel}, F. and {Borde}, S. and {Genova}, F. and {Jasniewicz}, G. and {Lalo{\"e}}, S. and {Lesteven}, S. and {Monier}, R.},
        title = "{The SIMBAD astronomical database. The CDS reference database for astronomical objects}",
      journal = {\aaps},
         year = 2000,
        month = apr,
       volume = {143},
        pages = {9-22},
          doi = {10.1051/aas:2000332},
archivePrefix = {arXiv},
       eprint = {astro-ph/0002110},
 primaryClass = {astro-ph},
       adsurl = {https://ui.adsabs.harvard.edu/abs/2000A&AS..143....9W}
}

@ARTICLE{savage2002,
       author = {{Savage}, B.~D. and {Sembach}, K.~R. and {Tripp}, T.~M. and {Richter}, P.},
        title = "{Far Ultraviolet Spectroscopic Explorer and Space Telescope Imaging Spectrograph Observations of Intervening O VI Absorption Line Systems in the Spectrum of PG 0953+415}",
      journal = {\apj},
         year = 2002,
        month = jan,
       volume = {564},
       number = {2},
        pages = {631-649},
          doi = {10.1086/324288},
       adsurl = {https://ui.adsabs.harvard.edu/abs/2002ApJ...564..631S}
}

@ARTICLE{Spitzer1956,
       author = {{Spitzer}, Jr., Lyman},
        title = "{On a Possible Interstellar Galactic Corona.}",
      journal = {\apj},
         year = 1956,
        month = jul,
       volume = {124},
        pages = {20},
          doi = {10.1086/146200},
       adsurl = {https://ui.adsabs.harvard.edu/abs/1956ApJ...124...20S}
}

@ARTICLE{Johnson2024,
       author = {{Johnson}, Sean D. and {Liu}, Zhuoqi (Will) and {Li}, Jennifer I. -Hsiu and {Schaye}, Joop and {Greene}, Jenny E. and {Cantalupo}, Sebastiano and {Rudie}, Gwen C. and {Qu}, Zhijie and {Chen}, Hsiao-Wen and {Rafelski}, Marc and {Muzahid}, Sowgat and {Chen}, Mandy C. and {Contini}, Thierry and {Kollatschny}, Wolfram and {Mishra}, Nishant and {Petitjean}, Patrick and {Rauch}, Michael and {Zahedy}, Fakhri S.},
        title = "{Discovery of Optically Emitting Circumgalactic Nebulae around the Majority of UV-luminous Quasars at Intermediate Redshift}",
      journal = {\apj},
         year = 2024,
        month = may,
       volume = {966},
       number = {2},
          eid = {218},
        pages = {218},
          doi = {10.3847/1538-4357/ad3911},
archivePrefix = {arXiv},
       eprint = {2404.00088},
 primaryClass = {astro-ph.GA},
       adsurl = {https://ui.adsabs.harvard.edu/abs/2024ApJ...966..218J}
}

@ARTICLE{Heckman2002,
       author = {{Heckman}, T.~M. and {Norman}, C.~A. and {Strickland}, D.~K. and {Sembach}, K.~R.},
        title = "{On the Physical Origin of O VI Absorption-Line Systems}",
      journal = {\apj},
         year = 2002,
        month = oct,
       volume = {577},
       number = {2},
        pages = {691-700},
          doi = {10.1086/342232},
archivePrefix = {arXiv},
       eprint = {astro-ph/0205556},
 primaryClass = {astro-ph},
       adsurl = {https://ui.adsabs.harvard.edu/abs/2002ApJ...577..691H}
}

@ARTICLE{FUSE,
       author = {{Moos}, H.~W. and {Cash}, W.~C. and {Cowie}, L.~L. and {Davidsen}, A.~F. and {Dupree}, A.~K. and {Feldman}, P.~D. and {Friedman}, S.~D. and {Green}, J.~C. and {Green}, R.~F. and {Gry}, C. and {Hutchings}, J.~B. and {Jenkins}, E.~B. and {Linsky}, J.~L. and {Malina}, R.~F. and {Michalitsianos}, A.~G. and {Savage}, B.~D. and {Shull}, J.~M. and {Siegmund}, O.~H.~W. and {Snow}, T.~P. and {Sonneborn}, G. and {Vidal-Madjar}, A. and {Willis}, A.~J. and {Woodgate}, B.~E. and {York}, D.~G. and {Ake}, T.~B. and {Andersson}, B. -G. and {Andrews}, J.~P. and {Barkhouser}, R.~H. and {Bianchi}, L. and {Blair}, W.~P. and {Brownsberger}, K.~R. and {Cha}, A.~N. and {Chayer}, P. and {Conard}, S.~J. and {Fullerton}, A.~W. and {Gaines}, G.~A. and {Grange}, R. and {Gummin}, M.~A. and {Hebrard}, G. and {Kriss}, G.~A. and {Kruk}, J.~W. and {Mark}, D. and {McCarthy}, D.~K. and {Morbey}, C.~L. and {Murowinski}, R. and {Murphy}, E.~M. and {Oegerle}, W.~R. and {Ohl}, R.~G. and {Oliveira}, C. and {Osterman}, S.~N. and {Sahnow}, D.~J. and {Saisse}, M. and {Sembach}, K.~R. and {Weaver}, H.~A. and {Welsh}, B.~Y. and {Wilkinson}, E. and {Zheng}, W.},
        title = "{Overview of the Far Ultraviolet Spectroscopic Explorer Mission}",
      journal = {\apjl},
         year = 2000,
        month = jul,
       volume = {538},
       number = {1},
        pages = {L1-L6},
          doi = {10.1086/312795},
archivePrefix = {arXiv},
       eprint = {astro-ph/0005529},
 primaryClass = {astro-ph},
       adsurl = {https://ui.adsabs.harvard.edu/abs/2000ApJ...538L...1M}
}

@ARTICLE{CalFUSE,
       author = {{Dixon}, W.~V. and {Sahnow}, D.~J. and {Barrett}, P.~E. and {Civeit}, T. and {Dupuis}, J. and {Fullerton}, A.~W. and {Godard}, B. and {Hsu}, J. -C. and {Kaiser}, M.~E. and {Kruk}, J.~W. and {Lacour}, S. and {Lindler}, D.~J. and {Massa}, D. and {Robinson}, R.~D. and {Romelfanger}, M.~L. and {Sonnentrucker}, P.},
        title = "{CalFUSE Version 3: A Data Reduction Pipeline for the Far Ultraviolet Spectroscopic Explorer}",
      journal = {\pasp},
         year = 2007,
        month = may,
       volume = {119},
       number = {855},
        pages = {527-555},
          doi = {10.1086/518617},
archivePrefix = {arXiv},
       eprint = {0704.0899},
 primaryClass = {astro-ph},
       adsurl = {https://ui.adsabs.harvard.edu/abs/2007PASP..119..527D}
}

@ARTICLE{Voit2018,
       author = {{Voit}, G. Mark},
        title = "{A Role for Turbulence in Circumgalactic Precipitation}",
      journal = {\apj},
         year = 2018,
        month = dec,
       volume = {868},
       number = {2},
          eid = {102},
        pages = {102},
          doi = {10.3847/1538-4357/aae8e2},
archivePrefix = {arXiv},
       eprint = {1803.06036},
 primaryClass = {astro-ph.GA},
       adsurl = {https://ui.adsabs.harvard.edu/abs/2018ApJ...868..102V}
}

@ARTICLE{Schmidt2021,
       author = {{Schmidt}, W. and {Schmidt}, J.~P. and {Grete}, P.},
        title = "{Turbulence in the intragroup and circumgalactic medium}",
      journal = {\aap},
         year = 2021,
        month = oct,
       volume = {654},
          eid = {A115},
        pages = {A115},
          doi = {10.1051/0004-6361/202140920},
archivePrefix = {arXiv},
       eprint = {2107.12125},
 primaryClass = {astro-ph.GA},
       adsurl = {https://ui.adsabs.harvard.edu/abs/2021A&A...654A.115S}
}

@ARTICLE{Koplitz2023,
       author = {{Koplitz}, Brad and {}, II, Edward Buie and {Scannapieco}, Evan},
        title = "{Constraining Circumgalactic Turbulence with QSO Absorption Line Measurements}",
      journal = {\apj},
         year = 2023,
        month = oct,
       volume = {956},
       number = {1},
          eid = {54},
        pages = {54},
          doi = {10.3847/1538-4357/acf4fc},
archivePrefix = {arXiv},
       eprint = {2308.12283},
 primaryClass = {astro-ph.GA},
       adsurl = {https://ui.adsabs.harvard.edu/abs/2023ApJ...956...54K}
}

@ARTICLE{Chen2023,
       author = {{Chen}, Hsiao-Wen and {Qu}, Zhijie and {Rauch}, Michael and {Chen}, Mandy C. and {Zahedy}, Fakhri S. and {Johnson}, Sean D. and {Schaye}, Joop and {Rudie}, Gwen C. and {Boettcher}, Erin and {Cantalupo}, Sebastiano and {Faucher-Gigu{\`e}re}, Claude-Andr{\'e} and {Greene}, Jenny E. and {Lopez}, Sebastian and {Simcoe}, Robert A.},
        title = "{The Cosmic Ultraviolet Baryon Survey: Empirical Characterization of Turbulence in the Cool Circumgalactic Medium}",
      journal = {\apjl},
         year = 2023,
        month = sep,
       volume = {955},
       number = {1},
          eid = {L25},
        pages = {L25},
          doi = {10.3847/2041-8213/acf85b},
archivePrefix = {arXiv},
       eprint = {2309.05699},
 primaryClass = {astro-ph.GA},
       adsurl = {https://ui.adsabs.harvard.edu/abs/2023ApJ...955L..25C}
}

@ARTICLE{Qu2022,
       author = {{Qu}, Zhijie and {Chen}, Hsiao-Wen and {Rudie}, Gwen C. and {Zahedy}, Fakhri S. and {Johnson}, Sean D. and {Boettcher}, Erin and {Cantalupo}, Sebastiano and {Chen}, Mandy C. and {Cooksey}, Kathy L. and {DePalma}, David and {Faucher-Gigu{\`e}re}, Claude-Andr{\'e} and {Rauch}, Michael and {Schaye}, Joop and {Simcoe}, Robert A.},
        title = "{The Cosmic Ultraviolet Baryon Survey (CUBS) V: on the thermodynamic properties of the cool circumgalactic medium at z {\ensuremath{\lesssim}} 1}",
      journal = {\mnras},
         year = 2022,
        month = nov,
       volume = {516},
       number = {4},
        pages = {4882-4897},
          doi = {10.1093/mnras/stac2528},
archivePrefix = {arXiv},
       eprint = {2209.01228},
 primaryClass = {astro-ph.GA},
       adsurl = {https://ui.adsabs.harvard.edu/abs/2022MNRAS.516.4882Q}
}

@ARTICLE{Buie2020,
       author = {{Buie}, II, Edward and {Gray}, William J. and {Scannapieco}, Evan and {Safarzadeh}, Mohammadtaher},
        title = "{Modeling Photoionized Turbulent Material in the Circumgalactic Medium. II. Effect of Turbulence within a Stratified Medium}",
      journal = {\apj},
         year = 2020,
        month = jun,
       volume = {896},
       number = {2},
          eid = {136},
        pages = {136},
          doi = {10.3847/1538-4357/ab9535},
archivePrefix = {arXiv},
       eprint = {2006.03066},
 primaryClass = {astro-ph.GA},
       adsurl = {https://ui.adsabs.harvard.edu/abs/2020ApJ...896..136B}
}

@INPROCEEDINGS{Chen2025,
       author = {{Chen}, Hsiao-Wen and {Zahedy}, Fakhri S.},
        title = "{The circumgalactic medium}",
    booktitle = {Encyclopedia of Astrophysics},
         year = 2026,
       volume = {4},
        month = jan,
        pages = {370-400},
          doi = {10.1016/B978-0-443-21439-4.00059-6},
       adsurl = {https://ui.adsabs.harvard.edu/abs/2026enap....4..370C}
}

@ARTICLE{CAFG2023,
       author = {{Faucher-Gigu{\`e}re}, Claude-Andr{\'e} and {Oh}, S. Peng},
        title = "{Key Physical Processes in the Circumgalactic Medium}",
      journal = {\araa},
         year = 2023,
        month = aug,
       volume = {61},
        pages = {131-195},
          doi = {10.1146/annurev-astro-052920-125203},
archivePrefix = {arXiv},
       eprint = {2301.10253},
 primaryClass = {astro-ph.GA},
       adsurl = {https://ui.adsabs.harvard.edu/abs/2023ARA&A..61..131F}
}

@ARTICLE{Ji2019,
       author = {{Ji}, Suoqing and {Oh}, S. Peng and {Masterson}, Phillip},
        title = "{Simulations of radiative turbulent mixing layers}",
      journal = {\mnras},
         year = 2019,
        month = jul,
       volume = {487},
       number = {1},
        pages = {737-754},
          doi = {10.1093/mnras/stz1248},
archivePrefix = {arXiv},
       eprint = {1809.09101},
 primaryClass = {astro-ph.GA},
       adsurl = {https://ui.adsabs.harvard.edu/abs/2019MNRAS.487..737J}
}

@ARTICLE{Gronke2022,
       author = {{Gronke}, Max and {Oh}, S. Peng and {Ji}, Suoqing and {Norman}, Colin},
        title = "{Survival and mass growth of cold gas in a turbulent, multiphase medium}",
      journal = {\mnras},
         year = 2022,
        month = mar,
       volume = {511},
       number = {1},
        pages = {859-876},
          doi = {10.1093/mnras/stab3351},
archivePrefix = {arXiv},
       eprint = {2107.13012},
 primaryClass = {astro-ph.GA},
       adsurl = {https://ui.adsabs.harvard.edu/abs/2022MNRAS.511..859G}
}

@ARTICLE{Tan2021,
       author = {{Tan}, Brent and {Oh}, S. Peng and {Gronke}, Max},
        title = "{Radiative mixing layers: insights from turbulent combustion}",
      journal = {\mnras},
         year = 2021,
        month = apr,
       volume = {502},
       number = {3},
        pages = {3179-3199},
          doi = {10.1093/mnras/stab053},
archivePrefix = {arXiv},
       eprint = {2008.12302},
 primaryClass = {astro-ph.GA},
       adsurl = {https://ui.adsabs.harvard.edu/abs/2021MNRAS.502.3179T}
}

@ARTICLE{Putman2012,
       author = {{Putman}, M.~E. and {Peek}, J.~E.~G. and {Joung}, M.~R.},
        title = "{Gaseous Galaxy Halos}",
      journal = {\araa},
         year = 2012,
        month = sep,
       volume = {50},
        pages = {491-529},
          doi = {10.1146/annurev-astro-081811-125612},
archivePrefix = {arXiv},
       eprint = {1207.4837},
 primaryClass = {astro-ph.GA},
       adsurl = {https://ui.adsabs.harvard.edu/abs/2012ARA&A..50..491P}
}

@ARTICLE{Tumlinson2017,
       author = {{Tumlinson}, Jason and {Peeples}, Molly S. and {Werk}, Jessica K.},
        title = "{The Circumgalactic Medium}",
      journal = {\araa},
         year = 2017,
        month = aug,
       volume = {55},
       number = {1},
        pages = {389-432},
          doi = {10.1146/annurev-astro-091916-055240},
archivePrefix = {arXiv},
       eprint = {1709.09180},
 primaryClass = {astro-ph.GA},
       adsurl = {https://ui.adsabs.harvard.edu/abs/2017ARA&A..55..389T}
}

@ARTICLE{Li2020,
       author = {{Li}, Yuan and {Gendron-Marsolais}, Marie-Lou and {Zhuravleva}, Irina and {Xu}, Siyao and {Simionescu}, Aurora and {Tremblay}, Grant R. and {Lochhaas}, Cassandra and {Bryan}, Greg L. and {Quataert}, Eliot and {Murray}, Norman W. and {Boselli}, Alessandro and {Hlavacek-Larrondo}, Julie and {Zheng}, Yong and {Fossati}, Matteo and {Li}, Miao and {Emsellem}, Eric and {Sarzi}, Marc and {Arzamasskiy}, Lev and {Vishniac}, Ethan T.},
        title = "{Direct Detection of Black Hole-driven Turbulence in the Centers of Galaxy Clusters}",
      journal = {\apjl},
         year = 2020,
        month = jan,
       volume = {889},
       number = {1},
          eid = {L1},
        pages = {L1},
          doi = {10.3847/2041-8213/ab65c7},
archivePrefix = {arXiv},
       eprint = {1911.06329},
 primaryClass = {astro-ph.GA},
       adsurl = {https://ui.adsabs.harvard.edu/abs/2020ApJ...889L...1L}
}

@ARTICLE{ChenC2023,
       author = {{Chen}, Mandy C. and {Chen}, Hsiao-Wen and {Rauch}, Michael and {Qu}, Zhijie and {Johnson}, Sean D. and {Li}, Jennifer I. -Hsiu and {Schaye}, Joop and {Rudie}, Gwen C. and {Zahedy}, Fakhri S. and {Boettcher}, Erin and {Cooksey}, Kathy L. and {Cantalupo}, Sebastiano},
        title = "{Empirical constraints on the turbulence in QSO host nebulae from velocity structure function measurements}",
      journal = {\mnras},
         year = 2023,
        month = jan,
       volume = {518},
       number = {2},
        pages = {2354-2372},
          doi = {10.1093/mnras/stac3193},
archivePrefix = {arXiv},
       eprint = {2209.04344},
 primaryClass = {astro-ph.GA},
       adsurl = {https://ui.adsabs.harvard.edu/abs/2023MNRAS.518.2354C}
}

@ARTICLE{Lv2024,
       author = {{Lv}, Alex and {Wang}, Lile and {Cen}, Renyue and {Ho}, Luis C.},
        title = "{Cloud Crushing and Dissipation of Uniformly Driven Adiabatic Turbulence in Circumgalactic Media}",
      journal = {\apj},
         year = 2024,
        month = dec,
       volume = {977},
       number = {2},
          eid = {274},
        pages = {274},
          doi = {10.3847/1538-4357/ad8f3d},
archivePrefix = {arXiv},
       eprint = {2406.18920},
 primaryClass = {astro-ph.GA},
       adsurl = {https://ui.adsabs.harvard.edu/abs/2024ApJ...977..274L}
}

@ARTICLE{Li2023,
       author = {{Li}, Yuan and {Luo}, Rongxin and {Fossati}, Matteo and {Sun}, Ming and {J{\'a}chym}, Pavel},
        title = "{Turbulence in the tail of a jellyfish galaxy}",
      journal = {\mnras},
         year = 2023,
        month = may,
       volume = {521},
       number = {3},
        pages = {4785-4791},
          doi = {10.1093/mnras/stad874},
archivePrefix = {arXiv},
       eprint = {2303.15500},
 primaryClass = {astro-ph.GA},
       adsurl = {https://ui.adsabs.harvard.edu/abs/2023MNRAS.521.4785L}
}

@ARTICLE{Chen2019,
       author = {{Chen}, Hsiao-Wen and {Boettcher}, Erin and {Johnson}, Sean D. and {Zahedy}, Fakhri S. and {Rudie}, Gwen C. and {Cooksey}, Kathy L. and {Rauch}, Michael and {Mulchaey}, John S.},
        title = "{A Giant Intragroup Nebula Hosting a Damped \{Ly\}\textbackslashalpha  Absorber at z = 0.313}",
      journal = {\apjl},
         year = 2019,
        month = jun,
       volume = {878},
       number = {2},
          eid = {L33},
        pages = {L33},
          doi = {10.3847/2041-8213/ab25ec},
archivePrefix = {arXiv},
       eprint = {1906.00005},
 primaryClass = {astro-ph.GA},
       adsurl = {https://ui.adsabs.harvard.edu/abs/2019ApJ...878L..33C}
}

@ARTICLE{Rauch1996,
       author = {{Rauch}, M. and {Sargent}, W.~L.~W. and {Womble}, D.~S. and {Barlow}, T.~A.},
        title = "{Temperature and Kinematics of C IV Absorption Systems}",
      journal = {\apjl},
         year = 1996,
        month = aug,
       volume = {467},
        pages = {L5},
          doi = {10.1086/310187},
archivePrefix = {arXiv},
       eprint = {astro-ph/9606041},
 primaryClass = {astro-ph},
       adsurl = {https://ui.adsabs.harvard.edu/abs/1996ApJ...467L...5R}
}

@ARTICLE{Rudie2019,
       author = {{Rudie}, Gwen C. and {Steidel}, Charles C. and {Pettini}, Max and {Trainor}, Ryan F. and {Strom}, Allison L. and {Hummels}, Cameron B. and {Reddy}, Naveen A. and {Shapley}, Alice E.},
        title = "{Column Density, Kinematics, and Thermal State of Metal-bearing Gas within the Virial Radius of z {\ensuremath{\sim}} 2 Star-forming Galaxies in the Keck Baryonic Structure Survey}",
      journal = {\apj},
         year = 2019,
        month = nov,
       volume = {885},
       number = {1},
          eid = {61},
        pages = {61},
          doi = {10.3847/1538-4357/ab4255},
archivePrefix = {arXiv},
       eprint = {1903.00004},
 primaryClass = {astro-ph.GA},
       adsurl = {https://ui.adsabs.harvard.edu/abs/2019ApJ...885...61R}
}

@ARTICLE{CUBSVII,
       author = {{Qu}, Zhijie and {Chen}, Hsiao-Wen and {Johnson}, Sean D. and {Rudie}, Gwen C. and {Zahedy}, Fakhri S. and {DePalma}, David and {Schaye}, Joop and {Boettcher}, Erin T. and {Cantalupo}, Sebastiano and {Chen}, Mandy C. and {Faucher-Gigu{\`e}re}, Claude-Andr{\'e} and {Li}, Jennifer I. -Hsiu and {Mulchaey}, John S. and {Petitjean}, Patrick and {Rafelski}, Marc},
        title = "{The Cosmic Ultraviolet Baryon Survey (CUBS). VII. On the Warm-hot Circumgalactic Medium Probed by O VI and Ne VIII at 0.4 {\ensuremath{\lesssim}} z {\ensuremath{\lesssim}} 0.7}",
      journal = {\apj},
         year = 2024,
        month = jun,
       volume = {968},
       number = {1},
          eid = {8},
        pages = {8},
          doi = {10.3847/1538-4357/ad410b},
archivePrefix = {arXiv},
       eprint = {2402.08016},
 primaryClass = {astro-ph.GA},
       adsurl = {https://ui.adsabs.harvard.edu/abs/2024ApJ...968....8Q}
}

@ARTICLE{CF3_edd,
       author = {{Kourkchi}, Ehsan and {Courtois}, H{\'e}l{\`e}ne M. and {Graziani}, Romain and {Hoffman}, Yehuda and {Pomar{\`e}de}, Daniel and {Shaya}, Edward J. and {Tully}, R. Brent},
        title = "{Cosmicflows-3: Two Distance-Velocity Calculators}",
      journal = {\aj},
         year = 2020,
        month = feb,
       volume = {159},
       number = {2},
          eid = {67},
        pages = {67},
          doi = {10.3847/1538-3881/ab620e},
archivePrefix = {arXiv},
       eprint = {1912.07214},
 primaryClass = {astro-ph.CO},
       adsurl = {https://ui.adsabs.harvard.edu/abs/2020AJ....159...67K}
}

@ARTICLE{Tully2016,
       author = {{Tully}, R. Brent and {Courtois}, H{\'e}l{\`e}ne M. and {Sorce}, Jenny G.},
        title = "{Cosmicflows-3}",
      journal = {\aj},
         year = 2016,
        month = aug,
       volume = {152},
       number = {2},
          eid = {50},
        pages = {50},
          doi = {10.3847/0004-6256/152/2/50},
archivePrefix = {arXiv},
       eprint = {1605.01765},
 primaryClass = {astro-ph.CO},
       adsurl = {https://ui.adsabs.harvard.edu/abs/2016AJ....152...50T}
}

@ARTICLE{Legacy_survey,
       author = {{Dey}, Arjun and {Schlegel}, David J. and {Lang}, Dustin and {Blum}, Robert and {Burleigh}, Kaylan and {Fan}, Xiaohui and {Findlay}, Joseph R. and {Finkbeiner}, Doug and {Herrera}, David and {Juneau}, St{\'e}phanie and {Landriau}, Martin and {Levi}, Michael and {McGreer}, Ian and {Meisner}, Aaron and {Myers}, Adam D. and {Moustakas}, John and {Nugent}, Peter and {Patej}, Anna and {Schlafly}, Edward F. and {Walker}, Alistair R. and {Valdes}, Francisco and {Weaver}, Benjamin A. and {Y{\`e}che}, Christophe and {Zou}, Hu and {Zhou}, Xu and {Abareshi}, Behzad and {Abbott}, T.~M.~C. and {Abolfathi}, Bela and {Aguilera}, C. and {Alam}, Shadab and {Allen}, Lori and {Alvarez}, A. and {Annis}, James and {Ansarinejad}, Behzad and {Aubert}, Marie and {Beechert}, Jacqueline and {Bell}, Eric F. and {BenZvi}, Segev Y. and {Beutler}, Florian and {Bielby}, Richard M. and {Bolton}, Adam S. and {Brice{\~n}o}, C{\'e}sar and {Buckley-Geer}, Elizabeth J. and {Butler}, Karen and {Calamida}, Annalisa and {Carlberg}, Raymond G. and {Carter}, Paul and {Casas}, Ricard and {Castander}, Francisco J. and {Choi}, Yumi and {Comparat}, Johan and {Cukanovaite}, Elena and {Delubac}, Timoth{\'e}e and {DeVries}, Kaitlin and {Dey}, Sharmila and {Dhungana}, Govinda and {Dickinson}, Mark and {Ding}, Zhejie and {Donaldson}, John B. and {Duan}, Yutong and {Duckworth}, Christopher J. and {Eftekharzadeh}, Sarah and {Eisenstein}, Daniel J. and {Etourneau}, Thomas and {Fagrelius}, Parker A. and {Farihi}, Jay and {Fitzpatrick}, Mike and {Font-Ribera}, Andreu and {Fulmer}, Leah and {G{\"a}nsicke}, Boris T. and {Gaztanaga}, Enrique and {George}, Koshy and {Gerdes}, David W. and {Gontcho}, Satya Gontcho A. and {Gorgoni}, Claudio and {Green}, Gregory and {Guy}, Julien and {Harmer}, Diane and {Hernandez}, M. and {Honscheid}, Klaus and {Huang}, Lijuan Wendy and {James}, David J. and {Jannuzi}, Buell T. and {Jiang}, Linhua and {Joyce}, Richard and {Karcher}, Armin and {Karkar}, Sonia and {Kehoe}, Robert and {Kneib}, Jean-Paul and {Kueter-Young}, Andrea and {Lan}, Ting-Wen and {Lauer}, Tod R. and {Le Guillou}, Laurent and {Le Van Suu}, Auguste and {Lee}, Jae Hyeon and {Lesser}, Michael and {Perreault Levasseur}, Laurence and {Li}, Ting S. and {Mann}, Justin L. and {Marshall}, Robert and {Mart{\'\i}nez-V{\'a}zquez}, C.~E. and {Martini}, Paul and {du Mas des Bourboux}, H{\'e}lion and {McManus}, Sean and {Meier}, Tobias Gabriel and {M{\'e}nard}, Brice and {Metcalfe}, Nigel and {Mu{\~n}oz-Guti{\'e}rrez}, Andrea and {Najita}, Joan and {Napier}, Kevin and {Narayan}, Gautham and {Newman}, Jeffrey A. and {Nie}, Jundan and {Nord}, Brian and {Norman}, Dara J. and {Olsen}, Knut A.~G. and {Paat}, Anthony and {Palanque-Delabrouille}, Nathalie and {Peng}, Xiyan and {Poppett}, Claire L. and {Poremba}, Megan R. and {Prakash}, Abhishek and {Rabinowitz}, David and {Raichoor}, Anand and {Rezaie}, Mehdi and {Robertson}, A.~N. and {Roe}, Natalie A. and {Ross}, Ashley J. and {Ross}, Nicholas P. and {Rudnick}, Gregory and {Safonova}, Sasha and {Saha}, Abhijit and {S{\'a}nchez}, F. Javier and {Savary}, Elodie and {Schweiker}, Heidi and {Scott}, Adam and {Seo}, Hee-Jong and {Shan}, Huanyuan and {Silva}, David R. and {Slepian}, Zachary and {Soto}, Christian and {Sprayberry}, David and {Staten}, Ryan and {Stillman}, Coley M. and {Stupak}, Robert J. and {Summers}, David L. and {Sien Tie}, Suk and {Tirado}, H. and {Vargas-Maga{\~n}a}, Mariana and {Vivas}, A. Katherina and {Wechsler}, Risa H. and {Williams}, Doug and {Yang}, Jinyi and {Yang}, Qian and {Yapici}, Tolga and {Zaritsky}, Dennis and {Zenteno}, A. and {Zhang}, Kai and {Zhang}, Tianmeng and {Zhou}, Rongpu and {Zhou}, Zhimin},
        title = "{Overview of the DESI Legacy Imaging Surveys}",
      journal = {\aj},
         year = 2019,
        month = may,
       volume = {157},
       number = {5},
          eid = {168},
        pages = {168},
          doi = {10.3847/1538-3881/ab089d},
archivePrefix = {arXiv},
       eprint = {1804.08657},
 primaryClass = {astro-ph.IM},
       adsurl = {https://ui.adsabs.harvard.edu/abs/2019AJ....157..168D}
}

@ARTICLE{SDSS-IV,
       author = {{Blanton}, Michael R. and {Bershady}, Matthew A. and {Abolfathi}, Bela and {Albareti}, Franco D. and {Allende Prieto}, Carlos and {Almeida}, Andres and {Alonso-Garc{\'\i}a}, Javier and {Anders}, Friedrich and {Anderson}, Scott F. and {Andrews}, Brett and {Aquino-Ort{\'\i}z}, Erik and {Arag{\'o}n-Salamanca}, Alfonso and {Argudo-Fern{\'a}ndez}, Maria and {Armengaud}, Eric and {Aubourg}, Eric and {Avila-Reese}, Vladimir and {Badenes}, Carles and {Bailey}, Stephen and {Barger}, Kathleen A. and {Barrera-Ballesteros}, Jorge and {Bartosz}, Curtis and {Bates}, Dominic and {Baumgarten}, Falk and {Bautista}, Julian and {Beaton}, Rachael and {Beers}, Timothy C. and {Belfiore}, Francesco and {Bender}, Chad F. and {Berlind}, Andreas A. and {Bernardi}, Mariangela and {Beutler}, Florian and {Bird}, Jonathan C. and {Bizyaev}, Dmitry and {Blanc}, Guillermo A. and {Blomqvist}, Michael and {Bolton}, Adam S. and {Boquien}, M{\'e}d{\'e}ric and {Borissova}, Jura and {van den Bosch}, Remco and {Bovy}, Jo and {Brandt}, William N. and {Brinkmann}, Jonathan and {Brownstein}, Joel R. and {Bundy}, Kevin and {Burgasser}, Adam J. and {Burtin}, Etienne and {Busca}, Nicol{\'a}s G. and {Cappellari}, Michele and {Delgado Carigi}, Maria Leticia and {Carlberg}, Joleen K. and {Carnero Rosell}, Aurelio and {Carrera}, Ricardo and {Chanover}, Nancy J. and {Cherinka}, Brian and {Cheung}, Edmond and {G{\'o}mez Maqueo Chew}, Yilen and {Chiappini}, Cristina and {Choi}, Peter Doohyun and {Chojnowski}, Drew and {Chuang}, Chia-Hsun and {Chung}, Haeun and {Cirolini}, Rafael Fernando and {Clerc}, Nicolas and {Cohen}, Roger E. and {Comparat}, Johan and {da Costa}, Luiz and {Cousinou}, Marie-Claude and {Covey}, Kevin and {Crane}, Jeffrey D. and {Croft}, Rupert A.~C. and {Cruz-Gonzalez}, Irene and {Garrido Cuadra}, Daniel and {Cunha}, Katia and {Damke}, Guillermo J. and {Darling}, Jeremy and {Davies}, Roger and {Dawson}, Kyle and {de la Macorra}, Axel and {Dell'Agli}, Flavia and {De Lee}, Nathan and {Delubac}, Timoth{\'e}e and {Di Mille}, Francesco and {Diamond-Stanic}, Aleks and {Cano-D{\'\i}az}, Mariana and {Donor}, John and {Downes}, Juan Jos{\'e} and {Drory}, Niv and {du Mas des Bourboux}, H{\'e}lion and {Duckworth}, Christopher J. and {Dwelly}, Tom and {Dyer}, Jamie and {Ebelke}, Garrett and {Eigenbrot}, Arthur D. and {Eisenstein}, Daniel J. and {Emsellem}, Eric and {Eracleous}, Mike and {Escoffier}, Stephanie and {Evans}, Michael L. and {Fan}, Xiaohui and {Fern{\'a}ndez-Alvar}, Emma and {Fernandez-Trincado}, J.~G. and {Feuillet}, Diane K. and {Finoguenov}, Alexis and {Fleming}, Scott W. and {Font-Ribera}, Andreu and {Fredrickson}, Alexander and {Freischlad}, Gordon and {Frinchaboy}, Peter M. and {Fuentes}, Carla E. and {Galbany}, Llu{\'\i}s and {Garcia-Dias}, R. and {Garc{\'\i}a-Hern{\'a}ndez}, D.~A. and {Gaulme}, Patrick and {Geisler}, Doug and {Gelfand}, Joseph D. and {Gil-Mar{\'\i}n}, H{\'e}ctor and {Gillespie}, Bruce A. and {Goddard}, Daniel and {Gonzalez-Perez}, Violeta and {Grabowski}, Kathleen and {Green}, Paul J. and {Grier}, Catherine J. and {Gunn}, James E. and {Guo}, Hong and {Guy}, Julien and {Hagen}, Alex and {Hahn}, ChangHoon and {Hall}, Matthew and {Harding}, Paul and {Hasselquist}, Sten and {Hawley}, Suzanne L. and {Hearty}, Fred and {Gonzalez Hern{\'a}ndez}, Jonay I. and {Ho}, Shirley and {Hogg}, David W. and {Holley-Bockelmann}, Kelly and {Holtzman}, Jon A. and {Holzer}, Parker H. and {Huehnerhoff}, Joseph and {Hutchinson}, Timothy A. and {Hwang}, Ho Seong and {Ibarra-Medel}, H{\'e}ctor J. and {da Silva Ilha}, Gabriele and {Ivans}, Inese I. and {Ivory}, KeShawn and {Jackson}, Kelly and {Jensen}, Trey W. and {Johnson}, Jennifer A. and {Jones}, Amy and {J{\"o}nsson}, Henrik and {Jullo}, Eric and {Kamble}, Vikrant and {Kinemuchi}, Karen and {Kirkby}, David and {Kitaura}, Francisco-Shu and {Klaene}, Mark and {Knapp}, Gillian R. and {Kneib}, Jean-Paul and {Kollmeier}, Juna A. and {Lacerna}, Ivan and {Lane}, Richard R. and {Lang}, Dustin and {Law}, David R. and {Lazarz}, Daniel and {Lee}, Youngbae and {Le Goff}, Jean-Marc and {Liang}, Fu-Heng and {Li}, Cheng and {Li}, Hongyu and {Lian}, Jianhui and {Lima}, Marcos and {Lin}, Lihwai and {Lin}, Yen-Ting and {Bertran de Lis}, Sara and {Liu}, Chao and {de Icaza Lizaola}, Miguel Angel C. and {Long}, Dan and {Lucatello}, Sara and {Lundgren}, Britt and {MacDonald}, Nicholas K. and {Deconto Machado}, Alice and {MacLeod}, Chelsea L. and {Mahadevan}, Suvrath and {Geimba Maia}, Marcio Antonio and {Maiolino}, Roberto and {Majewski}, Steven R. and {Malanushenko}, Elena and {Malanushenko}, Viktor and {Manchado}, Arturo and {Mao}, Shude and {Maraston}, Claudia and {Marques-Chaves}, Rui and {Masseron}, Thomas and {Masters}, Karen L. and {McBride}, Cameron K. and {McDermid}, Richard M. and {McGrath}, Brianne and {McGreer}, Ian D. and {Medina Pe{\~n}a}, Nicol{\'a}s and {Melendez}, Matthew},
        title = "{Sloan Digital Sky Survey IV: Mapping the Milky Way, Nearby Galaxies, and the Distant Universe}",
      journal = {\aj},
         year = 2017,
        month = jul,
       volume = {154},
       number = {1},
          eid = {28},
        pages = {28},
          doi = {10.3847/1538-3881/aa7567},
archivePrefix = {arXiv},
       eprint = {1703.00052},
 primaryClass = {astro-ph.GA},
       adsurl = {https://ui.adsabs.harvard.edu/abs/2017AJ....154...28B}
}

@ARTICLE{PS1,
       author = {{Chambers}, K.~C. and {Magnier}, E.~A. and {Metcalfe}, N. and {Flewelling}, H.~A. and {Huber}, M.~E. and {Waters}, C.~Z. and {Denneau}, L. and {Draper}, P.~W. and {Farrow}, D. and {Finkbeiner}, D.~P. and {Holmberg}, C. and {Koppenhoefer}, J. and {Price}, P.~A. and {Rest}, A. and {Saglia}, R.~P. and {Schlafly}, E.~F. and {Smartt}, S.~J. and {Sweeney}, W. and {Wainscoat}, R.~J. and {Burgett}, W.~S. and {Chastel}, S. and {Grav}, T. and {Heasley}, J.~N. and {Hodapp}, K.~W. and {Jedicke}, R. and {Kaiser}, N. and {Kudritzki}, R. -P. and {Luppino}, G.~A. and {Lupton}, R.~H. and {Monet}, D.~G. and {Morgan}, J.~S. and {Onaka}, P.~M. and {Shiao}, B. and {Stubbs}, C.~W. and {Tonry}, J.~L. and {White}, R. and {Ba{\~n}ados}, E. and {Bell}, E.~F. and {Bender}, R. and {Bernard}, E.~J. and {Boegner}, M. and {Boffi}, F. and {Botticella}, M.~T. and {Calamida}, A. and {Casertano}, S. and {Chen}, W. -P. and {Chen}, X. and {Cole}, S. and {Deacon}, N. and {Frenk}, C. and {Fitzsimmons}, A. and {Gezari}, S. and {Gibbs}, V. and {Goessl}, C. and {Goggia}, T. and {Gourgue}, R. and {Goldman}, B. and {Grant}, P. and {Grebel}, E.~K. and {Hambly}, N.~C. and {Hasinger}, G. and {Heavens}, A.~F. and {Heckman}, T.~M. and {Henderson}, R. and {Henning}, T. and {Holman}, M. and {Hopp}, U. and {Ip}, W. -H. and {Isani}, S. and {Jackson}, M. and {Keyes}, C.~D. and {Koekemoer}, A.~M. and {Kotak}, R. and {Le}, D. and {Liska}, D. and {Long}, K.~S. and {Lucey}, J.~R. and {Liu}, M. and {Martin}, N.~F. and {Masci}, G. and {McLean}, B. and {Mindel}, E. and {Misra}, P. and {Morganson}, E. and {Murphy}, D.~N.~A. and {Obaika}, A. and {Narayan}, G. and {Nieto-Santisteban}, M.~A. and {Norberg}, P. and {Peacock}, J.~A. and {Pier}, E.~A. and {Postman}, M. and {Primak}, N. and {Rae}, C. and {Rai}, A. and {Riess}, A. and {Riffeser}, A. and {Rix}, H.~W. and {R{\"o}ser}, S. and {Russel}, R. and {Rutz}, L. and {Schilbach}, E. and {Schultz}, A.~S.~B. and {Scolnic}, D. and {Strolger}, L. and {Szalay}, A. and {Seitz}, S. and {Small}, E. and {Smith}, K.~W. and {Soderblom}, D.~R. and {Taylor}, P. and {Thomson}, R. and {Taylor}, A.~N. and {Thakar}, A.~R. and {Thiel}, J. and {Thilker}, D. and {Unger}, D. and {Urata}, Y. and {Valenti}, J. and {Wagner}, J. and {Walder}, T. and {Walter}, F. and {Watters}, S.~P. and {Werner}, S. and {Wood-Vasey}, W.~M. and {Wyse}, R.},
        title = "{The Pan-STARRS1 Surveys}",
      journal = {arXiv e-prints},
         year = 2016,
        month = dec,
          eid = {arXiv:1612.05560},
        pages = {arXiv:1612.05560},
          doi = {10.48550/arXiv.1612.05560},
archivePrefix = {arXiv},
       eprint = {1612.05560},
 primaryClass = {astro-ph.IM},
       adsurl = {https://ui.adsabs.harvard.edu/abs/2016arXiv161205560C}
}

@ARTICLE{Chen2018,
       author = {{Chen}, Hsiao-Wen and {Zahedy}, Fakhri S. and {Johnson}, Sean D. and {Pierce}, Rebecca M. and {Huang}, Yun-Hsin and {Weiner}, Benjamin J. and {Gauthier}, Jean-Ren{\'e}},
        title = "{Characterizing circumgalactic gas around massive ellipticals at z {\ensuremath{\sim}} 0.4 - I. Initial results}",
      journal = {\mnras},
         year = 2018,
        month = sep,
       volume = {479},
       number = {2},
        pages = {2547-2563},
          doi = {10.1093/mnras/sty1541},
archivePrefix = {arXiv},
       eprint = {1805.07364},
 primaryClass = {astro-ph.GA},
       adsurl = {https://ui.adsabs.harvard.edu/abs/2018MNRAS.479.2547C}
}

@ARTICLE{Wakker2015,
       author = {{Wakker}, Bart P. and {Hernandez}, Audra K. and {French}, David M. and {Kim}, Tae-Sun and {Oppenheimer}, Benjamin D. and {Savage}, Blair D.},
        title = "{Nearby Galaxy Filaments and the Ly-alpha Forest: Confronting Simulations and the UV Background with Observations}",
      journal = {\apj},
         year = 2015,
        month = nov,
       volume = {814},
       number = {1},
          eid = {40},
        pages = {40},
          doi = {10.1088/0004-637X/814/1/40},
archivePrefix = {arXiv},
       eprint = {1504.02539},
 primaryClass = {astro-ph.CO},
       adsurl = {https://ui.adsabs.harvard.edu/abs/2015ApJ...814...40W}
}

@ARTICLE{Wakker2006,
       author = {{Wakker}, B.~P.},
        title = "{A FUSE Survey of High-Latitude Galactic Molecular Hydrogen}",
      journal = {\apjs},
         year = 2006,
        month = apr,
       volume = {163},
       number = {2},
        pages = {282-305},
          doi = {10.1086/500365},
archivePrefix = {arXiv},
       eprint = {astro-ph/0512444},
 primaryClass = {astro-ph},
       adsurl = {https://ui.adsabs.harvard.edu/abs/2006ApJS..163..282W}
}

@ARTICLE{LAB_HI,
       author = {{Kalberla}, P.~M.~W. and {Burton}, W.~B. and {Hartmann}, Dap and {Arnal}, E.~M. and {Bajaja}, E. and {Morras}, R. and {P{\"o}ppel}, W.~G.~L.},
        title = "{The Leiden/Argentine/Bonn (LAB) Survey of Galactic HI. Final data release of the combined LDS and IAR surveys with improved stray-radiation corrections}",
      journal = {\aap},
         year = 2005,
        month = sep,
       volume = {440},
       number = {2},
        pages = {775-782},
          doi = {10.1051/0004-6361:20041864},
archivePrefix = {arXiv},
       eprint = {astro-ph/0504140},
 primaryClass = {astro-ph},
       adsurl = {https://ui.adsabs.harvard.edu/abs/2005A&A...440..775K}
}

@ARTICLE{Huang2021,
       author = {{Huang}, Yun-Hsin and {Chen}, Hsiao-Wen and {Shectman}, Stephen A. and {Johnson}, Sean D. and {Zahedy}, Fakhri S. and {Helsby}, Jennifer E. and {Gauthier}, Jean-Ren{\'e} and {Thompson}, Ian B.},
        title = "{A complete census of circumgalactic Mg II at redshift z {\ensuremath{\lesssim}} 0.5}",
      journal = {\mnras},
         year = 2021,
        month = apr,
       volume = {502},
       number = {4},
        pages = {4743-4761},
          doi = {10.1093/mnras/stab360},
archivePrefix = {arXiv},
       eprint = {2009.12372},
 primaryClass = {astro-ph.GA},
       adsurl = {https://ui.adsabs.harvard.edu/abs/2021MNRAS.502.4743H}
}

@ARTICLE{Tumlinson2013,
       author = {{Tumlinson}, Jason and {Thom}, Christopher and {Werk}, Jessica K. and {Prochaska}, J. Xavier and {Tripp}, Todd M. and {Katz}, Neal and {Dav{\'e}}, Romeel and {Oppenheimer}, Benjamin D. and {Meiring}, Joseph D. and {Ford}, Amanda Brady and {O'Meara}, John M. and {Peeples}, Molly S. and {Sembach}, Kenneth R. and {Weinberg}, David H.},
        title = "{The COS-Halos Survey: Rationale, Design, and a Census of Circumgalactic Neutral Hydrogen}",
      journal = {\apj},
         year = 2013,
        month = nov,
       volume = {777},
       number = {1},
          eid = {59},
        pages = {59},
          doi = {10.1088/0004-637X/777/1/59},
archivePrefix = {arXiv},
       eprint = {1309.6317},
 primaryClass = {astro-ph.CO},
       adsurl = {https://ui.adsabs.harvard.edu/abs/2013ApJ...777...59T}
}

@ARTICLE{Zahedy2021,
       author = {{Zahedy}, Fakhri S. and {Chen}, Hsiao-Wen and {Cooper}, Thomas M. and {Boettcher}, Erin and {Johnson}, Sean D. and {Rudie}, Gwen C. and {Chen}, Mandy C. and {Cantalupo}, Sebastiano and {Cooksey}, Kathy L. and {Faucher-Gigu{\`e}re}, Claude-Andr{\'e} and {Greene}, Jenny E. and {Lopez}, Sebastian and {Mulchaey}, John S. and {Penton}, Steven V. and {Petitjean}, Patrick and {Putman}, Mary E. and {Rafelski}, Marc and {Rauch}, Michael and {Schaye}, Joop and {Simcoe}, Robert A. and {Walth}, Gregory L.},
        title = "{The Cosmic Ultraviolet Baryon Survey (CUBS) - III. Physical properties and elemental abundances of Lyman-limit systems at z < 1}",
      journal = {\mnras},
         year = 2021,
        month = sep,
       volume = {506},
       number = {1},
        pages = {877-902},
          doi = {10.1093/mnras/stab1661},
archivePrefix = {arXiv},
       eprint = {2106.04608},
 primaryClass = {astro-ph.GA},
       adsurl = {https://ui.adsabs.harvard.edu/abs/2021MNRAS.506..877Z}
}

@ARTICLE{Werk2013,
       author = {{Werk}, Jessica K. and {Prochaska}, J. Xavier and {Thom}, Christopher and {Tumlinson}, Jason and {Tripp}, Todd M. and {O'Meara}, John M. and {Peeples}, Molly S.},
        title = "{The COS-Halos Survey: An Empirical Description of Metal-line Absorption in the Low-redshift Circumgalactic Medium}",
      journal = {\apjs},
         year = 2013,
        month = feb,
       volume = {204},
       number = {2},
          eid = {17},
        pages = {17},
          doi = {10.1088/0067-0049/204/2/17},
archivePrefix = {arXiv},
       eprint = {1212.0558},
 primaryClass = {astro-ph.CO},
       adsurl = {https://ui.adsabs.harvard.edu/abs/2013ApJS..204...17W}
}

@ARTICLE{Chen2020,
       author = {{Chen}, Hsiao-Wen and {Zahedy}, Fakhri S. and {Boettcher}, Erin and {Cooper}, Thomas M. and {Johnson}, Sean D. and {Rudie}, Gwen C. and {Chen}, Mandy C. and {Walth}, Gregory L. and {Cantalupo}, Sebastiano and {Cooksey}, Kathy L. and {Faucher-Gigu{\`e}re}, Claude-Andr{\'e} and {Greene}, Jenny E. and {Lopez}, Sebastian and {Mulchaey}, John S. and {Penton}, Steven V. and {Petitjean}, Patrick and {Putman}, Mary E. and {Rafelski}, Marc and {Rauch}, Michael and {Schaye}, Joop and {Simcoe}, Robert A. and {Weiner}, Benjamin J.},
        title = "{The Cosmic Ultraviolet Baryon Survey (CUBS) - I. Overview and the diverse environments of Lyman limit systems at z < 1}",
      journal = {\mnras},
         year = 2020,
        month = sep,
       volume = {497},
       number = {1},
        pages = {498-520},
          doi = {10.1093/mnras/staa1773},
archivePrefix = {arXiv},
       eprint = {2005.02408},
 primaryClass = {astro-ph.GA},
       adsurl = {https://ui.adsabs.harvard.edu/abs/2020MNRAS.497..498C}
}

@ARTICLE{Zahedy2019,
       author = {{Zahedy}, Fakhri S. and {Chen}, Hsiao-Wen and {Johnson}, Sean D. and {Pierce}, Rebecca M. and {Rauch}, Michael and {Huang}, Yun-Hsin and {Weiner}, Benjamin J. and {Gauthier}, Jean-Ren{\'e}},
        title = "{Characterizing circumgalactic gas around massive ellipticals at z {\ensuremath{\sim}} 0.4 - II. Physical properties and elemental abundances}",
      journal = {\mnras},
         year = 2019,
        month = apr,
       volume = {484},
       number = {2},
        pages = {2257-2280},
          doi = {10.1093/mnras/sty3482},
archivePrefix = {arXiv},
       eprint = {1809.05115},
 primaryClass = {astro-ph.GA},
       adsurl = {https://ui.adsabs.harvard.edu/abs/2019MNRAS.484.2257Z}
}

@ARTICLE{Savage2014,
       author = {{Savage}, B.~D. and {Kim}, T. -S. and {Wakker}, B.~P. and {Keeney}, B. and {Shull}, J.~M. and {Stocke}, J.~T. and {Green}, J.~C.},
        title = "{The Properties of Low Redshift Intergalactic O VI Absorbers Determined from High S/N Observations of 14 QSOs with the Cosmic Origins Spectrograph}",
      journal = {\apjs},
         year = 2014,
        month = may,
       volume = {212},
       number = {1},
          eid = {8},
        pages = {8},
          doi = {10.1088/0067-0049/212/1/8},
archivePrefix = {arXiv},
       eprint = {1403.7542},
 primaryClass = {astro-ph.CO},
       adsurl = {https://ui.adsabs.harvard.edu/abs/2014ApJS..212....8S}
}

@ARTICLE{Kim2002,
       author = {{Kim}, T. -S. and {Carswell}, R.~F. and {Cristiani}, S. and {D'Odorico}, S. and {Giallongo}, E.},
        title = "{The physical properties of the Ly{\ensuremath{\alpha}} forest at z > 1.5}",
      journal = {\mnras},
         year = 2002,
        month = sep,
       volume = {335},
       number = {3},
        pages = {555-573},
          doi = {10.1046/j.1365-8711.2002.05599.x},
archivePrefix = {arXiv},
       eprint = {astro-ph/0205237},
 primaryClass = {astro-ph},
       adsurl = {https://ui.adsabs.harvard.edu/abs/2002MNRAS.335..555K}
}

@ARTICLE{Danforth2016,
       author = {{Danforth}, Charles W. and {Keeney}, Brian A. and {Tilton}, Evan M. and {Shull}, J. Michael and {Stocke}, John T. and {Stevans}, Matthew and {Pieri}, Matthew M. and {Savage}, Blair D. and {France}, Kevin and {Syphers}, David and {Smith}, Britton D. and {Green}, James C. and {Froning}, Cynthia and {Penton}, Steven V. and {Osterman}, Steven N.},
        title = "{An HST/COS Survey of the Low-redshift Intergalactic Medium. I. Survey, Methodology, and Overall Results}",
      journal = {\apj},
         year = 2016,
        month = feb,
       volume = {817},
       number = {2},
          eid = {111},
        pages = {111},
          doi = {10.3847/0004-637X/817/2/111},
archivePrefix = {arXiv},
       eprint = {1402.2655},
 primaryClass = {astro-ph.CO},
       adsurl = {https://ui.adsabs.harvard.edu/abs/2016ApJ...817..111D}
}

@ARTICLE{Rudie2012,
       author = {{Rudie}, Gwen C. and {Steidel}, Charles C. and {Pettini}, Max},
        title = "{The Temperature-Density Relation in the Intergalactic Medium at Redshift langzrang = 2.4}",
      journal = {\apjl},
         year = 2012,
        month = oct,
       volume = {757},
       number = {2},
          eid = {L30},
        pages = {L30},
          doi = {10.1088/2041-8205/757/2/L30},
archivePrefix = {arXiv},
       eprint = {1209.0005},
 primaryClass = {astro-ph.CO},
       adsurl = {https://ui.adsabs.harvard.edu/abs/2012ApJ...757L..30R}
}

@ARTICLE{Hui1997,
       author = {{Hui}, Lam and {Gnedin}, Nickolay Y.},
        title = "{Equation of state of the photoionized intergalactic medium}",
      journal = {\mnras},
         year = 1997,
        month = nov,
       volume = {292},
       number = {1},
        pages = {27-42},
          doi = {10.1093/mnras/292.1.27},
archivePrefix = {arXiv},
       eprint = {astro-ph/9612232},
 primaryClass = {astro-ph},
       adsurl = {https://ui.adsabs.harvard.edu/abs/1997MNRAS.292...27H}
}

@ARTICLE{Chen2001,
       author = {{Chen}, Hsiao-Wen and {Lanzetta}, Kenneth M. and {Webb}, John K. and {Barcons}, Xavier},
        title = "{The Gaseous Extent of Galaxies and the Origin of Ly{\ensuremath{\alpha}} Absorption Systems. V. Optical and Near-Infrared Photometry of Ly{\ensuremath{\alpha}}-absorbing Galaxies at z<1}",
      journal = {\apj},
         year = 2001,
        month = oct,
       volume = {559},
       number = {2},
        pages = {654-674},
          doi = {10.1086/322414},
archivePrefix = {arXiv},
       eprint = {astro-ph/0107137},
 primaryClass = {astro-ph},
       adsurl = {https://ui.adsabs.harvard.edu/abs/2001ApJ...559..654C}
}

@ARTICLE{Borthakur2015,
       author = {{Borthakur}, Sanchayeeta and {Heckman}, Timothy and {Tumlinson}, Jason and {Bordoloi}, Rongmon and {Thom}, Christopher and {Catinella}, Barbara and {Schiminovich}, David and {Dav{\'e}}, Romeel and {Kauffmann}, Guinevere and {Moran}, Sean M. and {Saintonge}, Amelie},
        title = "{Connection between the Circumgalactic Medium and the Interstellar Medium of Galaxies: Results from the COS-GASS Survey}",
      journal = {\apj},
         year = 2015,
        month = nov,
       volume = {813},
       number = {1},
          eid = {46},
        pages = {46},
          doi = {10.1088/0004-637X/813/1/46},
archivePrefix = {arXiv},
       eprint = {1504.01392},
 primaryClass = {astro-ph.GA},
       adsurl = {https://ui.adsabs.harvard.edu/abs/2015ApJ...813...46B}
}

@ARTICLE{Borthakur2024,
       author = {{Borthakur}, Sanchayeeta and {Padave}, Mansi and {Heckman}, Timothy and {Gim}, Hansung B. and {Olvera}, Alejandro J. and {Koplitz}, Brad and {Momjian}, Emmanuel and {Jansen}, Rolf A. and {Thilker}, David and {Kauffman}, Guinevere and {Fox}, Andrew J. and {Tumlinson}, Jason and {Kennicutt}, Robert C. and {Nelson}, Dylan and {Monckiewicz}, Jacqueline and {Naab}, Thorsten},
        title = "{DIISC Survey: Deciphering the Interplay Between the Interstellar Medium, Stars, and the Circumgalactic Medium Survey}",
      journal = {arXiv e-prints},
         year = 2024,
        month = sep,
          eid = {arXiv:2409.12554},
        pages = {arXiv:2409.12554},
          doi = {10.48550/arXiv.2409.12554},
archivePrefix = {arXiv},
       eprint = {2409.12554},
 primaryClass = {astro-ph.GA},
       adsurl = {https://ui.adsabs.harvard.edu/abs/2024arXiv240912554B}
}

@ARTICLE{Schaye1999,
       author = {{Schaye}, Joop and {Theuns}, Tom and {Leonard}, Anthony and {Efstathiou}, George},
        title = "{Measuring the equation of state of the intergalactic medium}",
      journal = {\mnras},
         year = 1999,
        month = nov,
       volume = {310},
       number = {1},
        pages = {57-70},
          doi = {10.1046/j.1365-8711.1999.02956.x},
archivePrefix = {arXiv},
       eprint = {astro-ph/9906271},
 primaryClass = {astro-ph},
       adsurl = {https://ui.adsabs.harvard.edu/abs/1999MNRAS.310...57S}
}

@ARTICLE{Chatzikos2023cloudy,
       author = {{Chatzikos}, M. and {Bianchi}, S. and {Camilloni}, F. and {Chakraborty}, P. and {Gunasekera}, C.~M. and {Guzm{\'a}n}, F. and {Milby}, J.~S. and {Sarkar}, A. and {Shaw}, G. and {van Hoof}, P.~A.~M. and {Ferland}, G.~J.},
        title = "{The 2023 Release of Cloudy}",
      journal = {\rmxaa},
         year = 2023,
        month = oct,
       volume = {59},
        pages = {327-343},
          doi = {10.22201/ia.01851101p.2023.59.02.12},
archivePrefix = {arXiv},
       eprint = {2308.06396},
 primaryClass = {astro-ph.GA},
       adsurl = {https://ui.adsabs.harvard.edu/abs/2023RMxAA..59..327C}
}

@ARTICLE{FG20,
       author = {{Faucher-Gigu{\`e}re}, Claude-Andr{\'e}},
        title = "{A cosmic UV/X-ray background model update}",
      journal = {\mnras},
         year = 2020,
        month = apr,
       volume = {493},
       number = {2},
        pages = {1614-1632},
          doi = {10.1093/mnras/staa302},
archivePrefix = {arXiv},
       eprint = {1903.08657},
 primaryClass = {astro-ph.CO},
       adsurl = {https://ui.adsabs.harvard.edu/abs/2020MNRAS.493.1614F}
}

@ARTICLE{HM12,
       author = {{Haardt}, Francesco and {Madau}, Piero},
        title = "{Radiative Transfer in a Clumpy Universe. IV. New Synthesis Models of the Cosmic UV/X-Ray Background}",
      journal = {\apj},
         year = 2012,
        month = feb,
       volume = {746},
       number = {2},
          eid = {125},
        pages = {125},
          doi = {10.1088/0004-637X/746/2/125},
       adsurl = {https://ui.adsabs.harvard.edu/abs/2012ApJ...746..125H}
}

@ARTICLE{KS19,
       author = {{Khaire}, Vikram and {Srianand}, Raghunathan},
        title = "{New synthesis models of consistent extragalactic background light over cosmic time}",
      journal = {\mnras},
         year = 2019,
        month = apr,
       volume = {484},
       number = {3},
        pages = {4174-4199},
          doi = {10.1093/mnras/stz174},
archivePrefix = {arXiv},
       eprint = {1801.09693},
 primaryClass = {astro-ph.GA},
       adsurl = {https://ui.adsabs.harvard.edu/abs/2019MNRAS.484.4174K}
}

@ARTICLE{Wang2025,
       author = {{Wang}, Jing and {Yang}, Dong and {Lin}, Xuchen and {Huang}, Qifeng and {Qu}, Zhijie and {Chen}, Hsiao-wen and {Guo}, Hong and {Ho}, Luis C. and {Jiang}, Peng and {Liang}, Zezhong and {P{\'e}roux}, C{\'e}line and {Staveley-Smith}, Lister and {Weng}, Simon},
        title = "{FEASTS: Radial Distribution of H I Surface Densities Down to 0.01 M$_{{\ensuremath{\odot}}}$ pc$^{‑2}$ of 35 Nearby Galaxies}",
      journal = {\apj},
         year = 2025,
        month = feb,
       volume = {980},
       number = {1},
          eid = {25},
        pages = {25},
          doi = {10.3847/1538-4357/ada95a},
archivePrefix = {arXiv},
       eprint = {2501.01289},
 primaryClass = {astro-ph.GA},
       adsurl = {https://ui.adsabs.harvard.edu/abs/2025ApJ...980...25W}
}

@ARTICLE{Sharma2017,
       author = {{Sharma}, Sanjib},
        title = "{Markov Chain Monte Carlo Methods for Bayesian Data Analysis in Astronomy}",
      journal = {\araa},
         year = 2017,
        month = aug,
       volume = {55},
       number = {1},
        pages = {213-259},
          doi = {10.1146/annurev-astro-082214-122339},
archivePrefix = {arXiv},
       eprint = {1706.01629},
 primaryClass = {astro-ph.IM},
       adsurl = {https://ui.adsabs.harvard.edu/abs/2017ARA&A..55..213S}
}

@ARTICLE{Stern2018,
       author = {{Stern}, Jonathan and {Faucher-Gigu{\`e}re}, Claude-Andr{\'e} and {Hennawi}, Joseph F. and {Hafen}, Zachary and {Johnson}, Sean D. and {Fielding}, Drummond},
        title = "{Does Circumgalactic O VI Trace Low-pressure Gas Beyond the Accretion Shock? Clues from H I and Low-ion Absorption, Line Kinematics, and Dust Extinction}",
      journal = {\apj},
         year = 2018,
        month = oct,
       volume = {865},
       number = {2},
          eid = {91},
        pages = {91},
          doi = {10.3847/1538-4357/aac884},
archivePrefix = {arXiv},
       eprint = {1803.05446},
 primaryClass = {astro-ph.GA},
       adsurl = {https://ui.adsabs.harvard.edu/abs/2018ApJ...865...91S}
}

@ARTICLE{Tchernyshyov2022,
       author = {{Tchernyshyov}, Kirill and {Werk}, Jessica K. and {Wilde}, Matthew C. and {Prochaska}, J. Xavier and {Tripp}, Todd M. and {Burchett}, Joseph N. and {Bordoloi}, Rongmon and {Howk}, J. Christopher and {Lehner}, Nicolas and {O'Meara}, John M. and {Tejos}, Nicolas and {Tumlinson}, Jason},
        title = "{The CGM$^{2}$ Survey: Circumgalactic O VI from Dwarf to Massive Star-forming Galaxies}",
      journal = {\apj},
         year = 2022,
        month = mar,
       volume = {927},
       number = {2},
          eid = {147},
        pages = {147},
          doi = {10.3847/1538-4357/ac450c},
archivePrefix = {arXiv},
       eprint = {2110.13167},
 primaryClass = {astro-ph.GA},
       adsurl = {https://ui.adsabs.harvard.edu/abs/2022ApJ...927..147T}
}

@ARTICLE{Bordoloi2017,
       author = {{Bordoloi}, Rongmon and {Wagner}, Alexander Y. and {Heckman}, Timothy M. and {Norman}, Colin A.},
        title = "{The Formation and Physical Origin of Highly Ionized Cooling Gas}",
      journal = {\apj},
         year = 2017,
        month = oct,
       volume = {848},
       number = {2},
          eid = {122},
        pages = {122},
          doi = {10.3847/1538-4357/aa8e9c},
archivePrefix = {arXiv},
       eprint = {1605.07187},
 primaryClass = {astro-ph.GA},
       adsurl = {https://ui.adsabs.harvard.edu/abs/2017ApJ...848..122B}
}

@ARTICLE{Goldner2025,
       author = {{Goldner}, Roy and {Stern}, Jonathan and {Fielding}, Drummond and {Faucher-Gigu{\`e}re}, Claude-Andr{\'e} and {Faerman}, Yakov and {Kakoly}, Aharon},
        title = "{Accretion-Driven Turbulence in the Circumgalactic Medium}",
      journal = {arXiv e-prints},
         year = 2025,
        month = oct,
          eid = {arXiv:2510.27678},
        pages = {arXiv:2510.27678},
          doi = {10.48550/arXiv.2510.27678},
archivePrefix = {arXiv},
       eprint = {2510.27678},
 primaryClass = {astro-ph.GA},
       adsurl = {https://ui.adsabs.harvard.edu/abs/2025arXiv251027678G}
}

@ARTICLE{Werk2016,
       author = {{Werk}, Jessica K. and {Prochaska}, J. Xavier and {Cantalupo}, Sebastiano and {Fox}, Andrew J. and {Oppenheimer}, Benjamin and {Tumlinson}, Jason and {Tripp}, Todd M. and {Lehner}, Nicolas and {McQuinn}, Matthew},
        title = "{The COS-Halos Survey: Origins of the Highly Ionized Circumgalactic Medium of Star-Forming Galaxies}",
      journal = {\apj},
         year = 2016,
        month = dec,
       volume = {833},
       number = {1},
          eid = {54},
        pages = {54},
          doi = {10.3847/1538-4357/833/1/54},
archivePrefix = {arXiv},
       eprint = {1609.00012},
 primaryClass = {astro-ph.GA},
       adsurl = {https://ui.adsabs.harvard.edu/abs/2016ApJ...833...54W}
}

@ARTICLE{McCourt2018,
       author = {{McCourt}, Michael and {Oh}, S. Peng and {O'Leary}, Ryan and {Madigan}, Ann-Marie},
        title = "{A characteristic scale for cold gas}",
      journal = {\mnras},
         year = 2018,
        month = feb,
       volume = {473},
       number = {4},
        pages = {5407-5431},
          doi = {10.1093/mnras/stx2687},
archivePrefix = {arXiv},
       eprint = {1610.01164},
 primaryClass = {astro-ph.GA},
       adsurl = {https://ui.adsabs.harvard.edu/abs/2018MNRAS.473.5407M}
}

@ARTICLE{Gronke2020,
       author = {{Gronke}, Max and {Oh}, S. Peng},
        title = "{Is multiphase gas cloudy or misty?}",
      journal = {\mnras},
         year = 2020,
        month = may,
       volume = {494},
       number = {1},
        pages = {L27-L31},
          doi = {10.1093/mnrasl/slaa033},
archivePrefix = {arXiv},
       eprint = {1912.07808},
 primaryClass = {astro-ph.GA},
       adsurl = {https://ui.adsabs.harvard.edu/abs/2020MNRAS.494L..27G}
}

@ARTICLE{Stern2020,
       author = {{Stern}, Jonathan and {Fielding}, Drummond and {Faucher-Gigu{\`e}re}, Claude-Andr{\'e} and {Quataert}, Eliot},
        title = "{The maximum accretion rate of hot gas in dark matter haloes}",
      journal = {\mnras},
         year = 2020,
        month = mar,
       volume = {492},
       number = {4},
        pages = {6042-6058},
          doi = {10.1093/mnras/staa198},
archivePrefix = {arXiv},
       eprint = {1909.07402},
 primaryClass = {astro-ph.GA},
       adsurl = {https://ui.adsabs.harvard.edu/abs/2020MNRAS.492.6042S}
}

@ARTICLE{Kumar2025,
       author = {{Kumar}, Suyash and {Chen}, Hsiao-Wen},
        title = "{Non-equilibrium ionization in the multiphase circumgalactic medium {\textendash} impact on quasar absorption-line analyses}",
      journal = {The Open Journal of Astrophysics},
         year = 2025,
        month = jul,
       volume = {8},
          eid = {98},
        pages = {98},
          doi = {10.33232/001c.142441},
archivePrefix = {arXiv},
       eprint = {2501.13170},
 primaryClass = {astro-ph.GA},
       adsurl = {https://ui.adsabs.harvard.edu/abs/2025OJAp....8E..98K}
}

@ARTICLE{Sutherland1993,
       author = {{Sutherland}, Ralph S. and {Dopita}, M.~A.},
        title = "{Cooling Functions for Low-Density Astrophysical Plasmas}",
      journal = {\apjs},
         year = 1993,
        month = sep,
       volume = {88},
        pages = {253},
          doi = {10.1086/191823},
       adsurl = {https://ui.adsabs.harvard.edu/abs/1993ApJS...88..253S}
}

@ARTICLE{White1978,
       author = {{White}, S.~D.~M. and {Rees}, M.~J.},
        title = "{Core condensation in heavy halos: a two-stage theory for galaxy formation and clustering.}",
      journal = {\mnras},
         year = 1978,
        month = may,
       volume = {183},
        pages = {341-358},
          doi = {10.1093/mnras/183.3.341},
       adsurl = {https://ui.adsabs.harvard.edu/abs/1978MNRAS.183..341W}
}

@ARTICLE{Gnat2017,
       author = {{Gnat}, Orly},
        title = "{Time-dependent Cooling in Photoionized Plasma}",
      journal = {\apjs},
         year = 2017,
        month = feb,
       volume = {228},
       number = {2},
          eid = {11},
        pages = {11},
          doi = {10.3847/1538-4365/228/2/11},
archivePrefix = {arXiv},
       eprint = {1706.09220},
 primaryClass = {astro-ph.GA},
       adsurl = {https://ui.adsabs.harvard.edu/abs/2017ApJS..228...11G}
}

@ARTICLE{Keres2009,
       author = {{Kere{\v{s}}}, Du{\v{s}}an and {Hernquist}, Lars},
        title = "{Seeding the Formation of Cold Gaseous Clouds in Milky Way-Size Halos}",
      journal = {\apjl},
         year = 2009,
        month = jul,
       volume = {700},
       number = {1},
        pages = {L1-L5},
          doi = {10.1088/0004-637X/700/1/L1},
archivePrefix = {arXiv},
       eprint = {0905.2186},
 primaryClass = {astro-ph.CO},
       adsurl = {https://ui.adsabs.harvard.edu/abs/2009ApJ...700L...1K}
}

@ARTICLE{Kirkman1997,
       author = {{Kirkman}, David and {Tytler}, David},
        title = "{Intrinsic Properties of the <z> = 2.7 Ly{\ensuremath{\alpha}} Forest from Keck Spectra of Quasar HS 1946+7658}",
      journal = {\apj},
         year = 1997,
        month = jul,
       volume = {484},
       number = {2},
        pages = {672-694},
          doi = {10.1086/304371},
archivePrefix = {arXiv},
       eprint = {astro-ph/9701209},
 primaryClass = {astro-ph},
       adsurl = {https://ui.adsabs.harvard.edu/abs/1997ApJ...484..672K}
}

@ARTICLE{Dave2001,
       author = {{Dav{\'e}}, Romeel and {Tripp}, Todd M.},
        title = "{The Statistical and Physical Properties of the Low-Redshift LY{\ensuremath{\alpha}} Forest Observed with the Hubble Space Telescope/STIS}",
      journal = {\apj},
         year = 2001,
        month = jun,
       volume = {553},
       number = {2},
        pages = {528-537},
          doi = {10.1086/320977},
archivePrefix = {arXiv},
       eprint = {astro-ph/0101419},
 primaryClass = {astro-ph},
       adsurl = {https://ui.adsabs.harvard.edu/abs/2001ApJ...553..528D}
}

\startlongtable
\begin{longrotatetable}
\begin{deluxetable*}{lcccccccccccccc}
\tablecaption{Summary of galaxy and absorber properties \label{tab:sample}}
\tablehead{
\colhead{} & \colhead{}& \colhead{} & \colhead{} & \colhead{Dist} &\colhead{} &
\colhead{$d_{\rm proj}$} & \colhead{$\log N_{\mathrm{tot},\rm HI}$} & \colhead{$\sigma_{v,\rm HI}$} & \colhead{$\log N_{\mathrm{tot},\rm OVI}$} & \colhead{$\sigma_{v,\rm OVI}$} \\
\colhead{Galaxy} & \colhead{RA}& \colhead{DEC} & \colhead{$z_{\rm gal}$} & \colhead{(Mpc)}& \colhead{AGN} &
\colhead{(kpc)} & \colhead{$/\cmjj$} & \colhead{($\kms$)} & \colhead{$/\cmjj$} & \colhead{($\kms$)} \\
\colhead{(1)} & \colhead{(2)} & \colhead{(3)} & \colhead{(4)} & \colhead{(5)} & \colhead{(6)} & \colhead{(7)} & \colhead{(8)} & \colhead{(9)} & \colhead{(10)} & \colhead{(11)} 
}

\startdata
PGC138064                & $00:02:48.30 $ & $+18:58:07.0 $ & $0.002$ & $8.7$ & MRK335         & $224.7$ & $<12.6$ & $...$ & $<12.5$ & $...$ \\
PGC1620667               & $00:05:29.16 $ & $+20:13:35.9 $ & $0.007$ & $23.1$ & MRK335         & $79.0$ & $13.79_{-0.01}^{+0.01}$ & $35.1_{-1.1}^{+1.0}$ & $<13.3$ & $...$ \\
LEDA1428758              & $00:29:15.39 $ & $+13:20:56.6 $ & $0.040$ & $175.9$ & PG0026+129     & $231.1$ & $14.25_{-0.03}^{+0.03}$ & $33.7_{-1.7}^{+1.9}$ & $<13.5$ & $...$ \\
J002909.2+131628         & $00:29:09.20 $ & $+13:16:28.0 $ & $0.033$ & $144.7$ & PG0026+129     & $46.9$ & $15.35_{-0.14}^{+0.14}$ & $23.0_{-1.0}^{+1.0}$ & $<14.1$ & $...$ \\
SDSSJ002843.85+131421.4  & $00:28:43.85 $ & $+13:14:21.4 $ & $0.031$ & $135.9$ & PG0026+129     & $278.6$ & $13.27_{-0.10}^{+0.09}$ & $16.4_{-4.7}^{+3.7}$ & $<14.2$ & $...$ \\
J040758.1-121224         & $04:07:58.10 $ & $-12:12:24.0 $ & $0.097$ & $443.1$ & PKS0405-123    & $267.1$ & $14.64_{-0.02}^{+0.02}$ & $31.8_{-1.5}^{+1.8}$ & $13.70_{-0.05}^{+0.04}$ & $20.4_{-2.6}^{+3.0}$ \\
J040748.4-1211369        & $04:07:43.20 $ & $-12:11:48.0 $ & $0.091$ & $418.3$ & PKS0405-123    & $132.0$ & $<11.9$ & $...$ & $<13.2$ & $...$ \\
J15-D7                   & $04:07:49.30 $ & $-12:12:16.0 $ & $0.092$ & $422.7$ & PKS0405-123    & $71.0$ & $14.49_{-0.01}^{+0.01}$ & $28.2_{-0.3}^{+0.3}$ & $<13.0$ & $...$ \\
SDSSJ083335.64+250847.1  & $08:33:35.65 $ & $+25:08:47.1 $ & $0.008$ & $39.7$ & PG0832+251     & $326.9$ & $13.94_{-0.06}^{+0.10}$ & $17.0_{-2.2}^{+2.0}$ & $<14.2$ & $...$ \\
LEDA1722581              & $08:35:37.08 $ & $+25:00:15.0 $ & $0.017$ & $82.6$ & PG0832+251     & $13.3$ & $>15.5$ & $74.1_{-16.6}^{+19.0}$ & $15.07_{-0.10}^{+0.14}$ & $83.8_{-16.8}^{+8.9}$ \\
SDSSJ083534.75+245901.9  & $08:35:34.76 $ & $+24:59:01.9 $ & $0.108$ & $501.0$ & PG0832+251     & $89.2$ & $15.44_{-0.09}^{+0.11}$ & $66.2_{-7.5}^{+7.8}$ & $<14.0$ & $...$ \\
2MASXJ08360739+2506457   & $08:36:07.41 $ & $+25:06:45.7 $ & $0.023$ & $109.3$ & PG0832+251     & $303.7$ & $13.57_{-0.07}^{+0.06}$ & $51.5_{-7.7}^{+10.7}$ & $<14.3$ & $...$ \\
UGC04527                 & $08:44:23.52 $ & $+76:55:01.6 $ & $0.002$ & $11.0$ & PG0838+770     & $7.1$ & $>15.6$ & $15.6_{-1.6}^{+1.7}$ & $13.82_{-0.13}^{+0.11}$ & $33.7_{-9.4}^{+13.4}$ \\
J085050.871+771540.17    & $08:50:50.87 $ & $+77:15:40.1 $ & $0.004$ & $17.3$ & PG0838+770     & $152.1$ & $13.40_{-0.04}^{+0.04}$ & $89.8_{-2.8}^{+3.1}$ & $<13.8$ & $...$ \\
J084027.434+770555.60    & $08:40:27.43 $ & $+77:05:55.6 $ & $0.007$ & $30.5$ & PG0838+770     & $168.9$ & $<12.4$ & $...$ & $<13.7$ & $...$ \\
LEDA24388                & $08:40:35.79 $ & $+76:30:28.9 $ & $0.007$ & $32.8$ & PG0838+770     & $252.4$ & $13.16_{-0.06}^{+0.06}$ & $34.0_{-5.5}^{+5.9}$ & $<13.2$ & $...$ \\
SDSSJ095821.73+322551.9  & $09:58:21.73 $ & $+32:25:51.9 $ & $0.080$ & $361.5$ & 3C232          & $165.3$ & $13.83_{-0.04}^{+0.05}$ & $53.0_{-9.0}^{+11.3}$ & $...$ & $...$ \\
NGC3067                  & $09:58:21.07 $ & $+32:22:12.0 $ & $0.005$ & $26.3$ & 3C232          & $13.9$ & $19.86_{-0.03}^{+0.03}$ & $...$ & $14.87_{-0.17}^{+0.15}$ & $123.3_{-36.5}^{+30.8}$ \\
Mrk412                   & $09:57:59.66 $ & $+32:14:22.9 $ & $0.015$ & $71.2$ & 3C232          & $214.2$ & $13.47_{-0.10}^{+0.08}$ & $96.0_{-6.4}^{+6.8}$ & $<14.2$ & $...$ \\
LAMOSTJ100413.72+130156.3 & $10:04:13.72 $ & $+13:01:56.3 $ & $0.003$ & $17.7$ & PG1004+130     & $249.2$ & $<12.4$ & $...$ & $<13.6$ & $...$ \\
PGC1418410               & $10:07:06.51 $ & $+12:53:51.3 $ & $0.009$ & $48.3$ & PG1004+130     & $94.6$ & $13.71_{-0.05}^{+0.05}$ & $33.1_{-3.9}^{+4.7}$ & $<13.8$ & $...$ \\
PGC2806981               & $10:07:33.16 $ & $+13:06:24.7 $ & $0.010$ & $52.9$ & PG1004+130     & $265.1$ & $13.69_{-0.05}^{+0.06}$ & $73.4_{-4.8}^{+5.5}$ & $<13.7$ & $...$ \\
SDSSJ100640.35+121900.4  & $10:06:40.35 $ & $+12:19:00.4 $ & $0.005$ & $30.2$ & PG1004+130     & $278.3$ & $<12.9$ & $...$ & $<14.0$ & $...$ \\
J100730.7+125350         & $10:07:30.74 $ & $+12:53:50.0 $ & $0.030$ & $138.8$ & PG1004+130     & $191.7$ & $14.59_{-0.08}^{+0.09}$ & $16.8_{-1.6}^{+1.6}$ & $<13.8$ & $...$ \\
SDSSJ101930.79+640708.4  & $10:19:30.80 $ & $+64:07:08.4 $ & $0.006$ & $28.7$ & Mrk141         & $76.7$ & $<14.7$ & $...$ & $<13.7$ & $...$ \\
NGC3442                  & $10:53:08.10 $ & $+33:54:37.3 $ & $0.006$ & $33.9$ & PG1048+342     & $176.6$ & $14.89_{-0.05}^{+0.10}$ & $94.6_{-9.8}^{+5.5}$ & $<14.4$ & $...$ \\
NGC3413                  & $10:51:20.71 $ & $+32:45:58.9 $ & $0.002$ & $11.5$ & PG1048+342     & $246.1$ & $<12.3$ & $...$ & $<13.9$ & $...$ \\
SINGGHIPASSJ1118-17      & $11:18:03.10 $ & $-17:38:31.0 $ & $0.004$ & $17.8$ & HE1115-1735    & $70.0$ & $<14.0$ & $...$ & $<13.8$ & $...$ \\
2MASXJ11184632-1751352   & $11:18:46.33 $ & $-17:51:35.8 $ & $0.026$ & $124.3$ & HE1115-1735    & $291.5$ & $14.93_{-0.13}^{+0.15}$ & $19.5_{-3.0}^{+3.4}$ & $<13.7$ & $...$ \\
ESO570-14                & $11:16:54.82 $ & $-17:53:11.9 $ & $0.012$ & $59.2$ & HE1115-1735    & $304.4$ & $14.96_{-0.07}^{+0.07}$ & $125.5_{-9.6}^{+10.1}$ & $<13.5$ & $...$ \\
HIPASSJ1119-17           & $11:19:34.20 $ & $-17:30:37.0 $ & $0.006$ & $32.2$ & HE1115-1735    & $269.9$ & $14.34_{-0.19}^{+0.16}$ & $38.4_{-14.3}^{+23.0}$ & $<13.8$ & $...$ \\
PGC1398872               & $11:22:23.40 $ & $+11:47:38.1 $ & $0.005$ & $32.3$ & MRK734         & $88.1$ & $<13.7$ & $...$ & $<13.6$ & $...$ \\
SDSSJ112135.62+114808.6  & $11:21:35.63 $ & $+11:48:08.6 $ & $0.038$ & $168.6$ & MRK734         & $216.3$ & $<13.9$ & $...$ & $<13.3$ & $...$ \\
2MASXJ11213641+1144142   & $11:21:36.42 $ & $+11:44:13.4 $ & $0.040$ & $176.0$ & MRK734         & $124.0$ & $14.46_{-0.13}^{+0.11}$ & $37.9_{-15.8}^{+16.9}$ & $...$ & $...$ \\
NGC3627                  & $11:20:15.03 $ & $+12:59:28.6 $ & $0.002$ & $12.6$ & MRK734         & $286.4$ & $...$ & $...$ & $<13.7$ & $...$ \\
SDSSJ114005.18+654801.2  & $11:40:05.18 $ & $+65:48:01.3 $ & $0.063$ & $283.7$ & 3C263          & $62.6$ & $15.27_{-0.03}^{+0.03}$ & $49.2_{-0.8}^{+0.9}$ & $14.39_{-0.03}^{+0.03}$ & $39.2_{-3.0}^{+3.1}$ \\
J121413.9+140331         & $12:14:13.95 $ & $+14:03:30.4 $ & $0.065$ & $290.2$ & PG1211+143     & $69.8$ & $15.21_{-0.02}^{+0.02}$ & $42.6_{-0.9}^{+1.0}$ & $14.15_{-0.04}^{+0.04}$ & $39.2_{-3.8}^{+4.6}$ \\
2MASXJ12140964+1404204   & $12:14:09.65 $ & $+14:04:20.4 $ & $0.051$ & $227.2$ & PG1211+143     & $133.8$ & $15.62_{-0.04}^{+0.04}$ & $38.5_{-1.6}^{+1.6}$ & $14.26_{-0.08}^{+0.08}$ & $38.3_{-8.4}^{+10.8}$ \\
NGC4203                  & $12:15:05.05 $ & $+33:11:50.3 $ & $0.004$ & $21.4$ & TON1480        & $13.0$ & $>17.7$ & $39.1_{-2.8}^{+3.8}$ & $14.13_{-0.17}^{+0.14}$ & $42.9_{-15.9}^{+18.1}$ \\
J121903.7+063343         & $12:19:03.75 $ & $+06:33:43.1 $ & $0.013$ & $61.8$ & PG1216+069     & $114.5$ & $14.04_{-0.02}^{+0.01}$ & $41.2_{-2.1}^{+4.6}$ & $13.99_{-0.25}^{+0.16}$ & $40.2_{-16.6}^{+18.1}$ \\
SDSSJ121923.43+063819.7  & $12:19:23.43 $ & $+06:38:19.7 $ & $0.124$ & $578.1$ & PG1216+069     & $89.0$ & $15.20_{-0.03}^{+0.03}$ & $129.5_{-1.8}^{+1.5}$ & $14.66_{-0.01}^{+0.02}$ & $147.0_{-2.7}^{+2.8}$ \\
NGC4319                  & $12:21:43.89 $ & $+75:19:21.0 $ & $0.005$ & $19.4$ & MRK205         & $4.0$ & $17.70_{-0.27}^{+0.22}$ & $14.8_{-1.8}^{+1.2}$ & $<13.4$ & $...$ \\
PGC1207185               & $12:28:15.96 $ & $+01:49:44.1 $ & $0.003$ & $13.1$ & 3C273          & $69.8$ & $14.22_{-0.01}^{+0.01}$ & $26.9_{-0.3}^{+0.3}$ & $13.25_{-0.09}^{+0.07}$ & $27.5_{-4.9}^{+5.5}$ \\
LEDA135803               & $12:27:46.10 $ & $+01:36:01.0 $ & $0.004$ & $24.0$ & 3C273          & $233.6$ & $<12.1$ & $...$ & $<13.1$ & $...$ \\
SDSSJ122910.05+020120.1  & $12:29:10.06 $ & $+02:01:20.1 $ & $0.125$ & $584.9$ & 3C273          & $268.4$ & $<11.9$ & $...$ & $<11.8$ & $...$ \\
SDSSJ122821.60+015645.9  & $12:28:21.57 $ & $+01:56:45.8 $ & $0.011$ & $54.3$ & 3C273          & $200.4$ & $<11.7$ & $...$ & $<12.8$ & $...$ \\
PGC1213772               & $12:29:50.58 $ & $+02:01:53.7 $ & $0.006$ & $32.6$ & 3C273          & $103.5$ & $15.92_{-0.07}^{+0.07}$ & $11.6_{-0.3}^{+0.3}$ & $<13.1$ & $...$ \\
LEDA41395                & $12:31:03.50 $ & $+01:40:32.2 $ & $0.004$ & $18.6$ & 3C273          & $198.5$ & $<11.7$ & $...$ & $<12.5$ & $...$ \\
2dFGRSTGN388Z087         & $12:30:46.77 $ & $+01:16:04.5 $ & $0.095$ & $435.9$ & QSOJ1230+0115  & $112.8$ & $14.40_{-0.01}^{+0.01}$ & $55.3_{-1.4}^{+1.1}$ & $<13.8$ & $...$ \\
LEDA1189825              & $12:30:34.42 $ & $+01:16:24.4 $ & $0.031$ & $136.1$ & QSOJ1230+0115  & $150.0$ & $14.01_{-0.01}^{+0.01}$ & $71.3_{-0.9}^{+1.1}$ & $14.44_{-0.08}^{+0.07}$ & $58.7_{-3.3}^{+3.6}$ \\
SDSSJ123047.60+011518.6  & $12:30:47.60 $ & $+01:15:18.6 $ & $0.078$ & $352.8$ & QSOJ1230+0115  & $53.3$ & $14.85_{-0.03}^{+0.03}$ & $55.6_{-0.8}^{+0.6}$ & $14.57_{-0.04}^{+0.04}$ & $60.5_{-3.2}^{+3.5}$ \\
SDSSJ123133.26+201928.3  & $12:31:33.26 $ & $+20:19:28.6 $ & $0.004$ & $26.7$ & PG1229+204     & $94.3$ & $14.07_{-0.04}^{+0.04}$ & $23.0_{-1.2}^{+1.4}$ & $<14.0$ & $...$ \\
PGC041463                & $12:31:42.37 $ & $+20:28:53.9 $ & $0.006$ & $34.4$ & PG1229+204     & $198.0$ & $13.93_{-0.04}^{+0.05}$ & $19.2_{-1.4}^{+1.2}$ & $<13.4$ & $...$ \\
SDSSJ122928.18+203348.8  & $12:29:28.18 $ & $+20:33:48.7 $ & $0.004$ & $20.8$ & PG1229+204     & $263.5$ & $<12.7$ & $...$ & $<13.7$ & $...$ \\
UGC07697                 & $12:32:51.62 $ & $+20:11:01.3 $ & $0.008$ & $43.8$ & PG1229+204     & $142.6$ & $13.85_{-0.03}^{+0.03}$ & $26.0_{-1.3}^{+1.4}$ & $<13.4$ & $...$ \\
SDSSJ122724.99+191548.8  & $12:27:24.99 $ & $+19:15:48.8 $ & $0.002$ & $9.6$ & PG1229+204     & $236.1$ & $<13.0$ & $...$ & $<13.6$ & $...$ \\
IC3436                   & $12:30:29.94 $ & $+19:40:22.9 $ & $0.003$ & $13.1$ & PG1229+204     & $138.0$ & $<12.5$ & $...$ & $<13.6$ & $...$ \\
SDSSJ123549.46+201755.0  & $12:35:49.47 $ & $+20:17:55.0 $ & $0.003$ & $11.5$ & PG1229+204     & $178.1$ & $<12.8$ & $...$ & $<13.5$ & $...$ \\
SDSSJ130114.75+590343.3  & $13:01:14.75 $ & $+59:03:43.3 $ & $0.046$ & $204.3$ & PG1259+593     & $88.5$ & $15.44_{-0.01}^{+0.01}$ & $41.9_{-0.6}^{+0.5}$ & $14.15_{-0.02}^{+0.02}$ & $40.2_{-1.2}^{+1.3}$ \\
SDSSJ125926.76+591735.0  & $12:59:26.79 $ & $+59:17:35.0 $ & $0.010$ & $47.8$ & PG1259+593     & $281.0$ & $13.36_{-0.02}^{+0.02}$ & $47.2_{-2.7}^{+3.1}$ & $<13.1$ & $...$ \\
UGC8146                  & $13:02:07.44 $ & $+58:41:53.8 $ & $0.002$ & $10.2$ & PG1259+593     & $63.5$ & $13.96_{-0.01}^{+0.01}$ & $48.6_{-1.7}^{+2.0}$ & $13.55_{-0.05}^{+0.05}$ & $41.9_{-5.6}^{+6.2}$ \\
J130532.1-103356         & $13:05:32.20 $ & $-10:33:58.0 $ & $0.094$ & $428.9$ & PKS1302-102    & $70.5$ & $16.79_{-0.07}^{+0.07}$ & $35.2_{-2.2}^{+3.0}$ & $13.70_{-0.07}^{+0.07}$ & $40.2_{-7.0}^{+8.5}$ \\
2MASXJ13052026-1036311   & $13:05:20.24 $ & $-10:36:30.7 $ & $0.043$ & $188.2$ & PKS1302-102    & $225.5$ & $14.86_{-0.05}^{+0.06}$ & $21.1_{-1.2}^{+1.2}$ & $14.28_{-0.03}^{+0.03}$ & $31.8_{-2.5}^{+2.9}$ \\
PGC986100                & $13:04:56.18 $ & $-09:48:49.7 $ & $0.005$ & $24.5$ & PKS1302-102    & $320.9$ & $15.26_{-0.14}^{+0.14}$ & $13.3_{-1.1}^{+1.8}$ & $<13.3$ & $...$ \\
J131014.0+081859         & $13:10:14.00 $ & $+08:18:59.0 $ & $0.034$ & $148.2$ & PG1307+085     & $271.2$ & $<12.5$ & $...$ & $<13.7$ & $...$ \\
2MASSJ13094426+0820039   & $13:09:44.27 $ & $+08:20:04.0 $ & $0.128$ & $598.2$ & PG1307+085     & $99.5$ & $<12.3$ & $...$ & $<13.4$ & $...$ \\
IC4213                   & $13:12:11.22 $ & $+35:40:10.7 $ & $0.003$ & $12.3$ & PG1309+355     & $88.4$ & $14.68_{-0.04}^{+0.06}$ & $104.3_{-7.1}^{+5.9}$ & $<13.8$ & $...$ \\
UGC08318                 & $13:14:30.64 $ & $+35:23:12.4 $ & $0.008$ & $38.4$ & PG1309+355     & $310.4$ & $<14.4$ & $...$ & $<14.3$ & $...$ \\
UGC08839                 & $13:55:24.95 $ & $+17:47:41.8 $ & $0.003$ & $13.0$ & PG1352+183     & $79.3$ & $13.10_{-0.09}^{+0.07}$ & $34.1_{-7.5}^{+8.1}$ & $<14.1$ & $...$ \\
SDSSJ154527.12+484642.2  & $15:45:27.12 $ & $+48:46:42.4 $ & $0.075$ & $339.8$ & PG1543+489     & $64.8$ & $19.20_{-0.01}^{+0.01}$ & $19.5_{-1.3}^{+1.0}$ & $<13.9$ & $...$ \\
J154530.3+4846093        & $15:45:19.85 $ & $+48:47:48.4 $ & $0.038$ & $168.4$ & PG1543+489     & $108.2$ & $<12.3$ & $...$ & $<14.0$ & $...$ \\
SDSSJ154535.86+484814.0  & $15:45:35.87 $ & $+48:48:14.0 $ & $0.097$ & $444.6$ & PG1543+489     & $245.3$ & $<12.8$ & $...$ & $<13.8$ & $...$ \\
SDSSJ162950.48+552129.1  & $16:29:50.47 $ & $+55:21:29.1 $ & $0.011$ & $48.7$ & PG1626+555     & $225.9$ & $<12.4$ & $...$ & $<12.9$ & $...$ \\
SDSSJ162742.56+552733.3  & $16:27:42.56 $ & $+55:27:33.4 $ & $0.009$ & $37.8$ & PG1626+555     & $58.2$ & $<12.0$ & $...$ & $<13.4$ & $...$ \\
SDSSJ162839.05+550547.6  & $16:28:39.06 $ & $+55:05:47.6 $ & $0.009$ & $41.3$ & PG1626+555     & $210.2$ & $<12.3$ & $...$ & $<13.4$ & $...$ \\
J170422.99+604332.4      & $17:04:22.99 $ & $+60:43:32.4 $ & $0.086$ & $392.1$ & 3C351          & $236.6$ & $12.89_{-0.11}^{+0.09}$ & $22.1_{-6.4}^{+7.0}$ & $<13.4$ & $...$ \\
J170420.25+604403.7      & $17:04:20.25 $ & $+60:44:03.7 $ & $0.098$ & $448.0$ & 3C351          & $283.5$ & $13.66_{-0.22}^{+0.15}$ & $27.0_{-11.3}^{+13.4}$ & $<13.2$ & $...$ \\
SDSSJ170434.33+604447.3  & $17:04:34.34 $ & $+60:44:47.4 $ & $0.092$ & $421.7$ & 3C351          & $93.1$ & $13.95_{-0.22}^{+0.12}$ & $59.8_{-6.5}^{+5.8}$ & $13.59_{-0.11}^{+0.10}$ & $21.4_{-3.4}^{+4.1}$ \\
J170441.3+60443035       & $17:04:39.44 $ & $+60:44:00.1 $ & $0.071$ & $321.8$ & 3C351          & $45.6$ & $14.29_{-0.04}^{+0.05}$ & $44.0_{-2.6}^{+3.3}$ & $<14.5$ & $...$ \\
J170423.43+604301.9      & $17:04:23.43 $ & $+60:43:01.9 $ & $0.070$ & $315.7$ & 3C351          & $212.1$ & $13.70_{-0.04}^{+0.04}$ & $43.7_{-3.8}^{+4.3}$ & $<13.9$ & $...$ \\
NGC6307                  & $17:07:40.47 $ & $+60:45:02.7 $ & $0.010$ & $43.7$ & 3C351          & $272.4$ & $<12.8$ & $...$ & $<13.8$ & $...$ \\
IC4889                   & $19:45:15.17 $ & $-54:20:38.9 $ & $0.009$ & $33.0$ & LEDA63618      & $56.6$ & $15.51_{-0.12}^{+0.17}$ & $52.5_{-7.5}^{+7.5}$ & $<13.3$ & $...$ \\
1XMMJ213758.7-143611     & $21:37:58.70 $ & $-14:36:10.0 $ & $0.052$ & $233.4$ & PKS2135-14     & $282.4$ & $13.50_{-0.03}^{+0.02}$ & $24.9_{-2.1}^{+2.2}$ & $<14.3$ & $...$ \\
LEDA923368               & $21:37:45.08 $ & $-14:32:06.2 $ & $0.075$ & $340.3$ & PKS2135-14     & $69.6$ & $14.07_{-0.01}^{+0.01}$ & $78.4_{-2.0}^{+2.2}$ & $<13.3$ & $...$ \\
LEDA190747               & $21:54:56.66 $ & $-09:18:08.5 $ & $0.052$ & $229.4$ & PHL1811        & $267.8$ & $13.85_{-0.01}^{+0.01}$ & $42.7_{-2.1}^{+2.1}$ & $13.95_{-0.08}^{+0.07}$ & $19.6_{-4.5}^{+5.0}$ \\
J215450.8-092233         & $21:54:50.87 $ & $-09:22:33.3 $ & $0.079$ & $356.6$ & PHL1811        & $234.1$ & $15.31_{-0.04}^{+0.04}$ & $27.1_{-1.2}^{+1.1}$ & $<12.5$ & $...$ \\
WISEAJ215447.57-092254.3 & $21:54:47.58 $ & $-09:22:54.3 $ & $0.077$ & $350.7$ & PHL1811        & $304.9$ & $15.85_{-0.05}^{+0.05}$ & $13.5_{-0.4}^{+0.4}$ & $<13.5$ & $...$ \\
2MASSJ21545996-0922249   & $21:54:59.96 $ & $-09:22:24.8 $ & $0.081$ & $367.1$ & PHL1811        & $34.7$ & $17.90_{-0.05}^{+0.06}$ & $10.4_{-0.3}^{+0.4}$ & $13.27_{-0.12}^{+0.11}$ & $18.2_{-7.7}^{+9.3}$ \\
J215506.5-092325         & $21:55:06.50 $ & $-09:23:25.0 $ & $0.132$ & $621.7$ & PHL1811        & $224.0$ & $14.66_{-0.01}^{+0.01}$ & $32.9_{-1.0}^{+1.1}$ & $14.09_{-0.02}^{+0.02}$ & $62.6_{-1.9}^{+1.8}$ \\
6dFGSgJ234020.6-545155   & $23:40:20.54 $ & $-54:51:53.2 $ & $0.005$ & $20.6$ & HE2336-5540    & $198.1$ & $<12.7$ & $...$ & $<13.7$ & $...$ \\
J233914.158-552344.63    & $23:39:14.16 $ & $-55:23:44.6 $ & $0.026$ & $108.5$ & HE2336-5540    & $4.8$ & $20.60_{-0.05}^{+0.05}$ & $...$ & $...$ & $...$ \\
\enddata
\end{deluxetable*}
\end{longrotatetable}

\startlongtable
    \begin{deluxetable*}{lcccccc}
    \tablecaption{Summary of Individual Absorption Components of \ion{H}{1} and \ion{O}{6}}
    \label{tab:idv}
\tablehead{
\colhead{Galaxy} & \colhead{$z_{\rm gal}$} & \colhead{QSO} &\colhead{Ion} & \colhead{$\log N/\cmjj$} & \colhead{$b ~(\rm\kms)$} &\colhead{$v_{\rm c} ~(\rm\kms)$}
}
\startdata
PGC1207185               & $0.003$ & 3C273          & HI       & $14.23_{-0.01}^{+0.01}$ & $38.0_{-0.4}^{+0.4}$ & $89.0_{-0.2}^{+0.2}$ \\
PGC1207185               & $0.003$ & 3C273          & OVI      & $13.25_{-0.08}^{+0.07}$ & $38.8_{-6.8}^{+7.8}$ & $98.2_{-5.4}^{+5.8}$ \\
LEDA41395                & $0.004$ & 3C273          & HI       & $<11.7$ & $...$ & $...$ \\
LEDA41395                & $0.004$ & 3C273          & OVI      & $<12.5$ & $...$ & $...$ \\
LEDA135803               & $0.004$ & 3C273          & HI       & $<12.1$ & $...$ & $...$ \\
LEDA135803               & $0.004$ & 3C273          & OVI      & $<13.1$ & $...$ & $...$ \\
PGC1213772               & $0.006$ & 3C273          & HI       & $13.62_{-0.15}^{+0.11}$ & $37.3_{-2.4}^{+2.5}$ & $64.5_{-6.1}^{+3.5}$ \\
PGC1213772               & $0.006$ & 3C273          & HI       & $15.91_{-0.07}^{+0.08}$ & $15.9_{-0.4}^{+0.3}$ & $80.0_{-0.3}^{+0.4}$ \\
PGC1213772               & $0.006$ & 3C273          & HI       & $12.34_{-0.13}^{+0.14}$ & $42.1_{-17.3}^{+22.4}$ & $191.3$ \\
PGC1213772               & $0.006$ & 3C273          & OVI      & $<13.1$ & $...$ & $...$ \\
SDSSJ122821.60+015645.9  & $0.011$ & 3C273          & HI       & $<11.8$ & $...$ & $...$ \\
SDSSJ122821.60+015645.9  & $0.011$ & 3C273          & OVI      & $<12.8$ & $...$ & $...$ \\
SDSSJ122910.05+020120.1  & $0.125$ & 3C273          & HI       & $<11.9$ & $...$ & $...$ \\
SDSSJ122910.05+020120.1  & $0.125$ & 3C273          & OVI      & $<11.8$ & $...$ & $...$ \\
PGC138064                & $0.002$ & MRK335         & HI       & $<12.6$ & $...$ & $...$ \\
PGC138064                & $0.002$ & MRK335         & OVI      & $<12.5$ & $...$ & $...$ \\
PGC1620667               & $0.007$ & MRK335         & HI       & $13.79_{-0.01}^{+0.01}$ & $49.4_{-1.4}^{+1.7}$ & $138.3_{-1.0}^{+1.0}$ \\
PGC1620667               & $0.007$ & MRK335         & OVI      & $<13.3$ & $...$ & $...$ \\
UGC8146                  & $0.002$ & PG1259+593     & HI       & $13.81_{-0.05}^{+0.04}$ & $69.8_{-4.4}^{+4.8}$ & $54.7_{-3.8}^{+2.4}$ \\
UGC8146                  & $0.002$ & PG1259+593     & HI       & $13.41_{-0.09}^{+0.09}$ & $33.9_{-5.8}^{+6.7}$ & $99.7_{-3.7}^{+3.8}$ \\
UGC8146                  & $0.002$ & PG1259+593     & OVI      & $13.55_{-0.05}^{+0.05}$ & $59.2_{-7.9}^{+8.7}$ & $54.7_{-4.8}^{+5.2}$ \\
SDSSJ125926.76+591735.0  & $0.010$ & PG1259+593     & HI       & $12.94_{-0.08}^{+0.09}$ & $47.0_{-10.0}^{+12.2}$ & $102.2_{-6.9}^{+11.5}$ \\
SDSSJ125926.76+591735.0  & $0.010$ & PG1259+593     & HI       & $13.15_{-0.07}^{+0.04}$ & $26.5_{-3.3}^{+3.1}$ & $182.9_{-2.4}^{+2.5}$ \\
SDSSJ125926.76+591735.0  & $0.010$ & PG1259+593     & OVI      & $<13.1$ & $...$ & $...$ \\
SDSSJ130114.75+590343.3  & $0.046$ & PG1259+593     & HI       & $15.37_{-0.02}^{+0.02}$ & $33.7_{-0.6}^{+0.5}$ & $29.9_{-0.7}^{+0.6}$ \\
SDSSJ130114.75+590343.3  & $0.046$ & PG1259+593     & HI       & $14.53_{-0.02}^{+0.02}$ & $32.2_{-1.8}^{+2.0}$ & $115.1_{-1.7}^{+1.5}$ \\
SDSSJ130114.75+590343.3  & $0.046$ & PG1259+593     & HI       & $13.62_{-0.03}^{+0.03}$ & $36.7_{-3.0}^{+3.1}$ & $205.1_{-2.4}^{+2.5}$ \\
SDSSJ130114.75+590343.3  & $0.046$ & PG1259+593     & OVI      & $13.88_{-0.04}^{+0.04}$ & $31.9_{-4.1}^{+4.9}$ & $46.8_{-2.3}^{+2.3}$ \\
SDSSJ130114.75+590343.3  & $0.046$ & PG1259+593     & OVI      & $13.82_{-0.05}^{+0.04}$ & $24.5_{-2.8}^{+3.3}$ & $116.5_{-2.0}^{+2.1}$ \\
2MASXJ12140964+1404204   & $0.051$ & PG1211+143     & HI       & $13.12_{-0.05}^{+0.06}$ & $29.7_{-6.0}^{+7.1}$ & $-105.5_{-3.6}^{+5.0}$ \\
2MASXJ12140964+1404204   & $0.051$ & PG1211+143     & HI       & $15.27_{-0.06}^{+0.05}$ & $39.0_{-1.3}^{+1.4}$ & $20.4_{-2.6}^{+1.9}$ \\
2MASXJ12140964+1404204   & $0.051$ & PG1211+143     & HI       & $15.33_{-0.08}^{+0.08}$ & $16.4_{-2.9}^{+2.4}$ & $42.7_{-2.5}^{+2.9}$ \\
2MASXJ12140964+1404204   & $0.051$ & PG1211+143     & HI       & $14.26_{-0.04}^{+0.04}$ & $44.3_{-4.0}^{+3.9}$ & $130.0_{-5.1}^{+3.6}$ \\
2MASXJ12140964+1404204   & $0.051$ & PG1211+143     & HI       & $13.59_{-0.02}^{+0.02}$ & $35.6_{-2.3}^{+3.0}$ & $260.0_{-2.0}^{+2.1}$ \\
2MASXJ12140964+1404204   & $0.051$ & PG1211+143     & OVI      & $14.26_{-0.08}^{+0.08}$ & $54.1_{-11.7}^{+15.5}$ & $12.0_{-5.3}^{+4.7}$ \\
J121413.9+140331         & $0.065$ & PG1211+143     & HI       & $15.19_{-0.02}^{+0.02}$ & $32.7_{-0.5}^{+0.6}$ & $106.5_{-0.6}^{+0.6}$ \\
J121413.9+140331         & $0.065$ & PG1211+143     & HI       & $13.74_{-0.05}^{+0.04}$ & $29.0_{-3.5}^{+3.8}$ & $221.0_{-1.9}^{+1.8}$ \\
J121413.9+140331         & $0.065$ & PG1211+143     & HI       & $12.92_{-0.17}^{+0.22}$ & $<36.6$ & $271.6_{-5.4}^{+4.0}$ \\
J121413.9+140331         & $0.065$ & PG1211+143     & HI       & $13.45_{-0.13}^{+0.12}$ & $79.7_{-16.5}^{+15.4}$ & $311.6_{-20.3}^{+18.0}$ \\
J121413.9+140331         & $0.065$ & PG1211+143     & OVI      & $14.15_{-0.04}^{+0.04}$ & $55.4_{-5.3}^{+6.4}$ & $135.8_{-4.4}^{+4.0}$ \\
J121413.9+140331         & $0.065$ & PG1211+143     & OVI      & $13.60_{-0.12}^{+0.11}$ & $24.1_{-5.5}^{+6.9}$ & $220.2_{-3.9}^{+3.5}$ \\
NGC4319                  & $0.005$ & MRK205         & HI       & $13.76_{-0.16}^{+0.13}$ & $<24.7$ & $2.9_{-3.1}^{+5.3}$ \\
NGC4319                  & $0.005$ & MRK205         & HI       & $17.32_{-0.25}^{+0.31}$ & $17.9_{-1.3}^{+1.4}$ & $68.2$ \\
NGC4319                  & $0.005$ & MRK205         & HI       & $>16.1$ & $15.8_{-2.0}^{+3.6}$ & $88.2$ \\
NGC4319                  & $0.005$ & MRK205         & HI       & $15.02_{-0.16}^{+0.19}$ & $32.7_{-3.7}^{+5.2}$ & $120.9_{-3.4}^{+3.7}$ \\
NGC4319                  & $0.005$ & MRK205         & OVI      & $<13.3$ & $...$ & $...$ \\
J040748.4-1211369        & $0.091$ & PKS0405-12     & HI       & $<11.8$ & $...$ & $...$ \\
J040748.4-1211369        & $0.091$ & PKS0405-12     & OVI      & $<13.2$ & $...$ & $...$ \\
J15\_D7                   & $0.092$ & PKS0405-12     & HI       & $14.48_{-0.01}^{+0.01}$ & $40.0_{-0.5}^{+0.5}$ & $-54.1_{-0.2}^{+0.2}$ \\
J15\_D7                   & $0.092$ & PKS0405-12     & OVI      & $<13.1$ & $...$ & $...$ \\
J040758.1-121224         & $0.097$ & PKS0405-12     & HI       & $14.56_{-0.02}^{+0.02}$ & $32.5_{-1.2}^{+1.1}$ & $145.6_{-0.5}^{+0.4}$ \\
J040758.1-121224         & $0.097$ & PKS0405-12     & HI       & $13.81_{-0.05}^{+0.06}$ & $89.1_{-10.8}^{+12.2}$ & $164.3_{-4.7}^{+6.6}$ \\
J040758.1-121224         & $0.097$ & PKS0405-12     & OVI      & $13.70_{-0.05}^{+0.04}$ & $28.9_{-3.7}^{+4.3}$ & $144.3_{-2.9}^{+2.8}$ \\
PGC986100                & $0.005$ & PKS1302-102    & HI       & $15.25_{-0.14}^{+0.13}$ & $17.2_{-1.5}^{+1.6}$ & $109.2_{-1.2}^{+1.0}$ \\
PGC986100                & $0.005$ & PKS1302-102    & HI       & $13.49_{-0.28}^{+0.30}$ & $47.4_{-11.6}^{+17.1}$ & $127.5_{-9.1}^{+24.2}$ \\
PGC986100                & $0.005$ & PKS1302-102    & OVI      & $<13.3$ & $...$ & $...$ \\
2MASXJ13052026-1036311   & $0.043$ & PKS1302-102    & HI       & $12.95_{-0.04}^{+0.04}$ & $13.9_{-2.1}^{+2.4}$ & $-21.9_{-1.6}^{+1.8}$ \\
2MASXJ13052026-1036311   & $0.043$ & PKS1302-102    & HI       & $14.73_{-0.06}^{+0.06}$ & $13.5_{-1.4}^{+1.1}$ & $58.7_{-1.2}^{+1.2}$ \\
2MASXJ13052026-1036311   & $0.043$ & PKS1302-102    & HI       & $14.20_{-0.04}^{+0.03}$ & $43.4_{-1.3}^{+1.4}$ & $80.7_{-1.2}^{+1.6}$ \\
2MASXJ13052026-1036311   & $0.043$ & PKS1302-102    & OVI      & $14.28_{-0.03}^{+0.03}$ & $44.9_{-3.5}^{+4.0}$ & $100.8_{-2.3}^{+2.1}$ \\
J130532.1-103356         & $0.094$ & PKS1302-102    & HI       & $14.98_{-0.04}^{+0.04}$ & $26.3_{-0.5}^{+0.7}$ & $266.8_{-0.6}^{+0.7}$ \\
J130532.1-103356         & $0.094$ & PKS1302-102    & HI       & $16.75_{-0.09}^{+0.09}$ & $23.9_{-1.4}^{+1.6}$ & $491.8_{-2.9}^{+2.3}$ \\
J130532.1-103356         & $0.094$ & PKS1302-102    & HI       & $15.61_{-0.12}^{+0.18}$ & $<21.1$ & $547.1_{-8.1}^{+5.9}$ \\
J130532.1-103356         & $0.094$ & PKS1302-102    & HI       & $14.76_{-0.39}^{+0.34}$ & $51.0_{-6.1}^{+9.6}$ & $504.1_{-3.2}^{+2.2}$ \\
J130532.1-103356         & $0.094$ & PKS1302-102    & OVI      & $13.70_{-0.07}^{+0.07}$ & $56.9_{-9.9}^{+11.4}$ & $505.8_{-6.2}^{+6.8}$ \\
LEDA190747               & $0.052$ & PHL1811        & HI       & $13.72_{-0.05}^{+0.03}$ & $64.4_{-2.9}^{+3.4}$ & $128.3_{-3.4}^{+2.4}$ \\
LEDA190747               & $0.052$ & PHL1811        & HI       & $13.23_{-0.12}^{+0.12}$ & $25.7_{-4.4}^{+5.0}$ & $148.1_{-2.4}^{+2.4}$ \\
LEDA190747               & $0.052$ & PHL1811        & OVI      & $13.95_{-0.08}^{+0.07}$ & $27.7_{-6.4}^{+7.1}$ & $174.3_{-3.7}^{+3.9}$ \\
J215450.8-092233         & $0.079$ & PHL1811        & HI       & $15.19_{-0.04}^{+0.04}$ & $20.5_{-0.9}^{+0.9}$ & $256.4_{-1.6}^{+1.5}$ \\
J215450.8-092233         & $0.079$ & PHL1811        & HI       & $14.68_{-0.08}^{+0.09}$ & $14.0_{-1.2}^{+1.2}$ & $311.4_{-1.8}^{+2.0}$ \\
J215450.8-092233         & $0.079$ & PHL1811        & OVI      & $<12.5$ & $...$ & $...$ \\
WISEAJ215447.57-092254.3 & $0.077$ & PHL1811        & HI       & $15.84_{-0.05}^{+0.05}$ & $18.5_{-0.5}^{+0.5}$ & $205.7_{-0.5}^{+0.5}$ \\
WISEAJ215447.57-092254.3 & $0.077$ & PHL1811        & HI       & $13.71_{-0.14}^{+0.16}$ & $53.0_{-6.9}^{+7.2}$ & $206.4_{-3.1}^{+2.7}$ \\
WISEAJ215447.57-092254.3 & $0.077$ & PHL1811        & OVI      & $<13.5$ & $...$ & $...$ \\
2MASSJ21545996-0922249   & $0.081$ & PHL1811        & HI       & $13.32_{-0.29}^{+0.34}$ & $49.8_{-14.9}^{+15.8}$ & $131.5_{-22.9}^{+30.3}$ \\
2MASSJ21545996-0922249   & $0.081$ & PHL1811        & HI       & $15.13_{-0.07}^{+0.06}$ & $24.1_{-2.2}^{+2.2}$ & $169.9_{-3.0}^{+3.5}$ \\
2MASSJ21545996-0922249   & $0.081$ & PHL1811        & HI       & $16.20_{-0.36}^{+0.19}$ & $17.5_{-3.1}^{+2.8}$ & $240.7_{-7.1}^{+4.5}$ \\
2MASSJ21545996-0922249   & $0.081$ & PHL1811        & HI       & $17.89_{-0.05}^{+0.06}$ & $11.6_{-1.2}^{+1.8}$ & $270.7_{-4.9}^{+3.6}$ \\
2MASSJ21545996-0922249   & $0.081$ & PHL1811        & OVI      & $13.26_{-0.12}^{+0.11}$ & $25.8_{-11.0}^{+13.0}$ & $166.1_{-5.2}^{+6.7}$ \\
J215506.5-092325         & $0.132$ & PHL1811        & HI       & $14.64_{-0.01}^{+0.02}$ & $31.6_{-0.5}^{+0.5}$ & $70.8_{-0.5}^{+0.4}$ \\
J215506.5-092325         & $0.132$ & PHL1811        & HI       & $13.25_{-0.04}^{+0.06}$ & $62.7_{-7.6}^{+11.4}$ & $189.1_{-6.9}^{+5.4}$ \\
J215506.5-092325         & $0.132$ & PHL1811        & OVI      & $13.93_{-0.03}^{+0.03}$ & $63.5_{-4.6}^{+5.2}$ & $113.9_{-3.6}^{+3.7}$ \\
J215506.5-092325         & $0.132$ & PHL1811        & OVI      & $13.57_{-0.05}^{+0.04}$ & $15.9_{-2.7}^{+3.3}$ & $221.4_{-1.5}^{+1.6}$ \\
J215506.5-092325         & $0.132$ & PHL1811        & OVI      & $13.70_{-0.03}^{+0.03}$ & $31.2_{-3.3}^{+3.6}$ & $307.3_{-1.9}^{+2.0}$ \\
J131014.0+081859         & $0.034$ & PG1307+085     & HI       & $<12.5$ & $...$ & $...$ \\
J131014.0+081859         & $0.034$ & PG1307+085     & OVI      & $<13.7$ & $...$ & $...$ \\
2MASSJ13094426+0820039   & $0.128$ & PG1307+085     & HI       & $<12.4$ & $...$ & $...$ \\
2MASSJ13094426+0820039   & $0.128$ & PG1307+085     & OVI      & $<13.4$ & $...$ & $...$ \\
LEDA1189825              & $0.031$ & QSOJ1230+0115  & HI       & $13.71_{-0.01}^{+0.01}$ & $42.7_{-1.4}^{+1.3}$ & $-65.7_{-0.9}^{+1.0}$ \\
LEDA1189825              & $0.031$ & QSOJ1230+0115  & HI       & $13.71_{-0.02}^{+0.01}$ & $62.1_{-2.8}^{+2.9}$ & $55.5_{-1.6}^{+1.7}$ \\
LEDA1189825              & $0.031$ & QSOJ1230+0115  & OVI      & $14.06_{-0.11}^{+0.10}$ & $27.8_{-6.3}^{+10.0}$ & $-52.7_{-5.1}^{+6.1}$ \\
LEDA1189825              & $0.031$ & QSOJ1230+0115  & OVI      & $14.19_{-0.10}^{+0.10}$ & $30.1_{-5.2}^{+7.2}$ & $58.9_{-5.4}^{+4.8}$ \\
SDSSJ123047.60+011518.6  & $0.078$ & QSOJ1230+0115  & HI       & $14.43_{-0.02}^{+0.02}$ & $36.4_{-0.8}^{+0.7}$ & $183.9_{-0.4}^{+0.4}$ \\
SDSSJ123047.60+011518.6  & $0.078$ & QSOJ1230+0115  & HI       & $14.64_{-0.04}^{+0.04}$ & $27.5_{-0.6}^{+0.7}$ & $288.8_{-0.3}^{+0.3}$ \\
SDSSJ123047.60+011518.6  & $0.078$ & QSOJ1230+0115  & OVI      & $14.11_{-0.09}^{+0.09}$ & $42.5_{-7.3}^{+11.6}$ & $178.8_{-7.4}^{+9.8}$ \\
SDSSJ123047.60+011518.6  & $0.078$ & QSOJ1230+0115  & OVI      & $14.37_{-0.07}^{+0.05}$ & $41.6_{-6.1}^{+7.2}$ & $290.0_{-5.2}^{+4.8}$ \\
2dFGRSTGN388Z087         & $0.095$ & QSOJ1230+0115  & HI       & $12.95_{-0.04}^{+0.05}$ & $68.8_{-5.9}^{+8.5}$ & $-166.3_{-5.5}^{+5.8}$ \\
2dFGRSTGN388Z087         & $0.095$ & QSOJ1230+0115  & HI       & $13.40_{-0.07}^{+0.08}$ & $36.5_{-3.6}^{+3.9}$ & $-47.9_{-3.7}^{+4.2}$ \\
2dFGRSTGN388Z087         & $0.095$ & QSOJ1230+0115  & HI       & $14.02_{-0.11}^{+0.07}$ & $31.2_{-4.1}^{+3.0}$ & $34.3_{-2.3}^{+1.7}$ \\
2dFGRSTGN388Z087         & $0.095$ & QSOJ1230+0115  & HI       & $14.04_{-0.09}^{+0.11}$ & $52.3_{-2.9}^{+3.6}$ & $45.4_{-3.7}^{+5.4}$ \\
2dFGRSTGN388Z087         & $0.095$ & QSOJ1230+0115  & OVI      & $<13.8$ & $...$ & $...$ \\
SDSSJ002843.85+131421.4  & $0.031$ & PG0026+129     & HI       & $13.27_{-0.09}^{+0.07}$ & $23.2_{-5.9}^{+4.8}$ & $73.3_{-4.6}^{+5.0}$ \\
SDSSJ002843.85+131421.4  & $0.031$ & PG0026+129     & OVI      & $<14.1$ & $...$ & $...$ \\
J002909.2+131628         & $0.033$ & PG0026+129     & HI       & $15.35_{-0.14}^{+0.14}$ & $33.1_{-1.5}^{+1.7}$ & $379.8_{-0.9}^{+1.0}$ \\
J002909.2+131628         & $0.033$ & PG0026+129     & OVI      & $<14.0$ & $...$ & $...$ \\
LEDA1428758              & $0.040$ & PG0026+129     & HI       & $14.25_{-0.03}^{+0.03}$ & $48.0_{-2.5}^{+2.3}$ & $24.5_{-1.2}^{+1.5}$ \\
LEDA1428758              & $0.040$ & PG0026+129     & OVI      & $<13.5$ & $...$ & $...$ \\
NGC3627                  & $0.002$ & MRK734         & OVI      & $<13.7$ & $...$ & $...$ \\
PGC1398872               & $0.005$ & MRK734         & HI       & $<13.7$ & $...$ & $...$ \\
PGC1398872               & $0.005$ & MRK734         & OVI      & $<13.6$ & $...$ & $...$ \\
SDSSJ112135.62+114808.6  & $0.038$ & MRK734         & HI       & $<13.9$ & $...$ & $...$ \\
SDSSJ112135.62+114808.6  & $0.038$ & MRK734         & OVI      & $<13.3$ & $...$ & $...$ \\
2MASXJ11213641+1144142   & $0.040$ & MRK734         & HI       & $14.46_{-0.13}^{+0.11}$ & $53.3_{-21.4}^{+24.2}$ & $260.6_{-9.7}^{+9.8}$ \\
SDSSJ114005.18+654801.2  & $0.063$ & 3C263          & HI       & $13.85_{-0.03}^{+0.02}$ & $25.9_{-1.5}^{+1.8}$ & $-37.6_{-1.4}^{+1.6}$ \\
SDSSJ114005.18+654801.2  & $0.063$ & 3C263          & HI       & $15.03_{-0.03}^{+0.03}$ & $56.4_{-1.3}^{+1.3}$ & $87.3_{-1.4}^{+1.4}$ \\
SDSSJ114005.18+654801.2  & $0.063$ & 3C263          & HI       & $14.86_{-0.07}^{+0.07}$ & $20.6_{-1.7}^{+1.8}$ & $140.0_{-2.3}^{+2.3}$ \\
SDSSJ114005.18+654801.2  & $0.063$ & 3C263          & OVI      & $14.39_{-0.03}^{+0.03}$ & $55.7_{-4.4}^{+4.3}$ & $115.3_{-2.8}^{+2.8}$ \\
SDSSJ122724.99+191548.8  & $0.002$ & PG1229+204     & HI       & $<13.0$ & $...$ & $...$ \\
SDSSJ122724.99+191548.8  & $0.002$ & PG1229+204     & OVI      & $<13.6$ & $...$ & $...$ \\
SDSSJ123549.46+201755.0  & $0.003$ & PG1229+204     & HI       & $<12.8$ & $...$ & $...$ \\
SDSSJ123549.46+201755.0  & $0.003$ & PG1229+204     & OVI      & $<13.5$ & $...$ & $...$ \\
IC3436                   & $0.003$ & PG1229+204     & HI       & $<12.5$ & $...$ & $...$ \\
IC3436                   & $0.003$ & PG1229+204     & OVI      & $<13.8$ & $...$ & $...$ \\
SDSSJ122928.18+203348.8  & $0.004$ & PG1229+204     & HI       & $<12.7$ & $...$ & $...$ \\
SDSSJ122928.18+203348.8  & $0.004$ & PG1229+204     & OVI      & $<13.7$ & $...$ & $...$ \\
SDSSJ123133.26+201928.3  & $0.004$ & PG1229+204     & HI       & $14.07_{-0.04}^{+0.04}$ & $32.6_{-1.8}^{+2.0}$ & $-25.9_{-1.1}^{+1.2}$ \\
SDSSJ123133.26+201928.3  & $0.004$ & PG1229+204     & OVI      & $<14.0$ & $...$ & $...$ \\
PGC041463                & $0.006$ & PG1229+204     & HI       & $13.93_{-0.04}^{+0.04}$ & $27.2_{-1.9}^{+1.9}$ & $76.9_{-1.2}^{+1.2}$ \\
PGC041463                & $0.006$ & PG1229+204     & OVI      & $<13.5$ & $...$ & $...$ \\
UGC07697                 & $0.008$ & PG1229+204     & HI       & $13.85_{-0.03}^{+0.03}$ & $36.6_{-2.1}^{+2.2}$ & $139.0_{-1.5}^{+1.4}$ \\
UGC07697                 & $0.008$ & PG1229+204     & OVI      & $<13.4$ & $...$ & $...$ \\
LAMOSTJ100413.72+130156.3 & $0.003$ & PG1004+130     & HI       & $<12.5$ & $...$ & $...$ \\
LAMOSTJ100413.72+130156.3 & $0.003$ & PG1004+130     & OVI      & $<13.6$ & $...$ & $...$ \\
SDSSJ100640.35+121900.4  & $0.005$ & PG1004+130     & HI       & $<13.0$ & $...$ & $...$ \\
SDSSJ100640.35+121900.4  & $0.005$ & PG1004+130     & OVI      & $<14.0$ & $...$ & $...$ \\
PGC1418410               & $0.009$ & PG1004+130     & HI       & $13.72_{-0.05}^{+0.04}$ & $45.8_{-5.0}^{+6.5}$ & $61.3_{-4.2}^{+4.2}$ \\
PGC1418410               & $0.009$ & PG1004+130     & OVI      & $<13.8$ & $...$ & $...$ \\
PGC2806981               & $0.010$ & PG1004+130     & HI       & $13.35_{-0.08}^{+0.06}$ & $21.6_{-4.2}^{+4.7}$ & $-54.8_{-3.0}^{+2.9}$ \\
PGC2806981               & $0.010$ & PG1004+130     & HI       & $13.44_{-0.09}^{+0.07}$ & $58.7_{-12.4}^{+13.1}$ & $79.3_{-9.1}^{+8.8}$ \\
PGC2806981               & $0.010$ & PG1004+130     & OVI      & $<13.7$ & $...$ & $...$ \\
J100730.7+125350         & $0.030$ & PG1004+130     & HI       & $14.59_{-0.08}^{+0.09}$ & $23.7_{-2.2}^{+2.3}$ & $329.1_{-2.3}^{+2.4}$ \\
J100730.7+125350         & $0.030$ & PG1004+130     & OVI      & $<13.8$ & $...$ & $...$ \\
SDSSJ162742.56+552733.3  & $0.009$ & PG1626+555     & HI       & $<12.0$ & $...$ & $...$ \\
SDSSJ162742.56+552733.3  & $0.009$ & PG1626+555     & OVI      & $<13.4$ & $...$ & $...$ \\
SDSSJ162839.05+550547.6  & $0.009$ & PG1626+555     & HI       & $<12.3$ & $...$ & $...$ \\
SDSSJ162839.05+550547.6  & $0.009$ & PG1626+555     & OVI      & $<13.4$ & $...$ & $...$ \\
SDSSJ162950.48+552129.1  & $0.011$ & PG1626+555     & HI       & $<12.4$ & $...$ & $...$ \\
SDSSJ162950.48+552129.1  & $0.011$ & PG1626+555     & OVI      & $<12.9$ & $...$ & $...$ \\
SINGGHIPASSJ1118-17      & $0.004$ & HE1115-1735    & HI       & $<14.0$ & $...$ & $...$ \\
SINGGHIPASSJ1118-17      & $0.004$ & HE1115-1735    & OVI      & $<13.8$ & $...$ & $...$ \\
HIPASSJ1119-17           & $0.006$ & HE1115-1735    & HI       & $14.34_{-0.18}^{+0.14}$ & $54.1_{-19.8}^{+33.3}$ & $163.7_{-11.6}^{+14.7}$ \\
HIPASSJ1119-17           & $0.006$ & HE1115-1735    & OVI      & $<13.9$ & $...$ & $...$ \\
ESO570-14                & $0.012$ & HE1115-1735    & HI       & $14.40_{-0.15}^{+0.12}$ & $74.1_{-24.4}^{+35.9}$ & $59.5_{-0.0}^{+0.0}$ \\
ESO570-14                & $0.012$ & HE1115-1735    & HI       & $14.35_{-0.14}^{+0.11}$ & $<46.2$ & $206.9_{-7.3}^{+5.4}$ \\
ESO570-14                & $0.012$ & HE1115-1735    & HI       & $14.63_{-0.09}^{+0.09}$ & $34.3_{-4.9}^{+6.0}$ & $347.7_{-4.4}^{+4.1}$ \\
ESO570-14                & $0.012$ & HE1115-1735    & OVI      & $<13.5$ & $...$ & $...$ \\
2MASXJ11184632-1751352   & $0.026$ & HE1115-1735    & HI       & $14.92_{-0.13}^{+0.15}$ & $27.6_{-4.3}^{+4.9}$ & $209.4_{-3.3}^{+3.8}$ \\
2MASXJ11184632-1751352   & $0.026$ & HE1115-1735    & OVI      & $<13.6$ & $...$ & $...$ \\
NGC4203                  & $0.004$ & TON1480        & HI       & $>17.7$ & $55.2_{-3.9}^{+5.3}$ & $97.2_{-6.0}^{+5.2}$ \\
NGC4203                  & $0.004$ & TON1480        & OVI      & $14.14_{-0.17}^{+0.13}$ & $60.7_{-22.6}^{+24.7}$ & $111.5_{-13.3}^{+18.5}$ \\
J121903.7+063343         & $0.013$ & PG1216+069     & HI       & $13.96_{-0.08}^{+0.03}$ & $44.2_{-4.7}^{+3.9}$ & $-104.8_{-2.2}^{+2.1}$ \\
J121903.7+063343         & $0.013$ & PG1216+069     & HI       & $13.02_{-0.25}^{+0.47}$ & $44.8_{-19.9}^{+49.9}$ & $-21.3_{-50.1}^{+17.6}$ \\
J121903.7+063343         & $0.013$ & PG1216+069     & OVI      & $13.98_{-0.25}^{+0.16}$ & $57.2_{-19.2}^{+24.1}$ & $-112.1_{-13.3}^{+15.7}$ \\
SDSSJ121923.43+063819.7  & $0.124$ & PG1216+069     & HI       & $14.51_{-0.04}^{+0.03}$ & $26.3_{-1.6}^{+1.2}$ & $147.3_{-1.8}^{+1.6}$ \\
SDSSJ121923.43+063819.7  & $0.124$ & PG1216+069     & HI       & $14.72_{-0.03}^{+0.04}$ & $26.5_{-1.3}^{+1.3}$ & $224.1_{-1.8}^{+1.7}$ \\
SDSSJ121923.43+063819.7  & $0.124$ & PG1216+069     & HI       & $14.32_{-0.21}^{+0.08}$ & $44.8_{-6.2}^{+3.2}$ & $419.5_{-13.8}^{+5.0}$ \\
SDSSJ121923.43+063819.7  & $0.124$ & PG1216+069     & HI       & $14.59_{-0.12}^{+0.12}$ & $16.2_{-3.0}^{+3.2}$ & $433.3_{-2.2}^{+1.6}$ \\
SDSSJ121923.43+063819.7  & $0.124$ & PG1216+069     & HI       & $14.13_{-0.06}^{+0.05}$ & $24.8_{-2.0}^{+1.8}$ & $502.5_{-2.5}^{+2.4}$ \\
SDSSJ121923.43+063819.7  & $0.124$ & PG1216+069     & OVI      & $13.44_{-0.10}^{+0.08}$ & $21.3_{-5.9}^{+9.3}$ & $60.8_{-4.3}^{+4.2}$ \\
SDSSJ121923.43+063819.7  & $0.124$ & PG1216+069     & OVI      & $14.14_{-0.04}^{+0.03}$ & $24.7_{-3.1}^{+3.0}$ & $155.3_{-2.5}^{+2.5}$ \\
SDSSJ121923.43+063819.7  & $0.124$ & PG1216+069     & OVI      & $14.18_{-0.03}^{+0.04}$ & $36.8_{-4.3}^{+5.3}$ & $229.6_{-3.6}^{+3.2}$ \\
SDSSJ121923.43+063819.7  & $0.124$ & PG1216+069     & OVI      & $13.68_{-0.29}^{+0.24}$ & $46.3_{-18.8}^{+16.3}$ & $441.0_{-18.8}^{+21.8}$ \\
SDSSJ121923.43+063819.7  & $0.124$ & PG1216+069     & OVI      & $13.97_{-0.20}^{+0.11}$ & $37.2_{-9.6}^{+11.6}$ & $499.5_{-7.4}^{+7.5}$ \\
NGC3413                  & $0.002$ & PG1048+342     & HI       & $<12.3$ & $...$ & $...$ \\
NGC3413                  & $0.002$ & PG1048+342     & OVI      & $<13.9$ & $...$ & $...$ \\
NGC3442                  & $0.006$ & PG1048+342     & HI       & $14.06_{-0.03}^{+0.02}$ & $31.2_{-2.3}^{+1.8}$ & $86.2_{-1.9}^{+1.4}$ \\
NGC3442                  & $0.006$ & PG1048+342     & HI       & $14.69_{-0.10}^{+0.13}$ & $47.3_{-7.2}^{+10.2}$ & $216.3_{-6.6}^{+7.2}$ \\
NGC3442                  & $0.006$ & PG1048+342     & HI       & $13.89_{-0.35}^{+0.30}$ & $40.9_{-15.6}^{+25.4}$ & $305.9_{-23.2}^{+14.5}$ \\
NGC3442                  & $0.006$ & PG1048+342     & HI       & $13.97_{-0.07}^{+0.05}$ & $86.6_{-9.2}^{+8.8}$ & $410.3_{-8.0}^{+11.3}$ \\
NGC3442                  & $0.006$ & PG1048+342     & OVI      & $<14.4$ & $...$ & $...$ \\
UGC08839                 & $0.003$ & PG1352+183     & HI       & $13.08_{-0.08}^{+0.07}$ & $48.2_{-10.7}^{+11.5}$ & $-66.5_{-6.4}^{+8.1}$ \\
UGC08839                 & $0.003$ & PG1352+183     & OVI      & $<14.1$ & $...$ & $...$ \\
NGC6307                  & $0.010$ & 3C351          & HI       & $<12.8$ & $...$ & $...$ \\
NGC6307                  & $0.010$ & 3C351          & OVI      & $<13.8$ & $...$ & $...$ \\
J170423.43+604301.9      & $0.070$ & 3C351          & HI       & $13.70_{-0.04}^{+0.03}$ & $62.6_{-5.9}^{+5.5}$ & $-30.8_{-3.0}^{+3.7}$ \\
J170423.43+604301.9      & $0.070$ & 3C351          & OVI      & $<13.9$ & $...$ & $...$ \\
J170441.3+60443035       & $0.071$ & 3C351          & HI       & $14.08_{-0.10}^{+0.10}$ & $34.3_{-6.5}^{+6.7}$ & $121.5_{-2.8}^{+2.8}$ \\
J170441.3+60443035       & $0.071$ & 3C351          & HI       & $13.91_{-0.22}^{+0.12}$ & $79.2_{-13.6}^{+7.9}$ & $151.9_{-12.2}^{+27.4}$ \\
J170441.3+60443035       & $0.071$ & 3C351          & OVI      & $<14.5$ & $...$ & $...$ \\
J170422.99+604332.4      & $0.086$ & 3C351          & HI       & $12.87_{-0.14}^{+0.11}$ & $28.0_{-8.2}^{+9.7}$ & $-277.6_{-7.3}^{+6.5}$ \\
J170422.99+604332.4      & $0.086$ & 3C351          & OVI      & $<13.3$ & $...$ & $...$ \\
SDSSJ170434.33+604447.3  & $0.092$ & 3C351          & HI       & $14.25_{-0.09}^{+0.07}$ & $45.0_{-4.8}^{+4.8}$ & $-156.5_{-4.1}^{+3.5}$ \\
SDSSJ170434.33+604447.3  & $0.092$ & 3C351          & HI       & $13.94_{-0.20}^{+0.11}$ & $84.8_{-9.1}^{+8.2}$ & $-114.4_{-10.1}^{+23.9}$ \\
SDSSJ170434.33+604447.3  & $0.092$ & 3C351          & OVI      & $13.59_{-0.11}^{+0.09}$ & $30.2_{-4.8}^{+5.9}$ & $-133.6_{-5.3}^{+4.7}$ \\
J170420.25+604403.7      & $0.098$ & 3C351          & HI       & $13.69_{-0.17}^{+0.15}$ & $37.1_{-12.5}^{+20.2}$ & $95.0_{-5.8}^{+8.3}$ \\
J170420.25+604403.7      & $0.098$ & 3C351          & OVI      & $<13.2$ & $...$ & $...$ \\
6dFGSgJ234020.6-545155   & $0.005$ & HE2336-5540    & HI       & $<12.8$ & $...$ & $...$ \\
6dFGSgJ234020.6-545155   & $0.005$ & HE2336-5540    & OVI      & $<13.7$ & $...$ & $...$ \\
J233914.158-552344.63    & $0.026$ & HE2336-5540    & OVI      & $<14.3$ & $...$ & $...$ \\
NGC3067                  & $0.005$ & 3C232          & HI       & $19.86_{-0.02}^{+0.03}$ & $...$ & $237.1_{-10.6}^{+9.0}$ \\
NGC3067                  & $0.005$ & 3C232          & OVI      & $14.87_{-0.16}^{+0.15}$ & $175.1_{-51.6}^{+46.4}$ & $262.8_{-69.4}^{+50.9}$ \\
Mrk412                   & $0.015$ & 3C232          & HI       & $13.12_{-0.12}^{+0.12}$ & $<29.5$ & $78.7_{-3.9}^{+3.0}$ \\
Mrk412                   & $0.015$ & 3C232          & HI       & $13.22_{-0.11}^{+0.10}$ & $55.7_{-16.5}^{+19.8}$ & $266.3_{-12.0}^{+11.1}$ \\
Mrk412                   & $0.015$ & 3C232          & OVI      & $<14.1$ & $...$ & $...$ \\
SDSSJ095821.73+322551.9  & $0.080$ & 3C232          & HI       & $13.82_{-0.04}^{+0.03}$ & $56.3_{-6.0}^{+7.0}$ & $163.2_{-4.2}^{+3.7}$ \\
SDSSJ095821.73+322551.9  & $0.080$ & 3C232          & HI       & $12.95_{-0.17}^{+0.13}$ & $<69.9$ & $280.0$ \\
SDSSJ095821.73+322551.9  & $0.080$ & 3C232          & OVI      & $<14.2$ & $...$ & $...$ \\
J154530.3+4846093        & $0.038$ & PG1543+489     & HI       & $<12.3$ & $...$ & $...$ \\
J154530.3+4846093        & $0.038$ & PG1543+489     & OVI      & $<14.0$ & $...$ & $...$ \\
SDSSJ154527.12+484642.2  & $0.075$ & PG1543+489     & HI       & $19.19_{-0.01}^{+0.01}$ & $27.5_{-1.8}^{+1.5}$ & $-22.6_{-2.2}^{+2.3}$ \\
SDSSJ154527.12+484642.2  & $0.075$ & PG1543+489     & OVI      & $<14.0$ & $...$ & $...$ \\
SDSSJ154535.86+484814.0  & $0.097$ & PG1543+489     & HI       & $<12.8$ & $...$ & $...$ \\
SDSSJ154535.86+484814.0  & $0.097$ & PG1543+489     & OVI      & $<13.8$ & $...$ & $...$ \\
IC4213                   & $0.003$ & PG1309+355     & HI       & $13.07_{-0.44}^{+0.34}$ & $<49.9$ & $26.6_{-9.5}^{+14.3}$ \\
IC4213                   & $0.003$ & PG1309+355     & HI       & $13.94_{-0.14}^{+0.12}$ & $62.0_{-12.7}^{+18.1}$ & $99.0_{-18.3}^{+18.8}$ \\
IC4213                   & $0.003$ & PG1309+355     & HI       & $14.41_{-0.09}^{+0.10}$ & $45.8_{-8.7}^{+11.6}$ & $200.9_{-5.2}^{+5.2}$ \\
IC4213                   & $0.003$ & PG1309+355     & HI       & $13.95_{-0.12}^{+0.09}$ & $82.8_{-19.6}^{+12.0}$ & $308.1_{-17.6}^{+17.9}$ \\
IC4213                   & $0.003$ & PG1309+355     & HI       & $13.63_{-0.10}^{+0.07}$ & $37.2_{-6.9}^{+7.2}$ & $432.5_{-4.2}^{+5.8}$ \\
IC4213                   & $0.003$ & PG1309+355     & OVI      & $<13.8$ & $...$ & $...$ \\
UGC08318                 & $0.008$ & PG1309+355     & HI       & $<14.4$ & $...$ & $...$ \\
UGC08318                 & $0.008$ & PG1309+355     & OVI      & $<14.3$ & $...$ & $...$ \\
1XMMJ213758.7-143611     & $0.052$ & PKS2135-14     & HI       & $13.50_{-0.03}^{+0.02}$ & $35.3_{-3.0}^{+3.0}$ & $294.7_{-1.8}^{+2.1}$ \\
1XMMJ213758.7-143611     & $0.052$ & PKS2135-14     & OVI      & $<14.3$ & $...$ & $...$ \\
LEDA923368               & $0.075$ & PKS2135-14     & HI       & $13.04_{-0.05}^{+0.05}$ & $30.2_{-5.0}^{+5.7}$ & $-89.3_{-2.6}^{+4.0}$ \\
LEDA923368               & $0.075$ & PKS2135-14     & HI       & $13.92_{-0.01}^{+0.01}$ & $37.8_{-1.6}^{+1.8}$ & $89.5_{-0.8}^{+0.9}$ \\
LEDA923368               & $0.075$ & PKS2135-14     & HI       & $13.37_{-0.03}^{+0.03}$ & $37.7_{-2.8}^{+3.4}$ & $203.2_{-2.5}^{+1.9}$ \\
LEDA923368               & $0.075$ & PKS2135-14     & OVI      & $<13.3$ & $...$ & $...$ \\
SDSSJ083335.64+250847.1  & $0.008$ & PG0832+251     & HI       & $13.94_{-0.06}^{+0.09}$ & $24.0_{-3.0}^{+3.0}$ & $85.1_{-1.6}^{+1.6}$ \\
SDSSJ083335.64+250847.1  & $0.008$ & PG0832+251     & OVI      & $<14.2$ & $...$ & $...$ \\
LEDA1722581              & $0.017$ & PG0832+251     & HI       & $12.83_{-0.18}^{+0.17}$ & $<45.2$ & $-86.1_{-6.2}^{+8.8}$ \\
LEDA1722581              & $0.017$ & PG0832+251     & HI       & $>14.1$ & $82.7_{-15.0}^{+10.5}$ & $128.0_{-45.2}^{+36.1}$ \\
LEDA1722581              & $0.017$ & PG0832+251     & HI       & $>14.9$ & $67.0_{-8.1}^{+7.4}$ & $267.4_{-32.8}^{+29.7}$ \\
LEDA1722581              & $0.017$ & PG0832+251     & OVI      & $14.56_{-0.12}^{+0.11}$ & $46.6_{-17.0}^{+14.8}$ & $-0.6_{-10.6}^{+8.8}$ \\
LEDA1722581              & $0.017$ & PG0832+251     & OVI      & $14.88_{-0.11}^{+0.17}$ & $51.5_{-9.0}^{+10.9}$ & $166.4_{-6.5}^{+6.4}$ \\
2MASXJ08360739+2506457   & $0.023$ & PG0832+251     & HI       & $13.58_{-0.07}^{+0.06}$ & $73.1_{-11.4}^{+15.3}$ & $83.8_{-8.2}^{+9.2}$ \\
2MASXJ08360739+2506457   & $0.023$ & PG0832+251     & OVI      & $<14.3$ & $...$ & $...$ \\
SDSSJ083534.75+245901.9  & $0.108$ & PG0832+251     & HI       & $14.68_{-0.26}^{+0.29}$ & $16.9_{-2.9}^{+3.7}$ & $-79.9_{-3.7}^{+3.7}$ \\
SDSSJ083534.75+245901.9  & $0.108$ & PG0832+251     & HI       & $15.34_{-0.09}^{+0.09}$ & $59.3_{-3.4}^{+3.8}$ & $61.8_{-4.8}^{+4.2}$ \\
SDSSJ083534.75+245901.9  & $0.108$ & PG0832+251     & OVI      & $<13.9$ & $...$ & $...$ \\
UGC04527                 & $0.002$ & PG0838+770     & HI       & $16.21_{-0.33}^{+0.36}$ & $20.8_{-1.9}^{+2.0}$ & $93.5_{-3.5}^{+2.6}$ \\
UGC04527                 & $0.002$ & PG0838+770     & HI       & $14.54_{-0.14}^{+0.17}$ & $32.9_{-2.7}^{+2.8}$ & $119.2$ \\
UGC04527                 & $0.002$ & PG0838+770     & OVI      & $13.83_{-0.13}^{+0.11}$ & $47.7_{-13.3}^{+18.6}$ & $114.0_{-9.9}^{+10.1}$ \\
J085050.871+771540.17    & $0.004$ & PG0838+770     & HI       & $13.02_{-0.07}^{+0.06}$ & $41.4_{-8.1}^{+11.4}$ & $136.3_{-5.5}^{+4.8}$ \\
J085050.871+771540.17    & $0.004$ & PG0838+770     & HI       & $13.16_{-0.04}^{+0.04}$ & $28.0_{-3.2}^{+4.8}$ & $311.4_{-2.3}^{+2.1}$ \\
J085050.871+771540.17    & $0.004$ & PG0838+770     & OVI      & $<13.8$ & $...$ & $...$ \\
J084027.434+770555.60    & $0.007$ & PG0838+770     & HI       & $<12.4$ & $...$ & $...$ \\
J084027.434+770555.60    & $0.007$ & PG0838+770     & OVI      & $<13.7$ & $...$ & $...$ \\
LEDA24388                & $0.007$ & PG0838+770     & HI       & $13.16_{-0.05}^{+0.05}$ & $47.9_{-7.6}^{+8.5}$ & $92.4_{-4.7}^{+5.2}$ \\
LEDA24388                & $0.007$ & PG0838+770     & OVI      & $<13.2$ & $...$ & $...$ \\
SDSSJ101930.79+640708.4  & $0.006$ & Mrk141         & HI       & $<14.7$ & $...$ & $...$ \\
SDSSJ101930.79+640708.4  & $0.006$ & Mrk141         & OVI      & $<13.8$ & $...$ & $...$ \\
IC4889                   & $0.009$ & LEDA63618      & HI       & $14.43_{-0.11}^{+0.09}$ & $55.0_{-18.9}^{+14.6}$ & $50.4_{-11.2}^{+8.4}$ \\
IC4889                   & $0.009$ & LEDA63618      & HI       & $15.51_{-0.12}^{+0.17}$ & $74.4_{-10.5}^{+10.4}$ & $236.4_{-5.5}^{+5.2}$ \\
IC4889                   & $0.009$ & LEDA63618      & OVI      & $<13.4$ & $...$ & $...$ \\
J233914.158-552344.63    & $0.026$ & HE2336-5540    & HI       & $20.60_{-0.05}^{+0.05}$ & $...$ & $0.0_{-1.0}^{+1.0}$
\enddata
    \end{deluxetable*}



\end{document}